\documentclass[12pt]{article}

\usepackage[utf8]{inputenc}
\usepackage[T1]{fontenc}
\usepackage{tgtermes}
\usepackage{lmodern}

\usepackage{amssymb}

\usepackage{scalefnt,letltxmacro}
\LetLtxMacro{\oldtextsc}{\textsc}
\renewcommand{\textsc}[1]{\oldtextsc{\scalefont{1.10}#1}}
\usepackage[ttdefault=true]{AnonymousPro}

\usepackage{array}
\usepackage{amsmath}
\usepackage{bm}

\usepackage[
  paper  = letterpaper,
  left   = 1.25in,
  right  = 1.25in,
  top    = 1.0in,
  bottom = 1.0in,
  ]{geometry}
\usepackage[english]{babel}
\usepackage[parfill]{parskip}
\usepackage{afterpage}
\usepackage{enumitem}
\usepackage{framed}
\usepackage{xspace}

\usepackage{lineno}

\usepackage{ragged2e}

\newcounter{parcount}

\usepackage[usenames,dvipsnames]{xcolor}
\definecolor{shadecolor}{gray}{0.9}

\renewcommand{\labelitemi}{{\color{black!67}\textbullet}}

\usepackage{caption}
\usepackage{graphicx}
\usepackage[labelfont=bf]{caption}
\usepackage[format=hang]{subcaption}

\usepackage{booktabs, array}

\usepackage{listings}
\usepackage{fancyvrb}
\fvset{fontsize=\small}

\usepackage[numbers,sort]{natbib}

\usepackage[colorlinks,linktoc=all]{hyperref}
\usepackage[all]{hypcap}
\hypersetup{citecolor=MidnightBlue}
\hypersetup{linkcolor=MidnightBlue}
\hypersetup{urlcolor=MidnightBlue}
\usepackage[nameinlink]{cleveref}
\creflabelformat{equation}{#2\textup{#1}#3}  

\usepackage{algorithm,algpseudocode}

\usepackage
[acronym,smallcaps,nowarn,section,nogroupskip,nonumberlist]{glossaries}
\glsdisablehyper{}

\lstdefinestyle{mystyle}{
    commentstyle=\color{OliveGreen},
    numberstyle=\tiny\color{black!60},
    stringstyle=\color{BrickRed},
    basicstyle=\ttfamily\scriptsize,
    breakatwhitespace=false,
    breaklines=true,
    captionpos=b,
    keepspaces=true,
    numbers=none,
    numbersep=5pt,
    showspaces=false,
    showstringspaces=false,
    showtabs=false,
    tabsize=2
}
\usepackage{tikz}
\usepackage{pgfplots}
\usepackage{pgfplotstable}

\usepgfplotslibrary{external}
\pgfplotsset{compat=newest}
\pgfplotsset{
  cycle list={CornflowerBlue\\Dandelion\\ForestGreen\\BrickRed\\},
}

\newacronym{VI}{vi}{variational inference}
\newacronym{KL}{kl}{Kullback-Leibler}
\newacronym{ELBO}{elbo}{\emph{evidence lower bound}}
\newacronym{MCMC}{mcmc}{Markov chain Monte Carlo}

\usepackage{tikz}

\usetikzlibrary{bayesnet} 
\usetikzlibrary{external}

\pgfdeclarelayer{edgelayer}
\pgfdeclarelayer{nodelayer}
\pgfsetlayers{edgelayer,nodelayer,main}

\definecolor{hexcolor0xbfbfbf}{rgb}{0.749,0.749,0.749}

\tikzset{>=latex}
\tikzstyle{none}   = [inner sep=0pt]
\tikzstyle{line}   = [ -, thick, shorten <=1pt, shorten >=1pt ]
\tikzstyle{arrow}  = [ ->, thick, shorten <=1pt, shorten >=1pt ]
\tikzstyle{ardash} = [ dashed, ->, thick, shorten <=1pt, shorten >=1pt ]

\tikzstyle{empty}=[circle,opacity=0.0,text opacity=1.0,inner sep=0pt]
\tikzstyle{box}=[rectangle,fill=White,draw=Black]
\tikzstyle{filled}=[circle,thick,fill=hexcolor0xbfbfbf,draw=Black]
\tikzstyle{hollow}=[circle,thick,fill=White,draw=Black]
\tikzstyle{param}=[rectangle,fill=Black,draw=Black,inner sep=0pt,minimum width=4pt,minimum height=4pt]
\tikzstyle{paramhollow}=[rectangle,thick,fill=White,draw=Black,inner sep=0pt,minimum
width=4pt,minimum height=4pt]

\usepackage{pgfplots}                               
\pgfplotsset{compat=newest}
\pgfplotsset{plot coordinates/math parser=false}
\usepgfplotslibrary{statistics}
\usepgfplotslibrary{colorbrewer}                    
\newlength\figureheight
\newlength\figurewidth
\newlength\figureheightsmall
\newlength\figurewidthsmall
\definecolor{POSTcolor}{rgb}{0.48, 0.20, 0.58} 
\definecolor{Qcolor}{rgb}{0.00, 0.53, 0.22} 

\usepackage{array}
\usepackage{hyperref}

\usepackage{pgfplots}
\usetikzlibrary{fit,positioning,arrows,automata,calc}
\usepackage{verbatim}
\usepackage{placeins}
\usepackage{titlesec}

\usepackage{graphicx}
\usepackage{caption}
\usepackage{cleveref}

\newcommand{\be}{\begin{equation}}
\newcommand{\ee}{\end{equation}}
\newcommand{\ec}{\end{center}}
\newcommand{\bc}{\begin{center}}
\newcommand{\eea}{\end{eqnarray}}
\newcommand{\bea}{\begin{eqnarray}}
\newcommand{\bd}{\begin{description}}
\newcommand{\ed}{\end{description}}
\newcommand{\bi}{\begin{itemize}}
\newcommand{\ei}{\end{itemize}}
\newcommand{\pa}{\partial}
\newcommand{\bs}{\boldsymbol}
\def\RR{ \mathbb R}
\newcommand{\refeq}[1]{Equation (\ref{#1})}

\newcommand{\by}{\bs{y}}
\newcommand{\bY}{\bm{Y}}

\newcommand{\bV}{\bs{V}}
\newcommand{\bv}{\bs{v}}
\newcommand{\bx}{\bs{x}}

\newcommand{\eqq}{\mathbb{E}_{q(\by;\bm\phi)}}
\newcommand{\err}{\mathbb{E}_{r(\by;\bm\xi)}}

\title{%
\textbf{Beyond black-boxes in Bayesian inverse problems and model validation:
applications in solid mechanics of elastography}
}

\author{
Lukas Bruder \\
Technical University of Munich \\
\\
Phaedon-Stelios Koutsourelakis \\
Technical University of Munich \\
}

\begin{document}

\maketitle
\bigskip

\begin{abstract}
 The present paper is motivated by one of the most fundamental challenges in  inverse problems, that of quantifying model discrepancies and errors. While significant strides have been made in calibrating model parameters, the overwhelming majority of pertinent methods is based on the assumption of a perfect model. Motivated by problems in solid mechanics which, as all problems in continuum thermodynamics, are described by conservation laws and phenomenological constitutive closures, we argue that in order to quantify model uncertainty in a physically meaningful manner, one should break open the black-box forward model. In particular we  propose formulating an undirected probabilistic model that explicitly accounts for the governing equations and their validity. This recasts the solution of both forward and inverse problems as probabilistic inference tasks where the problem's state variables should not only be compatible with the data but also with the governing equations as well. 
Even though the probability densities involved do not contain any black-box terms, they live in much higher-dimensional spaces. In combination with the intractability of the  normalization constant of the undirected model employed, this poses significant challenges which we propose to address with a linearly-scaling, double-layer of Stochastic Variational Inference. We demonstrate the capabilities and efficacy of the proposed model in synthetic forward and inverse problems (with and without model error)  in elastography.

\end{abstract}


\clearpage

\tableofcontents
\clearpage


\section{Introduction}
\label{sec:introduction}

The extensive use of large-scale computational models  poses several challenges in {\bf model 
calibration and validation} \cite{biegler_large-scale_2010,cullen_large_2013}. Traditionally in 
numerical simulations, the emphasis  has been placed on decreasing the truncation/discretization  
errors. Nevertheless, the fidelity of the  predictions of such simulations depends strongly on 
assigning proper values to the  model parameters  as well as utilizing high-fidelity  models. This in turn 
necessitates a data-driven approach where elaborate computational models are fused with data, 
 originating either from experiments/measurements or from  models of 
higher 
fidelity (e.g. molecular dynamics). This process is naturally fraught with 
significant 
uncertainties.
One such source is obviously the noise in the data which constitutes probabilistic estimates more 
rational. This is particularly important when multiple hypotheses are consistent with the data or 
the level of confidence in the estimates produced needs to be quantified. Another source of 
uncertainty, which is largely  unaccounted for, is {\bf \em model  uncertainty} 
\cite{ohagan_uncertainty_1999,draper_assessment_1995,higdon_combining_2004}. Namely, the parameters 
which are calibrated are associated with a particular forward model    (in our case a system of 
(discretized) PDEs consisting of conservation laws  and constitutive equations or closures) but one 
cannot be certain about the validity of the model employed. In general, there will be deviations 
between the physical reality where measurements are made, and the idealized 
mathematical/computational description.

An application that motivates this work comes from biomechanics and the  
identification of the mechanical properties of  biological materials, in the context of 
non-invasive  medical diagnosis (\textbf{elastography}).
While in certain cases mechanical properties can also be 
measured directly by excising multiple tissue samples, non-invasive procedures offer obvious 
advantages in terms of ease, cost and reducing the risk of complications to the patient.  Rather 
than x-ray techniques which capture variations in density, the identification of stiffness, or 
mechanical properties in general, can potentially  lead to earlier and more accurate diagnosis 
\cite{ISI:000263259100006,liver2006},  provide valuable insights that  differentiate between 
modalities of the same pathology \cite{cur12gen}, monitor the progress of treatments and  
ultimately lead to patient-specific treatment strategies.

All elastographic  techniques consist of the following three basic steps 
\cite{doyley_model-based_2012} : 1) excite the tissue using a (quasi-)static, harmonic or transient 
source, 2) (indirectly) measure tissue deformation (e.g. displacements, velocities)  using an 
imaging technique such as ultrasound  \cite{Ophir:1991}, magnetic resonance \cite{mut95mag} or 
optical tomography \cite{khalil_tissue_2005}, and
 3) infer the mechanical properties from this data using a suitable continuum mechanical model of 
the tissue's deformation.  {\em Indirect or iterative or model-based} methods  for solving the 
latter problem (in contrast 
to direct methods \cite{ISI:000275699600006}) admit an inverse problem formulation where the 
discrepancy 
\cite{ISI:000240849100006}) 
 between observed and model-predicted deformations is minimized with respect to the material 
fields of interest 
\cite{ISI:000223500200013,doyley_enhancing_2006,ISI:000275756200016,ISI:000280774700004,
doyley_model-based_2012}. 
 While these approaches utilize directly the raw data, they generally imply an increased 
computational cost as the forward problem, and potentially parametric derivatives, have to be 
solved/computed 
several times. This effort is amplified when stochastic/statistical formulations are employed as those arising 
from the Bayesian paradigm.

The solution of such  model calibration (and validation) problems in the Bayesian framework is 
hampered by two main difficulties. The first pertains to their computational efficiency and stems 
from the poor scaling of traditional Bayesian inference tools  with respect to the dimensionality 
of the unknown parameter vector - another instance of the {\em curse-of-dimensionality}. In 
elastography,  the model parameters of interest (i.e. material 
properties) exhibit 
spatial variability which requires fine-discretization  in order to be captured. This variability 
can also span different scales \cite{koutsourelakis_multi-resolution_2009,ellam_bayesian_2016}. 
Standard Markov Chain Monte Carlo (MCMC, \cite{green_bayesian_2015}) techniques require an 
exorbitant number of likelihood evaluations (i.e. solutions of the forward model) in order to reach 
convergence 
\cite{roberts_exponential_1996,roberts_optimal_1998,mattingly_diffusion_2012,pillai_optimal_2012}. 
As each of these calls implies the solution of very large systems of (non)linear and potentially 
transient,  equations, it is critical to minimize their number  particularly in 
time-sensitive applications. 
Advanced sampling schemes, involving adaptive MCMC 
\cite{lee_markov_2002,holloman_multi-resolution_2006,chopin_free_2012} and Sequential Monte Carlo 
(SMC, \cite{moral_sequential_2006,koutsourelakis_multi-resolution_2009,del_moral_adaptive_2012}), 
exploit the physical  insight and the use of multi-fidelity solvers in order to expedite the 
inference process. Nevertheless, they fail to address fundamental challenges as the number of 
forward calls can still be in the order of tens of thousands.
Several attempts have also been directed towards using emulators, surrogates or reduced-order 
models of various kinds 
\cite{marzouk_stochastic_2007,bui-thanh_model_2008,rosic_sampling-free_2012,bilionis_solution_2014, dodwell_hierarchical_2015,
lan_emulation_2016} but such a task is severely hindered by the input 
dimensionality. 
The use of first and second-order derivatives has also been advocated either in a standard MCMC 
format or by developing advanced sampling strategies. These are generally available by solving 
appropriate {\em adjoint problems}  which are well-understood in the context of deterministic 
formulations 
\cite{flath_fast_2011,girolami_riemann_2011,martin_stochastic_2012,bui-thanh_solving_2014,
petra_computational_2014}. 
More recent treatments, attempt to exploit the (potentially) lower intrinsic dimensionality
 of the target posterior by 
identifying subspaces where either most of the probability mass is contained 
\cite{franck_sparse_2016} or where maximal sensitivity is observed 
\cite{cui_likelihood-informed_2014,spantini_optimal_2015,cui_scalable_2015,
cui_dimension-independent_2016}. This enables inference tasks to be performed on much 
lower dimensions which are not hampered by the aforementioned difficulties. 
Generally all such 
schemes construct such approximations around the MAP point by employing local information (e.g.  
gradients) and therefore are not suitable for multi-modal or highly non-Gaussian  posteriors.

The second challenge facing Bayesian inverse problems pertains to the model structure  itself. 
This is especially important in the context of biomechanics applications where 
inferring  model 
parameters associated with an incorrect model  can lead to incorrect  
diagnosis and prognosis\footnote{"I remember my friend Johnny von Neumann used to say, "With four 
parameters I can fit an elephant and with five I can make him wiggle his trunk." {\em A meeting 
with Enrico Fermi}, Nature 427, 297; 2004.}.
Despite progress in Bayesian model  calibration tasks \cite{higdon_combining_2004}, the issue of 
model validation still poses many open questions, which, if not adequately addressed, can lead to  
biased and over-confident parameter estimates
and predictions \cite{brynjarsdottir_learning_2014}.  Most efforts have been directed towards  
providing quantitative comparisons between competing models. In the latter context, Bayes factors 
\cite{kass_bayes_1995} provide a rigorous means of model comparison. Nevertheless, apart from the  
computational difficulties  involved, they do not  reveal the key driver behind model uncertainty 
nor do they quantify its effect on the  predictions of the model.
 Alternatively, the most widely-adopted statistical strategy 
involves augmenting the relation between data and model predictions with a Gaussian process, which 
accounts in an additive fashion, for the model error's contribution 
\cite{kennedy_bayesian_2001,higdon_computer_2008}. 
However, this explicit additive term may violate physical constraints (e.g. conservation  of mass, 
energy), can get entangled with the measurement error, is not   physically interpretable and 
cumbersome or impractical to infer when it depends on a large number of input parameters 
\cite{bayarri_framework_2007,berliner_modeling_2008,koutsourelakis_novel_2012,strong_when_2014,
sargsyan_statistical_2015}.

\begin{figure}[!t]
\centering
  \begin{minipage}[b]{0.45\textwidth}
   \hspace{-1cm} \vspace{1cm}
   \includegraphics[width=8.cm,height=4cm]{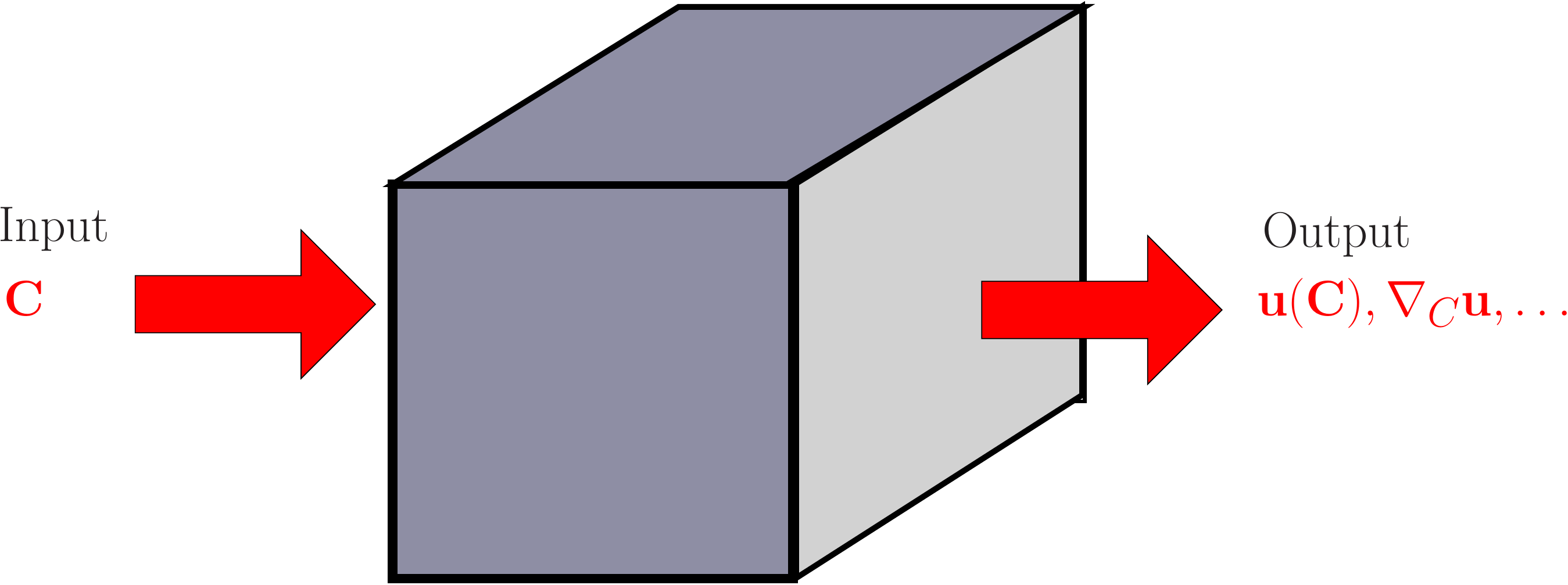}
   \caption{Black-box setting for Bayesian model  calibration.}
   \label{fig:bb}
  \end{minipage}
\hfill
 \begin{minipage}[b]{0.45\textwidth}
   \includegraphics[width=8cm,height=5.5cm]{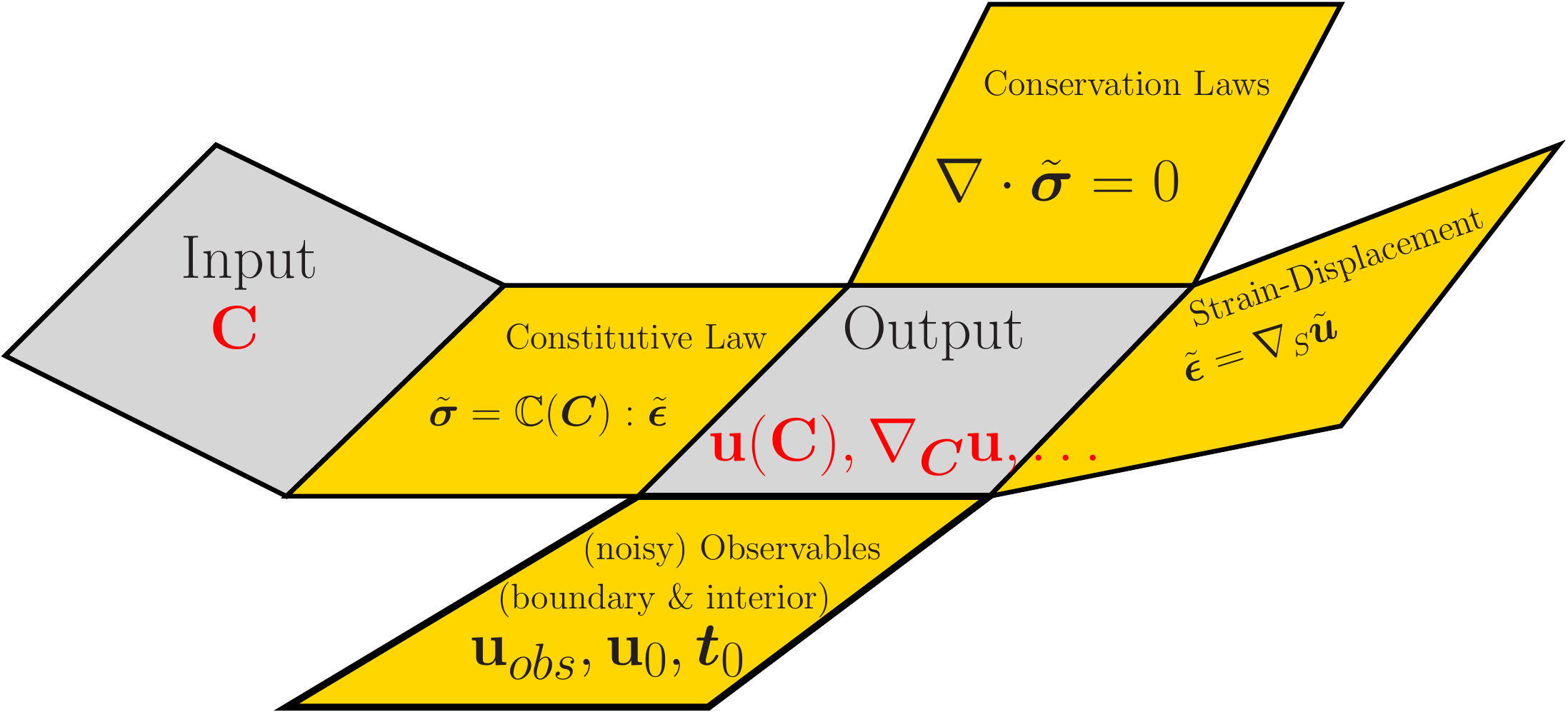}
   \caption{Proposed framework of {\em unfolding} the black-box and revealing {\em all} model 
equations.}
   \label{fig:ubb}
  \end{minipage}
%
\end{figure}

The present paper extends previous work \cite{koutsourelakis_novel_2012} towards developing  a novel modeling framework and a set of scalable  
algorithms that will address the two main challenges in model calibration and validation in 
PDE-based models, i.e. a) the significant computational cost in problems with an expensive, black-box forward model, and b) the quantification of structural, model uncertainty and 
its effect on model calibration  and predictive estimates.
This paper advocates a new paradigm that goes beyond the standard black-box setting 
(Figure \ref{fig:bb}) employed thus far. Such a shift is necessitated by the need to quantify model 
discrepancies in a physically relevant manner. It is based on bringing all model equations in the 
forefront and, in a  Bayesian fashion,   quantifying their relative validity using  the language of 
 probabilities. 
 As we show in the next sections, this recasts the solution of, even deterministic, forward problems as problems of probabilistic inference. More importantly, in the context of inverse problems, it  leads to an augmented posterior that involves all of the state  
variables of the forward model. Nevertheless, it untangles their complex relations into local terms 
(Figure  \ref{fig:ubb}) and yields  a well-posed inverse problem, even in cases when the forward problem is not  (e.g. due to incomplete boundary conditions).
We demonstrate the potential of this framework  in the context of biomechanics where the solution of the 
aforementioned issues can  significantly impact  progress in the non-invasive, diagnostic 
capabilities and assist in the development of patient-specific treatment strategies. 

The paper is organized as follows. In~\cref{sec:probabilistic_cm}, we present the proposed probabilistic framework in the context of  linear elastostatics. We introduce an undirected probabilistic model that  encapsulates all model equations  and  discuss the recasting of forward and inverse problems as probabilistic inference tasks. In   \cref{sec:variational_approx}, we present an efficient computational framework for carrying out the aforementioned inference tasks  based on Stochastic Variational Inference. The scheme  introduced employs a twofold variational approximation which is capable of dealing with the intractable normalization constant in the model-informed prior density.
 Lastly, the feasibility of the framework is demonstrated by several numerical examples in~\cref{sec:examples}.

\clearpage

\section{Proposed modeling framework}\label{sec:probabilistic_cm}

  We begin the elaboration of the  proposed model in the context of linear, elliptic PDEs as those arising in small deformation, linear elasticity. Similar models appear for example in heat diffusion or Darcy flow and can be treated similarly even though the physical meaning of the state variables is different. Furthermore, the extension to nonlinear elliptic PDEs as those for example appearing in the context of  large deformation, nonlinear elasticity (geometric and 
constitutive nonlinearities) requires some technical alterations which are not discussed. Further extensions to time-dependent, nonlinear models, which are also of interest (e.g. harmonic and transient elastography), are deferred to a  future paper.   With regards to the notations adopted we generally adhere to the following rules:
\bi
\item with  boldface, we denote vectors/vector fields.
\item with $~~\tilde{}~~$ we denote  scalar/vector fields.
\item with upper-case, we denote  random vectors/variables.
\item with lower-case, we denote values taken by vectors/variables.
\ei

Almost all   problems  in continuum thermodynamics  share a common structure which consists of:

\begin{list}{\labelitemi}{\leftmargin=1.1em \itemsep=2mm}

\item[1)]  \textbf{\em conservation law(s)} that arise from physical principles and are generally well-founded and trusted. In the case of biomechanics this amounts to the conservation of linear momentum:
\be
\nabla \cdot \tilde{\bm\sigma}(\bm{x})=0 \quad \forall \bs{x} \in \Omega, \quad 
\label{eq:conservation}
\ee
where $\tilde{\bm\sigma}(\bs{x})$ is the Cauchy  stress tensor at a point $\bs{x}$ in the problem domain $\Omega$.
Conservation laws serve as the skeleton  in many other problems i.e. the  conservation of mass/energy in the  case of diffusion/advection of mass or heat  flow through a (porous) medium. Discretized versions of the aforementioned PDEs are employed, which naturally introduce \textbf{\em discretization error}.

\item[2)]   \textbf{\em equation(s) of state/constitutive law(s)/closure(s)}. Constitutive equations  refer in general to relations between conjugate thermodynamic variables, such as   the stress and strain  tensors in bio/solid mechanics. In linear elasticity such relations can be  expressed as:
\be
 \tilde{\bm\sigma}(\bm{x})= \mathbb{C}(\bm{x}) : \tilde{\bm\epsilon}(\bm{x}) \hspace{0.5cm} \forall \bm{x} \in \Omega,
\label{eq:constelast}
\ee
where $\mathbb{C}(\bm{x})$ is the elasticity tensor and  $\tilde{\bm\epsilon}(\bm{x})$ the inifnitesimal strain tensor, which relates to the displacements $\tilde{\bm{u}}(\bm{x})$ as follows:
\be
\tilde{\bm\epsilon}(\bm{x}) = \cfrac{1}{2} \left( \nabla \tilde{\bm{u}}(\bm{x}) + (\nabla \tilde{\bm{u}}(\bm{x}))^T \right) = \nabla_{S} \tilde{\bm{u}}(\bm{x}).
\label{eq:straindisp}
\ee
Similar constitutive relations have been established between velocity and pressure in flow through permeable media, or flux and temperature in heat diffusion, involving permeability or   conductivity tensors.
{\em In the context of elastography, our goal is to estimate $\mathbb{C}$ and its spatial variability.}
Constitutive models are largely phenomenological and therefore represent  the primary source of \textbf{\em model error/uncertainty}. 
\item[3)] \textbf{\em observables} in the form of boundary/interior conditions or initial conditions in time-dependent problems. In a typical, static {\em forward}  problem of biomechanics these consist of boundary displacements (Dirichlet boundary conditions) on the part of the boundary $\Gamma_{DBC} \subseteq \pa \Omega$:
\be
		\tilde{\bm{u}}(\bm{x}) = \bm{u}_0(\bm{x}) \hspace{0.5cm} \forall \bm{x} \in \Gamma_{DBC}
\label{eq:DBC_boundary_conditions}		
\ee
and, potentially also, traction (Neumann) boundary conditions   on the part of the boundary $\Gamma_{NBC} \subset \pa \Omega$:
\be
		\tilde{\bm\sigma}(\bm{x}) \cdot \bm{n}(\bm{x}) = \bm{t}_0(\bm{x}) \hspace{0.5cm} \forall \bm{x} \in \Gamma_{NBC},
		\label{eq:NBC_boundary_conditions}
\ee		
where $\bm{n}(\bm{x})$ denotes the unit outward normal vector. These boundary tractions/displacements might be specified deterministically or stochastically and the well-posedness of the {\em forward} problem necessitates that $\Gamma_{DBC} \cup \Gamma_{NBC}=\pa \Omega$ and  $\Gamma_{DBC} \cap \Gamma_{NBC}= \emptyset$. In the context of the inverse problems considered these are complemented by observations of interior displacements which we denoted by $\bs{u}_{obs}$. These are obtained by employing image processing techniques to the undeformed and deformed (e.g. ultrasound) images of the tissue (\cite{sarvazyan_elasticity_2011}).
In general, the observables $\bs{u}_{obs}$ (and potentially also $\bm{u}_0,\bm{t}_0$)  are   contaminated by noise and represent the primary source of  \textbf{\em observation error}.
\end{list}

Existing deterministic or stochastic (Bayesian) strategies for  the solution  of the associated inverse problem are based on a premise of a perfect model (up to discretization errors) and  a well-posed forward problem i.e. the specification of boundary conditions that ensure the existence/uniqueness of solution.  In  bio/solid mechanics, a forward solver is usually obtained, upon discretization of the governing equations,  in terms of displacements $\tilde{\bs{u}}(\bs{x}) \xrightarrow[discretize]{}\bs{u}$  and of the  constitutive parameters in $\mathbb{C}(\bm{x}) \xrightarrow[discretize]{}  \bm{C}$. 
Traditional Bayesian formulations \cite{kennedy_bayesian_2001,higdon_computer_2008} postulate a 
relation  for the   observed displacements $\bs{u}_{obs}$ of the form:
\be
\bs{u}_{obs}=\bs{u}(\bm{C})+\bs{\eta}, \quad \bs{\eta} \sim \mathcal{N}(\bs{0}, \sigma_{\eta}^2 \bs{I}),
\label{eq:ko}
\ee
where  $\bs{\eta}$ is the observation noise. Solution strategies treat the forward solver that  computes the parameter-to-state map $\bs{u}(\bm{C})$ as {\em a black-box} (Figure \ref{fig:bb}). Often,  gradients $\nabla_{\bm{C}} \bs{u}(\bm{C})$ are available through adjoint formulations \cite{oberai_evaluation_2004}. Nevertheless, the number of forward solutions required can become extremely large and scales poorly with the dimension of the unknowns $\bm{C}$ as discussed in the  introduction.   In addition to these deficiencies, such a formulation lacks the ability to quantify model errors and while estimates for $\bm{C}$ can always be obtained, these are obviously invalid if they are based on an incorrect model. The use of an additional term of the form $\bs{\delta}(\bm{C})$ or $\bs{\delta}(\bs{x})$ in \refeq{eq:ko} as in \cite{kennedy_bayesian_2001} may violate physical constraints, can get entangled with the measurement error, is not physically interpretable
and cumbersome or impracticable when  $\bm{C}$ is high-dimensional as in cases of practical interest \cite{moser_validation_2015}.
Furthermore, when uncertainty is present with regards to the boundary conditions, these must also be included in the vector of unknowns and be inferred from the data, which introduces additional difficulties.


In contrast, our strategy attempts to break open the black-box model 
(Figure \ref{fig:ubb})
 and bring to the surface all quantities/fields involved in the mathematical description of the 
physical process of deformation. Under this framework, the solution of both forward and inverse 
problems is recast as one  of statistical inference 
\cite{diaconis_bayesian_1988,hennig_probabilistic_2014,
hennig_probabilistic_2015,chkrebtii_bayesian_2016} and  we attempt to find  {\em all} latent, 
unobserved quantities that are compatible not only with the observables, but 
with the physics-based-model 
equations as well. Their reliability, or absence thereof, is expressed in terms of probabilities, 
allowing one to quantify all error sources such as  {\em discretization} and {\em structural, model} 
errors.

The formulation advocated consists of three pivotal components which  are discussed in the sequel, namely:
\bi
\item[1)] an augmented  {\em prior} and {\em posterior}  distribution dictated by the model equations  (section \ref{sec:augp}).
\item[2)] the representation (discretization) of the unknown state variables (section \ref{sec:rep}).
\item[3)] the solution  as probabilistic inference (section \ref{sec:inf}).
\ei

\subsection{ Augmented prior/posterior densities}
\label{sec:augp}

\begin{figure}
 \centering
 \includegraphics[width=.75\textwidth]{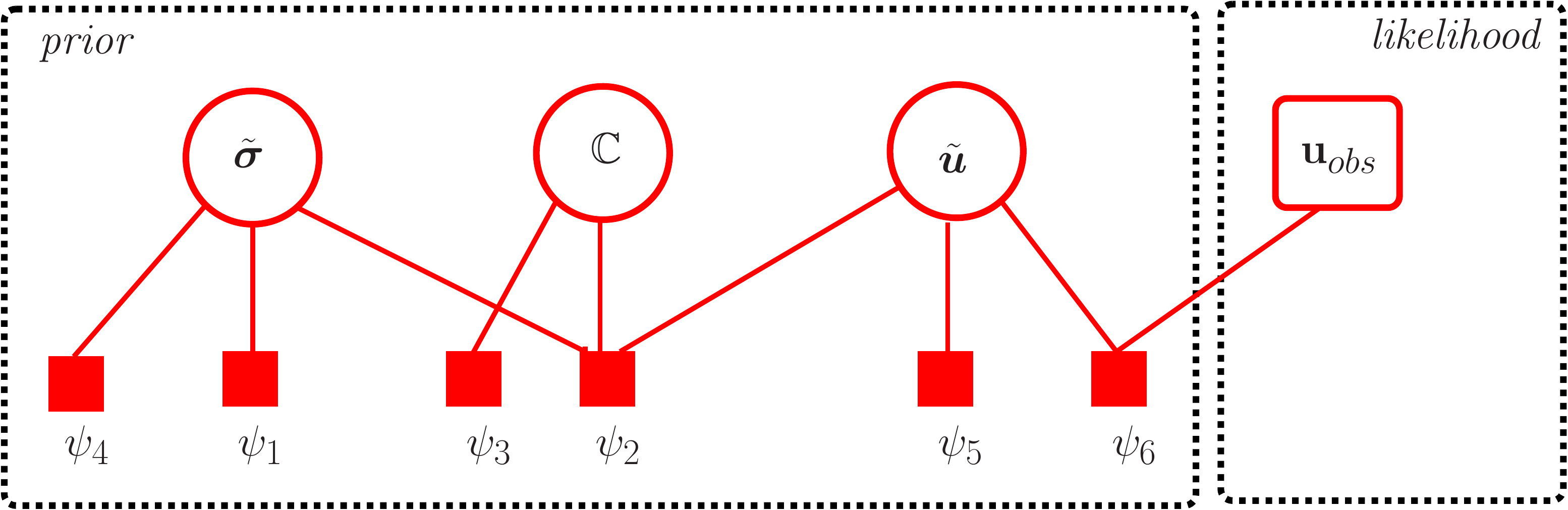}
  \caption{Factor graph exhibiting the augmented state space  in the case of 
linear elasticity. Round nodes represent variables, square nodes represent 
factors. There is an edge from each variable to every factor that mentions it.  
The factor potentials  $\psi_i$ arise from model equations (conservation \& constitutive  laws) as well as observations. 
}
  \label{fig:factor}
 \end{figure}

In contrast to existing Bayesian formulations which prescribe prior densities merely encapsulating beliefs  (in our case, the elastic material properties $\bs{C}$), we advocate a prior model that incorporates  the relations and dependencies between system states  as implied by the governing equations, i.e. conservation \& constitutive laws.
In particular, we view each of those equations as a {\em source of information} 
\footnote{It is due to this different perspective in the interpretation of solutions of 
PDEs that we refrain from the classical mathematical treatment of trial and weighting 
function spaces.}  and we employ a 
model interrogation scheme \cite{chkrebtii_bayesian_2016,cockayne_probabilistic_2016} in order to extract it.
For the {\em conservation of linear momentum} (\refeq{eq:conservation}) and without loss of generality, these interrogations can be performed using the Method of Weighted Residuals (MWR,  \cite{finlayson_method_1972}), according to which, for each weighting function $\bs{w}^{(i_e)}(\bs{x})$ such that $\bs{w}^{(i_e)}(\bs{x})=0$ on $\Gamma_{DBC}$, the weighted residual $r_e^{(i_e)} (\tilde{\bm\sigma}(\bm{x})  )$ is (using indicial notation):
\be
\begin{array}{ll}
r_e^{(i_e)} (\tilde{\bs{\sigma}}(\bx) )& =\int_{\Omega} \bs{w}^{(i_e)}_j~\tilde{\bm\sigma}_{ji,i} ~d V  \\
& = \int_{\Gamma_{NBC}} ~ \bs{w}^{(i_e)}_j \tilde{\bm\sigma}_{ji} \bm{n}_i~d\Gamma - \int_{\Omega} \bs{w}^{(i_e)}_{j,i} \tilde{\bm\sigma}_{ji}~d V   \\
& =  \int_{\Gamma_{NBC}} ~ \bs{w}^{(i_e)}_j \bm{t}_{0,j}~d\Gamma - \int_{\Omega} \bs{w}^{(i_e)}_{j,i} \tilde{\bm\sigma}_{ji}~d V  \quad \textrm{(from \refeq{eq:NBC_boundary_conditions})}.
\end{array}
\label{eq:rese}
\ee
\textbf{ Rather than setting each of these residuals $\{r_e^{(i_e)}\}_{i_e=1}^{N_e}$ to zero, we use 
them to induce   a probability measure on candidate stress fields $\tilde{\bm\sigma}(\bm{x})$}. Such 
measures should contract to a  Dirac measure  around the true, but unknown, stress field, as $N_e 
\to \infty$. Our strategy 
  resembles conceptually  the emerging field known as Probabilistic Numerics
 \cite{hennig_probabilistic_2014,cockayne_bayesian_2017}.
In particular, if $\epsilon$ is analogous to the numerical tolerance employed in deterministic schemes to enforce zero residuals, we can define a Gaussian probability density  $\mathcal{N}(0, \epsilon^2)$ for  $r_e^{(i_e)}$ which in turn implies a probability measure on the space of candidate solutions $\tilde{\bm\sigma}(\bs{x})$.
The combination of $N_e$ such relations yields a density for  $\tilde{\bm\sigma}(\bm{x})$ of the form \footnote{In the following, we abuse notation by defining probability densities on functions/fields in order to simplify the presentation. In computational implementations, these fields are discretized as discussed in section \ref{sec:variational_approx}.} (we omit dependence on $\epsilon$ for simplicity):
\be
\begin{array}{llr}
p_e(\tilde{\bm\sigma}(\bs{x}))  & \propto \psi_1(\tilde{\bm\sigma}(\bs{x}))  &  \textrm{(see Figure \ref{fig:factor})}\\
 & = \prod_{i_e=1}^{N_e} \mathcal{ N}(r_e^{(i_e)} (\tilde{\bm\sigma}(\bs{x}))~; 0,\epsilon^2). & 
\end{array}
\label{eq:priore}
\ee
This scores candidate stress fields $\tilde{\bm\sigma}(\bs{x})$ according to their satisfaction of the governing equation (in the MWR form). Fields that yield $0$ residuals get the highest score and,  the more weighting 
functions $N_e$ are used, the narrower becomes the region in the space of trial solutions where  the 
probability mass is concentrated. 
We note at this stage that the MWR considered is not at all restrictive as (at least) six methods (i.e. collocation, sub-domain, least-squares, (Petrov)-Galerkin, moments) can be considered as special cases by appropriate selection of the  $\bs{w}^{(i)}(\bs{x})$.

Enforcing any {\em reliable} model equation (e.g. strain-displacement relation, incompressibility) can also be done in a similar manner by artificially defining a probability measure as above which can degenerate to a Dirac measure as the precision parameter $\epsilon$ decays. 
This is particularly useful when it comes to the second piece in the puzzle which arises from the {\em constitutive equations} and requires the specification  of the strain tensor $\tilde{\bm\epsilon}(\bm{x})$. This can be indirectly expressed  by a probabilistic enforcement of the strain-displacements relation in \refeq{eq:straindisp} or directly  in terms of the displacement field $\tilde{\bm{u}}(\bx)$. 
If one enforces \refeq{eq:constelast} point-wise at $\{ \bs{x}^{(i_c)} \}_{i_c=1}^{N_c}$, then at each of those points, the residual $\bs{r}_c^{(i_c)}$:
\be
\begin{array}{ll}
\bs{r}_c^{(i_c)} \left( \tilde{\bm\sigma}(\bm{x}), \tilde{\bs{u}}(\bm{x}), \mathbb{C}( \bs{x}) \right)& = \tilde{\bm\sigma}( \bs{x}^{(i_c)})-\mathbb{C}( \bs{x}^{(i_c)}) : \nabla_S \tilde{\bs{u}}(\bs{x}^{(i_c)}),
\end{array}
\label{eq:resc}
\ee
which quantifies the discrepancy  between the {\em actual stresses} and the {\em constitutive model 
predicted stresses}, is zero only  in the case of a perfect model. As the validity of 
the model is unknown a priori, we propose a hierarchical prior of the form:
\be
p(\bs{r}_c^{(i_c)}| \Lambda_{i_c})=\mathcal{N}(\bs{0}, \Lambda_{i_c}^{-1} \bs{I}).
\label{eq:priorci}
\ee
The simplest such model utilizes a single hyper-parameter  $\Lambda_{i_c}$ per  point 
$\bs{x}^{(i_c)}$, which reflects the magnitude of the {\em constitutive model error}. Multivariate 
representations  as well as more complex  model-error distributions are also possible. Large  values of $\Lambda_{i_c}^{-1}$ indicate locations  where the model 
error is high and vice versa. We emphasize  that nonzero values of $\bs{r}_c^{(i_c)}$ do not arise 
due to variability in the material parameters $\mathbb{C}( \bs{x})$, but rather due to  \textbf{the inadequacy of 
the constitutive model to provide sufficient closures to the governing equations}. 
As discussed earlier,  \refeq{eq:priorci} employed over all enforcement points $N_c$ implies a (joint) probability density over the fields  $\tilde{\bm\sigma}(\bm{x}), \tilde{\bs{u}}(\bm{x}), \mathbb{C}( \bs{x})$ (see Figure \ref{fig:factor}):
\be
\begin{array}{ll}
p_c \left(\tilde{\bm\sigma}(\bm{x}), \tilde{\bs{u}}(\bm{x}), \mathbb{C}( \bs{x}) | ~\bs{\Lambda}  \right)   & \propto \psi_2 \left( \tilde{\bm\sigma}(\bm{x}), \tilde{\bs{u}}(\bm{x}), \mathbb{C}( \bs{x}) \right) , \quad \bs{\Lambda}=\{ \Lambda_{i_c} \}_{i_c=1}^{N_c}   \\
& = \prod_{i_c=1}^{N_c} \mathcal{N}( \bs{r}_c^{(i_c)} \left( \tilde{\bm\sigma}(\bm{x}), \tilde{\bs{u}}(\bm{x}), \mathbb{C}( \bs{x}) \right); ~~\bs{0}, \Lambda_{i_c}^{-1} \bs{I} ).
\end{array}
\label{eq:priorc}
\ee
The aforementioned densities can be straightforwardly  complemented by ``traditional'' priors employed in canonical Bayesian formulations \cite{bardsley_gaussian_2013}, e.g.:
\be
p_{C}(\mathbb{C}( \bs{x}) | \bm{\theta}_C) \propto \psi_3(\mathbb{C}( \bs{x}) ) \quad \textrm{(see Figure \ref{fig:factor})}
\label{eq:priormaterial}
\ee
for the unknown material property field $\mathbb{C}( \bs{x})$ with hyperparameters $\bm{\theta}_C$. 
Similarly, if other prior information is available about the other state variables $\tilde{\bm\sigma}(\bm{x}), \tilde{\bs{u}}(\bm{x})$, these can be incorporated e.g. as:
\be
p_{\sigma}(\tilde{\bm\sigma}(\bm{x}) | \bm{\theta}_{\sigma}) \propto \psi_4(\tilde{\bm\sigma}(\bm{x}) )
\label{eq:priorsigma}
\ee
and:
\be
p_{u}(\tilde{\bs{u}}(\bm{x}) | \bm{\theta}_{u}) \propto \psi_5(\tilde{\bs{u}}(\bm{x})).
\label{eq:prioru}
\ee
The combination of Equations (\ref{eq:priore}), (\ref{eq:priorc}) as well as (\ref{eq:priorc}), (\ref{eq:priorsigma}), (\ref{eq:prioru}),    define a 
joint {\rm prior} for the state variables of the following form (we omit hyperparameters $\bm{\theta}_{C},  \bm{\theta}_{\sigma},  \bm{\theta}_{u}$  for simplicity): 
\be
\def\arraystretch{2.}
\begin{array}{ll}
 p \left(  \tilde{\bm\sigma}(\bm{x}), \tilde{\bs{u}}(\bm{x}), \mathbb{C}( \bs{x}) | \bs{\Lambda} \right) & = \cfrac{1}{Z(\bs{\Lambda})} ~\prod_{k=1}^5 \psi_k \left( \tilde{\bm\sigma}(\bm{x}), \tilde{\bs{u}}(\bm{x}), \mathbb{C}( \bs{x}) \right) \\
 & = \cfrac{ \pi_{\bm\Lambda} \left( \tilde{\bm\sigma}(\bm{x}), \tilde{\bs{u}}(\bm{x}), \mathbb{C}( \bs{x}) \right) }{ Z(\bs{\Lambda}) }\\
 & =\cfrac{1}{Z(\bs{\Lambda})} ~p_e \left( \tilde{\bm\sigma}(\bm{x}) \right)~  ~p_c \left( \tilde{\bm\sigma}(\bm{x}), \tilde{\bs{u}}(\bm{x}), \mathbb{C}( \bs{x}) \right) \\
 & \times ~~p_{C}(\mathbb{C}( \bs{x}) | \bm{\theta}_C) p_{\sigma}(\tilde{\bm\sigma}(\bm{x}) | \bm{\theta}_{\sigma}) p_{u}(\tilde{\bs{u}}(\bm{x}) | \bm{\theta}_{u}), 
\end{array}
\label{eq:priorjoint}
\ee
where $Z(\bs{\Lambda})$ is the normalization constant \footnote{In the absence of specific information, very vague Gaussians can be used for the priors in Equations (\ref{eq:priorc}), (\ref{eq:priorsigma}), (\ref{eq:prioru}) to ensure integrability, i.e. $Z(\bs{\Lambda})<+\infty$}.  

We note that the prior model implies an \textbf{undirected probabilistic graphical model} \cite{koller_probabilistic_2009}  which takes the form of a {\em factor graph} between state variables (Figure \ref{fig:factor}).
We emphasize that this does not rely on the availability of a well-posed forward 
model (e.g. one would notice the absence  of boundary conditions since these are 
data and will be treated in the likelihood), but rather  invokes (in a 
probabilistic fashion) {\em relationships} between the system's states as 
suggested  by the  model's equations.
More importantly, it relieves the formulation of the black-box,  input-output relation 
$\bs{u}(\bm{C})$  as in \refeq{eq:ko}. Dependencies between all state variables appear 
\textbf{explicitly} in the factors  making up \refeq{eq:priorjoint}. As explained in section 
\ref{sec:inf},  this will 
allow significant accelerations in the inference task,  despite the augmented state space in 
comparison to standard deterministic or 
Bayesian formulations. 
We should also note that the prior model effectively depends on  the number of interrogations $N_e$ 
and $N_c$ of the governing equations. 
The larger $N_e, N_c$ become, the more tightly  should the prior concentrate around  
the continuous, model-dictated values. In contrast, the smaller $N_e,N_c$ are, the flatter the 
prior becomes as  the number of equations enforced is smaller. This provides an added advantage in 
the inference task (section \ref{sec:inf}), as these numbers define effectively a tempering  
schedule, from simpler to more complex densities \cite{koutsourelakis_multi-resolution_2009}.

The incorporation of various error sources  in the prior model enables one to properly account for the {\em observation} error in the data. Given  
 displacement data (interior or boundary) $\bs{u}_{obs} = \{ \bs{u}_{obs,i_o} \}_{i_o=1}^{N_o}$ at $N_o$ locations $\bs{x}^{(i_o)}$, then:
 \be
\bs{u}_{obs,i_o}=\tilde{\bs{u}}(\bs{x}^{(i_o)})+\eta_{i_o}, \quad  \eta_{i_o}\sim \mathcal{N}(0, \sigma_{\eta}^2), \quad i_0=1,\ldots,N_o,
\label{eq:uobs}
\ee
where $\sigma_{\eta}^2$ is the variance of the observation noise. We emphasize here a fundamental difference with the analogous \refeq{eq:ko} used in traditional Bayesian formulations. Therein, the $\eta$'s express the difference between observables and  displacements predicted by the model which is assumed to be perfect, whereas here the model-related equations have been sequestered in the prior (\refeq{eq:priorjoint}). The Gaussian assumption for the $\eta_{i_o}$ is not restrictive as in \refeq{eq:ko}, where it must simultaneously account for model errors.
The equation above yields a {\em likelihood}:
\be
\begin{array}{ll}
 p( \bs{u}_{obs} | \tilde{\bs{u}}(\bs{x}))  & \propto \psi_6(\tilde{\bs{u}}(\bs{x})) \\
        & = \prod_{i_o=1}^{N_o} ~\mathcal{N} (\bs{u}_{obs,i_o}~; \tilde{\bs{u}}(\bs{x}^{(i_o)}), \sigma_{\eta}^2),
\end{array}
\label{eq:like}
\ee
which, as is suggested in Figure \ref{fig:factor}, gives rise to an {\em augmented posterior} density over all state variables as follows:
\be
\def\arraystretch{2.}
\begin{array}{ll}
p(  \tilde{\bm\sigma}(\bm{x}), \tilde{\bs{u}}(\bm{x}), \mathbb{C}( \bs{x})  |\bs{u}_{obs},  \bs{\Lambda}) & \propto  \prod_{k=1}^6 \psi_k(\tilde{\bm\sigma}(\bm{x}), \tilde{\bs{u}}(\bm{x}), \mathbb{C}( \bs{x})  ) \\
& \propto \underbrace{ p( \bs{u}_{obs} | \tilde{\bs{u}}(\bs{x}) ) }_{\textrm{likelihood \refeq{eq:like}} }~\underbrace{ p(  \tilde{\bm\sigma}(\bm{x}), \tilde{\bs{u}}(\bm{x}), \mathbb{C}( \bs{x}) | \bs{\Lambda})}_{\textrm{prior  \refeq{eq:priorjoint}} }
\end{array} 
\label{eq:post}
\ee

Despite the superficial complexity  of the  
posterior density above, we note that this 
consists of large products, of tractable, local terms nevertheless. The absence of 
any "black-box" terms is the key to  facilitating the inference process.

\subsection{Representation of the unknown state variables}
\label{sec:rep}

The presentation of the augmented prior/posterior  densities in the previous section was done (with some abuse of
notation) in infinite dimensions, i.e. without presupposing a particular discretization of the latent state fields. A possibility for carrying out inference tasks is the utilization of  Gaussian Processes that are frequently employed for nonparametric representations (e.g. 
\cite{chkrebtii_bayesian_2016}). We believe though that in the current setting such a choice would  be impracticable due to the 
non-Gaussian   terms in \refeq{eq:post}. In order to carry computations, we advocate  
 discretized representations  which will unavoidably introduce discretization errors.
We note here that such a representation of the latent fields $ \tilde{\bm\sigma}(\bm{x}), \tilde{\bs{u}}(\bm{x}), \mathbb{C}( \bs{x})$, is not dependent 
 on the discretization of the governing  equations, which is controlled by the number of 
weighting functions $N_e$ (\refeq{eq:priore}) and the number of (effectively collocation) 
points $N_c$, where the constitutive equation is interrogated (\refeq{eq:priorc}).
This decoupling is an 
additional advantage of the proposed strategy that can lead to {\em  adaptive, 
multi-resolution inference} as discussed in the conclusions. 

In the framework proposed, we seek all problem fields, i.e.  the stresses  $\tilde{\bm\sigma}(\bm{x})$, 
displacements $\tilde{\bs{u}}(\bm{x})$, and constitutive model 
parameters $\mathbb{C}( \bs{x})$ and  employ  a (potentially overcomplete) dictionary of feature  functions $\bm{f}(\bx)=\{ f_{u,j}(\bx), f_{C,j}(\bx), f_{\sigma,j}(\bx) \}$ such 
 that:
\be
\begin{array}{ll}
\tilde{\bs{u}}(\bm{x}) & =\sum_{j=1}^{M_u}  \bs{U}_j~f_{u,j}(\bs{x}), \\
 \tilde{\mathbb{C}}( \bs{x}) & =\sum_{j=1}^{M_{C}} \mathbb{C}_j~f_{C,j}(\bs{x}), \\
\tilde{\bm\sigma}(\bm{x}) & =\sum_{j=1}^{M_{\sigma}}  \bm{\Sigma}_j~f_{\sigma,j}(\bs{x}),
\end{array}
\label{eq:sparse_disp}
\ee 
where, in general,  $ \bs{U}_j$ are vector-valued,  $ \bm{\Sigma}_j$  second-order-tensor-valued and $\mathbb{C}_j$ fourth-order-tensor-valued  coefficients.
For a fixed set of functions $\bm{f}(\bx)$ this would imply an {\em augmented posterior} on the state vector $\{\bs{U},\bs{C}, \bs{\Sigma} \}$, where $\bs{U}=\{ \bs{U}_j \}_{j=1}^{M_u},~ \bs{C}=\{  \mathbb{C}_j\}_{j=1}^{M_{C}}, ~\bs{\Sigma}=\{ \bm{\sigma}_j\}_{j=1}^{M_{\sigma}}$ arising from the substitution of \refeq{eq:sparse_disp} into the 
respective equations, e.g. \refeq{eq:priorjoint} or \refeq{eq:post}.

On one hand, it is desirable to maximize the expressivity of  such representations in 
order to minimize the discretization error. On the other, by increasing the dimensionality 
of the unknown parameters in  the representation, not only one impedes computations, but  
can potentially lead to a multi-modal posterior, especially when the number of unknowns 
is much larger than the data $N_o$ and  $N_e+N_c$.

This poses an interesting model selection problem. One option  to address this is to initiate the search with a small number of feature functions  and 
progressively add more. These can be selected from a pool of candidates by employing appropriate 
criteria \cite{bilionis_free_2012, della_pietra_inducing_1997}.
Another alternative is to preselect a large, overcomplete set of feature functions and prune them by enforcing {\em sparsity}. In the Bayesian setting advocated, this can be readily achieved by employing sparsity-promoting hierarchical priors (e.g. spike-and-slab \cite{ishwaran_spike_2005}, automatic relevance determination \cite{mackay_information_2003,wipf_new_2008}, horseshoe \cite{carvalho_handling_2009}). 
The underlying assumption is that the fields of interest are highly compressible, i.e. most of the 
coefficients $\bs{U}_j,  \mathbb{C}_j, \bs{\Sigma}_j$ are zero 
\cite{donoho_compressed_2006,candes_robust_2006,lewicki_learning_2000}.
 Sparsity as a principle of parsimony  has successfully been applied in Bayesian formulations in various contexts  such  the representation of natural images \cite{olshausen_sparse_1997} or the solution of PDE-based inverse problems \cite{franck_sparse_2016} and a host of algorithms have been developed, not only for finding such representations, but also an appropriate dictionary for achieving this goal \cite{lewicki_learning_2000}.
 
 In this work, we adopt a simpler strategy that employs feature functions from traditional finite element analysis (FEA). In particular we employ irregular  triangulations $\mathcal{T}=\cup_{e}^{n_{el}} \Omega_e$ (in 2D)  of the problem domain and feature functions $f_{C,j}(\bx),f_{\sigma,j}(\bx)$ that are constant within each triangle $\Omega_e$ i.e.:
 \be
 f_{C,j}(\bx)=\left\{ \begin{array}{ll}
                          1, & \textrm{if $\bx \in \Omega_j$} \\
                          0, & \textrm{otherwise},
                         \end{array}
\right. \qquad 
f_{\sigma,j}(\bx)=\left\{ \begin{array}{ll}
                          1, & \textrm{if $\bx \in \Omega_j$} \\
                          0, & \textrm{otherwise}.
                         \end{array}
\right.
\ee
As for the feature function $f_{u,j}$, we employ the piece-wise linear shape functions corresponding to linear triangular elements from finite element analysis \cite{hughes_finite_2000}.
A direct implication of this choice is that the coefficients $\bs{U}_j$ correspond to the displacements at the vertices of each triangle (also called nodal displacements). Furthermore, the corresponding strain tensor $\tilde{\bm\epsilon}$ computed from such a displacement field though \refeq{eq:straindisp} would also be piece-wise constant (as $\tilde{\bm\sigma}$ and $ \tilde{\mathbb{C}}$). As a result, enforcement of the constitutive law as in \refeq{eq:priorc}, can be effectively done on an element-by-element basis (i.e. the number of such contributions $N_c$ is at most equal to the number of elements/triangles $n_{el}$). 
We do not claim that such a representation is optimal or the sparsest possible and that it unavoidably introduces a discretization error. We note  though that given a sufficiently fine triangulation, one can approximate arbitrary close the true stess, displacement and material parameter fields, which are generally not piece-wise constant, nor piece-wise linear. We assume that the discretization in the problems considered are sufficiently fine to ignore the discretization error which would be smaller than the other contributions.

The weighting functions $\bm{w}(\bx)$ employed in the MWR as in \refeq{eq:rese} are also given by the piece-wise linear shape functions of linear triangular elements (as for $f_{u,j}$). Hence, each residual in \refeq{eq:rese} corresponds to enforcing equilibrium of forces at each vertex/node.
The locality of feature and weighting functions endows the formulation with the computational conveniences of  FEA, particularly with the computation of residuals which can also be done on an element basis and make use of existing computer codes.
In particular, we note that based on the aforementioned discussion:
\bi
\item  we denote with $\bs{U} \in \RR^{n_u}$ the vector for the discretized representation of the displacement field $\tilde{\bs{u}}(\bx)$. It consists of the displacements at each of the $n_V$ vertices/nodes, i.e. $\bs{U}= \{ \bs{U}_j \}_{j=1}^{n_V}$. The strains obtained from \refeq{eq:straindisp} will be piece-wise constant. We denote by $\bs{E}_e$ the strains corresponding to each element $e$. These can be readily computed from the nodal displacements of the element, say $\bs{U}_e$, by employing the well-known in FEA strain-displacement matrix $\bs{B}_e$ as $\bs{E}_e=\bs{B}_e \bs{U}_e$. 
\item we denote with $\bs{C} \in \RR^{n_C}$ the vector for the discretized representation of the elastic tensor field $\mathbb{C}(\bx)$. It can also be partitioned on an element/triangle basis as $\bs{C}=\{ \bs{C}_e\}_{e=1}^{n_{el}}$.
\item we denote with $\bs{\Sigma} \in \RR^{n_{\sigma}}$ the vector for the discretized representation of the stress field $\tilde{\bs{\sigma}}(\bx)$. Given the piece-wise constant nature of this representation, it can also be partitioned on an element/triangle basis as $\bs{\Sigma}=\{ \bs{\Sigma}_e\}_{e=1}^{n_{el}}$.
\item the enforcement of the conservation of linear momentum as in \refeq{eq:rese} for each weighting function $\bs{w}^{(i_e)}(\bx)$ can be expressed as:
\be
r_e^{(i_e)} (\tilde{\bs{\sigma}}(\bx))={f}_{i_e}-\bs{b}_{i_e}^T \bs{\Sigma},
\label{eq:reseind}
\ee
where ${f}_{i_e}= \int_{\Gamma_{NBC}} ~ \bs{w}^{(i_e)}_j \bm{t}_{0,j}~d\Gamma$ and $\bs{b}_{i_e}^T \bs{\Sigma}=\int_{\Omega} \bs{w}^{(i_e)}_{j,i} \tilde{\bm\sigma}_{ji}~d V$. The vector $\bs{b}_{i_e}$ is constant due to the discretization choices made and can be pre-computed  along with the scalar ${f}_{i_e}$. The assembly of all such residuals for $i_e=1, \ldots, N_e$ as in \refeq{eq:priore} gives rises to the following factor potential $\psi_1$:
\be
\psi_1(\bs{\Sigma})= e^{-\frac{1}{2 \epsilon^2} || \bs{f}- \bs{B}^T \bs{\Sigma} ||^2}.
\label{eq:psi1}
\ee
\item the enforcement of the constitutive law as in \refeq{eq:resc} over each triangle/element $e$ can be expressed as:
\be
\bs{r}_c^{(e)} \left( \tilde{\bm\sigma}(\bm{x}), \tilde{\bs{u}}(\bm{x}), \mathbb{C}( \bs{x}) \right) = \bs{\Sigma}_e -\bs{C}_e \bs{B}_e \bs{U}_e.
\label{eq:rescind}
\ee
 The assembly of all such residuals for $e=1, \ldots, n_{el}$ as in \refeq{eq:priorc} gives rise to the following factor potential $\psi_2$:
\be
\psi_2(\bs{\Sigma}, \bs{C}, \bs{U}; ~\bs{\Lambda})=\prod_{e=1}^{n_{el}} e^{-\frac{\Lambda_e}{2} || \bs{\Sigma}_e -\bs{C}_e \bs{B}_e \bs{U}_e ||^2}.
\label{eq:psi2}
\ee
\item the priors on the material parameters will be specialized in the applications. We denote by $\psi_3(\bs{C})$ the corresponding factor potential as in \refeq{eq:priormaterial}.
\item we employ vague Gaussian priors on $\bs{\Sigma}$ and $\bs{U}$ as in \refeq{eq:priorsigma} and \refeq{eq:prioru} respectively. Their role is not to reflect any prior information (which is unavoidably problem dependent), but rather to ensure the integrability of the corresponding augmented prior in \refeq{eq:priorjoint}. The corresponding factor potentials are:
\be
\begin{array}{ll}
 \psi_4(\bs{\Sigma}) & = e^{-\frac{\theta_{\sigma}}{2} || \bs{\Sigma} ||^2}, \\
\psi_5(\bs{U}) & = e^{-\frac{\theta_{u}}{2} || \bs{U} ||^2},
\end{array}
\label{eq:psi45}
\ee
where $\theta_{\sigma}, \theta_{u} \to 0$.

\ei
Adaptations to one or three dimensions are straightforward where one can employ the machinery of linear or trilinear hexahedral elements respectively.

\subsection{Rephrasing forward/inverse problems as probabilistic inference tasks}
\label{sec:inf}

In this section, we recapitulate the framework introduced in section \ref{sec:augp} in view of the discretizations adopted in section \ref{sec:rep}, and, more importantly, demonstrate how forward and inverse problems can be solved as probabilistic inference tasks, without ever resorting to the black-box PDE solver.  

Consider first a typical \textbf{forward problem} where one is given:
\bi

\item the conservation law of \refeq{eq:conservation}, which is enforced as in \refeq{eq:rese} with $\epsilon \to 0$ (in practice, very small values are prescribed).
\item the constitutive law of \refeq{eq:constelast}, which is enforced as in \refeq{eq:resc} and for the purpose of solving the forward problem is assumed to be valid (i.e. $\Lambda_{i_c}=\lambda_{i_c} \to \infty$). In addition, the elasticity tensor $\mathbb{C}(\bx)$ is prescribed, i.e. in the discretized representation $\bs{C}=\bs{c}$ where $\bs{c}$ is known.     
\item boundary conditions as in \refeq{eq:DBC_boundary_conditions} or \refeq{eq:NBC_boundary_conditions}. For the sake of simplicity, suppose only Dirichlet boundary conditions are prescribed (i.e. $\bs{u}_0(\bx)$ in \refeq{eq:DBC_boundary_conditions}), which in the discretized represenation of the displacement field implies that part of the vector $\bs{U}$ is known. If $\bs{U}=(\bs{U}_i, \bs{U}_b)$ is partitioned to interior $\bs{U}_i$ and boundary $\bs{U}_b$ displacements, then $\bs{U}_b=\bs{u}_0$ where $\bs{u}_0$ is known. 

\ei

The augmented prior presented in \refeq{eq:priorjoint} and adapted in section \ref{sec:rep}  can be expressed as:
\be
\begin{array}{ll}
 p \left(  \bs{\Sigma}, \bs{C}, \bs{U}_i, \bs{U}_b | \bs{\Lambda} \right) & = \cfrac{1}{Z(\bs{\Lambda})} ~\psi_1(\bs{\Sigma}) \psi_2(\bs{\Sigma}, \bs{C}, \bs{U}_i, \bs{U}_b ; ~\bs{\Lambda})  \psi_3(\bs{C})  \psi_4(\bs{\Sigma}) \psi_5(\bs{U}_i, \bs{U}_b )  \\
 & = \cfrac{ \pi_{\bs{\Lambda}}(\bs{\Sigma}, \bs{C}, \bs{U}_i, \bs{U}_b)}{Z(\bs{\Lambda})}, 
\end{array}
\label{eq:priorjointf}
\ee
where:
\be
\pi_{\bs{\Lambda}}(\bs{\Sigma}, \bs{C}, \bs{U}_i, \bs{U}_b)= \psi_1(\bs{\Sigma}) \psi_2(\bs{\Sigma}, \bs{C}, \bs{U}_i, \bs{U}_b ; ~\bs{\Lambda})  \psi_3(\bs{C})  \psi_4(\bs{\Sigma}) \psi_5(\bs{U}_i, \bs{U}_b )
\label{eq:pi}
\ee
and:
\be
Z(\bs{\lambda})=\int \psi_1(\bs{\sigma}) \psi_2(\bs{\sigma}, \bs{c}, \bs{u}_i, \bs{u}_b ; ~\bs{\lambda})  \psi_3(\bs{c}) \psi_4(\bs{\sigma}) \psi_5(\bs{u}_i, \bs{u}_b ) ~d\bs{\sigma} d\bs{c}~d\bs{u}_i~d\bs{u}_b.
\label{eq:normZ}
\ee
Under the specifications of the forward problem (i.e. $\bs{C}=\bs{c}$, $\bs{U}_b=\bs{u}_0$ and  given $\bs{\Lambda}=\bs{\lambda}$ ), the goal is to write the conditional density of the undetermined latent random variables $ \bs{\Sigma}, \bs{U}_i$, i.e.:
\be
\def\arraystretch{2.}
\begin{array}{ll}
 p \left( \bs{\Sigma}, \bs{U}_i | \bs{C}=\bs{c}, \bs{U}_b =\bs{u}_0, \bs{\lambda} \right) & \propto \cfrac{ p \left(  \bs{\Sigma}, \bs{C}=\bs{c}, \bs{U}_i,  \bs{U}_b=\bs{u}_0 | \bs{\lambda} \right)  }{ p \left(  \bs{C}=\bs{c},\bs{U}_b=\bs{u}_0 | \bs{\lambda} \right) } \\
 & = \cfrac{ p \left(  \bs{\Sigma}, \bs{C}=\bs{c}, \bs{U}_i,  \bs{U}_b=\bs{u}_0 | \bs{\lambda} \right)  } { \int p \left(  \bs{\sigma}, \bs{C}=\bs{c}, \bs{u}_i,  \bs{U}_b=\bs{u}_0 | \bs{\lambda} \right) ~d\bs{\sigma} ~d\bs{u}_i } \\
 & = \cfrac{ \pi_{\bm\lambda} \left(  \bs{\Sigma}, \bs{C}=\bs{c}, \bs{U}_i,  \bs{U}_b=\bs{u}_0 \right) }{ \int \pi_{\bm\lambda} \left(  \bs{\sigma}, \bs{C}=\bs{c}, \bs{u}_i,  \bs{U}_b=\bs{u}_0 \right)~d\bs{\sigma} ~d\bs{u}_i }.
\end{array}
\label{eq:priorjointc}
\ee
Hence, identifying the latent variables $\bs{\Sigma}, \bs{U}_i$ (i.e. stresses and interior displacements) is equivalent to finding the conditional augmented prior above. This, for example, can be performed by sampling (e.g. using MCMC or SMC), which, as can be seen from the form of the factor of potentials, involves explicit terms, i.e. no black-boxes and no calls to a PDE solver. Similarly, derivatives (first and second order) with respect to $\bs{\Sigma}, \bs{U}_i$ can be readily computed and used to facilitate the sampling/inference process.
In the following section, we demonstrate the use of Stochastic Variational Inference tools, which can be used for that purpose and efficiently provide approximations of the sought conditional density.

Other types of boundary conditions can be readily treated in the same fashion. Furthermore, uncertainty propagation tasks can also be dealt efficiently. If, for example, random material properties $\bs{C}$ were assumed and a density was prescribed, then one could readily sample values, say $\bs{c}$, for $\bs{C}$ from this density and subsequently identify the conditional as in \refeq{eq:priorjointc} for each of these $\bs{c}$\footnote{In such a scenario, the presence of the prior $p_C(\bs{C}| \bm{\theta}_C)$ in \refeq{eq:priorjoint} would be superfluous.}.
We finally note the possibility of solving ill-posed forward problems, e.g. problems where boundary conditions are under- or over-specified. This is because the corresponding augmented density is still well-defined and therefore inference can be readily performed. 

Consider now a typical \textbf{inverse problem} where one is given:
\bi

\item the conservation law of \refeq{eq:conservation}, which is enforced as in \refeq{eq:rese} with $\epsilon \to 0$ (in practice, very small values are prescribed).
\item the constitutive law of \refeq{eq:constelast}, which is enforced as in \refeq{eq:resc}, but the validity of which is unknown (i.e.  $\Lambda_{i_c}$  are unspecified). In addition, the constitutive parameters $\bs{C}$ are unspecified, in which case $p_C(\bs{C} | \bm{\theta}_C)$ in \refeq{eq:priormaterial}      plays the role of the prior in canonical Bayesian formulations.
\item boundary conditions as in \refeq{eq:DBC_boundary_conditions} or \refeq{eq:NBC_boundary_conditions}. For the sake of simplicity suppose only Dirichlet boundary conditions are prescribed (i.e. $\bs{u}_0(\bx)$ in \refeq{eq:DBC_boundary_conditions}), which in the discretized represenation of the displacement field implies that $\bs{U}_b=\bs{u}_0$. 

\item {\em noisy} interior displacements $\bs{u}_{obs}$ as in \refeq{eq:uobs}, which give rise to a likelihood $p(\bs{u}_{obs} | \bs{U}_i)$ as in \refeq{eq:like} (upon discretization).

\ei

In this scenario, the unknown state variables consist of $\bs{\Sigma}, \bs{U}_i$ and $\bs{C}$. In addition one must  also identify the constitutive model error hyper-parameters $\bs{\Lambda}$, which are effectively the connecting threads in the probabilistic web of relationships 
constructed -  the same way constitutive laws are necessary to provide closure to the 
deterministic, PDE-based model.  Rather than a hard enforcement though, they control the  
validity of the constitutive law at each interrogation point $\bs{x}^{i_c}$ and must be  
{\em learned}, simultaneously with the other latent state variables. Clearly, such a problem is under-determined unless stress and strain data  were simultaneously available. Critical to learning  meaningful values  for these  
hyperparameters is therefore the selection of an appropriate prior model.  A vague such 
model would unnecessarily relax the threads  connecting  the state variables in  
the constitutive model, resulting in a posterior that is also artificially vague. To 
address this ill-posedness, \textbf{we invoke an additional postulate of parsimony, i.e. that the 
constitutive law is correct, unless strong evidence in the data/observables suggests 
differently}. This suggests that, ceteris paribus,  solutions with  more constitutive law residuals $\bs{r}_c^{(i_c)}$ in \refeq{eq:resc} 
 being zero, should be favored.  To that end, we advocate the use of Automatic Relevance Determination  (ARD, \cite{mackay_information_2003,wipf_new_2008}), which is a heavy-tailed, hierarchical prior model, independently  for each $\Lambda_{i_c}$ of the following form:
 \be
 p( \Lambda_{i_c} | \alpha_0, \beta_0 ) = Gamma( \Lambda_{i_c}; \alpha_0, \beta_0)=\cfrac{\Gamma(\alpha_0)}{\beta_0^{\alpha_0}} \Lambda_{i_c}^{\alpha_0-1} ~e^{-\beta_0  \Lambda_{i_c}}.
 \label{eq:ard}
 \ee
 Very small values for $\alpha_0, \beta_0$ (in the numerical investigations $\alpha_0= \beta_0=10^{-10}$ was used) promote robust sparsity patterns \cite{bishop_variational_2000}.
%
 
The combination of the aforementioned  hierarchical prior with the augmented posterior of \refeq{eq:post} yields the following density (based on Bayes' rule):
\be
\begin{array}{ll}
p( \bs{\Sigma}, \bs{C}, \bs{U}_i, \bs{\Lambda} | \bs{u}_{obs}, \bs{U}_b=\bs{u}_0) & = \cfrac{ p\left( \bs{u}_{obs} | \bs{\Sigma}, \bs{C}, \bs{U}_i, \bs{\Lambda},  \bs{U}_b=\bs{u}_0 \right) ~p\left(\bs{\Sigma}, \bs{C}, \bs{U}_i, \bs{\Lambda} |  \bs{U}_b=\bs{u}_0\right)}{ p(\bs{u}_{obs}) } \\
& = \cfrac{   p(\bs{u}_{obs} | \bs{U}_i)~ p(\bs{\Sigma}, \bs{C}, \bs{U}_i | \bs{\Lambda}, \bs{U}_b=\bs{u}_0) ~p(\bs{\Lambda}) }{  p(\bs{u}_{obs}) },
\end{array}
\label{eq:fullpost}
\ee
where $p(\bs{u}_{obs} | \bs{U}_i)$ is the aforementioned likelihood, $p(\bs{\Lambda})$ is the ARD defined in \refeq{eq:ard} (we omit dependence on $\alpha_0, \beta_0$ for simplicity) and $ p(\bs{\Sigma}, \bs{C}, \bs{U}_i | \bs{\Lambda}, \bs{U}_b=\bs{u}_0)$ can be found from the augmented prior of \refeq{eq:priorjoint} as:
\be
\begin{array}{ll}
 p(\bs{\Sigma}, \bs{C}, \bs{U}_i | \bs{\Lambda}, \bs{U}_b=\bs{u}_0) & = \cfrac{ p(\bs{\Sigma}, \bs{C}, \bs{U}_i, \bs{U}_b=\bs{u}_0 | \bs{\Lambda}) }{ p( \bs{U}_b=\bs{u}_0 | \bs{\Lambda})} \\
 & = \cfrac{ p(\bs{\Sigma}, \bs{C}, \bs{U}_i, \bs{U}_b=\bs{u}_0 | \bs{\Lambda}) }{ \int p(\bs{\sigma}, \bs{c}, \bs{u}_i, \bs{U}_b=\bs{u}_0 | \bs{\Lambda}) ~d\bs{\sigma}~d\bs{c}~d\bs{u}_i }\\
 & = \cfrac{ \pi_{\bm\Lambda} (\bs{\Sigma}, \bs{C}, \bs{U}_i, \bs{U}_b=\bs{u}_0 ) }{ \int \pi_{\bm\Lambda} (\bs{\sigma}, \bs{c}, \bs{u}_i, \bs{U}_b=\bs{u}_0 ) ~d\bs{\sigma}~d\bs{c}~d\bs{u}_i }.
\end{array}
\label{eq:postjointc}
\ee
We discuss suitable inference tools based on Stochastic Variational Inference in the next section.

 \clearpage

\section{Variational Inference}
\label{sec:variational_approx}

In the previous section, a probabilistic reformulation of the forward and inverse problems for linear elastostatics was proposed. The solution of canonical such problems amounts to  inference 
 tasks with respect to appropriate densities such as the ones in \refeq{eq:priorjointc} and \refeq{eq:postjointc}, over the unknown state variables and model parameters. While these densities do not require calls to any black-box, PDE solver, they are analytically intractable. We advocate the use of Variational Inference (VI) \cite{beal_variational_2003, bishop_pattern_2006} techniques, which, in contrast to Monte Carlo-based ones, are approximate albeit highly efficient.
  Such methods have risen into prominence for probabilistic inference tasks in the machine learning community    \cite{jordan_introduction_1999, attias_variational_2000,wainwright_graphical_2008,paisley_variational_2012}. They yield  approximations to the target density, which are obtained    by solving an optimization problem over a family of appropriately selected probability densities  with the objective of minimizing the Kullback-Leibler divergence \cite{cover_elements_1991} (subsections \ref{sec:vif} and \ref{sec:vii}). The success of such an approach  hinges upon the selection of appropriate densities that have the capacity of providing good approximations while enabling efficient optimization with regards to their parameters (subsection \ref{sec:formqr}). In the  sequel  we discuss the application  
   of VI to the problems of interest and in the last subsection \ref{sec:optim} present the proposed algorithms. 
   
  \subsection{VI for forward problems}
  \label{sec:vif}
  
  We first review  the basic aspects of the method in the context of solving a canonical  \textbf{forward problem} as described in section \ref{sec:inf}, which reduces to the density in \refeq{eq:priorjointc}. For economy of notation we denote with $\bY$ the latent variables and by $\bs{V}$ the given/observed ones. Hence, in this case $\bY=\{\bs{\Sigma}, \bs{U}_i\}$ and $\bV=\{  \bs{C}, \bs{U}_b \}$. Let also $\bv$ denote the given/observed value of $\bV$, which according to \refeq{eq:priorjointc}  is $\bv=\{ \bs{c}, \bs{u}_0 \}$. The goal is to approximate the intractable density:
  \be
  p( \bY | \bV=\bv, \bm\lambda) =  \cfrac{ \pi_{\bm\lambda} \left( \bY, \bV=\bv \right) }{ \int \pi_{\bm\lambda} \left( \bY, \bV=\bv \right)  ~d\bY}.
\ee
 Let $q(\bY; ~\bm\phi)$ a family of approximating densities parametrized by $\bm \phi$. We seek to determine the best possible approximation to $ p( \bY | \bV=\bv, \bm\lambda)$ by:
 \be
 \min_{\bm\phi} KL\left( q(\bY; ~\bm\phi)~|| ~p( \bY | \bV=\bv, \bm\lambda)\right)
 \ee
 By employing Jensen's inequality one can establish that:
 \be
 \begin{array}{ll}
   0 \le KL\left( q(\bY; ~\bm\phi)~|| ~p( \bY | \bV=\bv, \bm\lambda)\right) & = \int q(\by; ~\bm\phi)~\log  \cfrac{ q(\by; ~\bm\phi)}{p( \by | \bV=\bv, \bm\lambda)} ~d\by \\ 
   & = \int  q(\by; ~\bm\phi)~ \log \cfrac{ q(\by; ~\bm\phi)}{ \cfrac{ \pi_{\bm\lambda} \left( \by, \bV=\bv \right) }{ \int \pi_{\bm\lambda} \left( \by, \bV=\bv \right)  ~d\by}} ~d\by \\ 
    & = \int  q(\by; ~\bm\phi)~ \log \cfrac{ q(\by; ~\bm\phi)}{\pi_{\bm\lambda} \left( \by, \bV=\bv \right) }~d\by \\
    & ~~+\log \int \pi_{\bm\lambda} \left( \by, \bV=\bv \right)  ~d\by \\
    & = -\mathcal{F}_{for}(q(\by; ~\bm\phi))+\log \int \pi_{\bm\lambda} \left( \by, \bV=\bv \right)  ~d\by.
 \end{array}
\ee
Hence, minimizing the KL-divergence is equivalent to maximizing the Evidence Lower Bound (ELBO, \cite{blei_variational_2017}) $\mathcal{F}_{for} (q(\by; ~\bm\phi)) \le \log \int \pi_{\bm\lambda} \left( \by, \bV=\bv \right)  ~d\by$.

Based on the form of $\pi_{\bs{\lambda}}$ (\refeq{eq:pi}) one can obtain the following expression for the ELBO:
\be
\begin{array}{ll}
\mathcal{F}_{for} (q(\by; ~\bm\phi))  & = \int q(\by; ~\bm\phi) \log \cfrac{\pi_{\bm\lambda} \left( \by, \bV=\bv \right) }{q(\by; ~\bm\phi)}~d\by \\ 
  & =  \mathbb{E}_{q(\by;\bm\phi)} \left[ \log \psi_1(\bs{\sigma})  \right]+ \eqq \left[ \log \psi_2(\bs{\sigma}, \bs{u}_i, \bs{C}=\bs{c}, \bs{U}_b=\bs{u}_0; \bs{\lambda})\right]  \\
  &~~ + \eqq \left[ \log \psi_4(\bs{\sigma}) \right] + \eqq \left[ \log \psi_5(\bs{u}_i)  \right] +\mathbb{H} \left[ q(\bm{y};\bm\phi) \right] \\
  & = \mathcal{L}_{for}(\bm \phi),
\end{array}
\label{eq:elbof}
\ee
where $\mathbb{H} \left[ q(\bm{y};\bm\phi) \right]$ is the entropy of $q$. The term corresponding to $\psi_3(\bs{C})$ was omitted as it does not depend on $q$.
We discuss the form of the $q$ employed in   subsection~\ref{sec:formqr} and the associated algorithmic steps (Algorithm \ref{alg:forward_problem}) as well as provide  additional details for the derivatives  of the objective $\mathcal{L}_{for}$  with respect to $\bm \phi$ in  \ref{sec:appA}.

\subsection{VI for inverse problems}
  \label{sec:vii}
We consider now the application of Variational Inference for the solution of the \textbf{inverse problem}, as reformulated in section~\ref{sec:inf}. This involves the augmented posterior of \refeq{eq:postjointc}. We denote again with $\bY$ the latent state variables which in this case consist of $\{\bs{\Sigma}, \bs{C}, \bs{U}_i\}$. The observed/known state variables $\bV$ entail only $\bs{U}_b$. Furthermore, the model parameters $\bs{\Lambda}$ controlling constitutive model errors are unknown. Due to the undirected nature \citep{wainwright_new_2002,murray_bayesian_2004,silva_bayesian_2006,silva_hidden_2009} of the proposed graphical model and the dependence of the normalization constant ${Z}$ on $\bs{\Lambda}$ (\refeq{eq:normZ}), we adopt a hybrid strategy where the full posterior of $\bY$ is approximated and point-estimates, corresponding to the Maximum-A-Posteriori (MAP) value, for $\bs{\Lambda}$ are computed. 

In particular,  we consider the {\em marginal} posterior $p(\bs{\Lambda} | \bs{V}=\bs{v}, \bs{u}_{obs})$ of $\bs{\Lambda}$  and seek to maximize it using a Variational Expectation-Maximization scheme \cite{beal_variational_2003} that is based on the following lower-bound:
\be
\begin{array}{ll}
 \log p(\bs{\lambda} |  \bs{U}_b=\bs{u}_0, \bs{u}_{obs}) & = \log \int p(\bs{y}, \bs{\lambda}|  \bs{U}_b=\bs{u}_0, \bs{u}_{obs}) ~d\by \\ 
 & = \log \int q(\bm{y};\bm\phi) \cfrac{p(\bs{y}, \bs{\lambda}|  \bs{U}_b=\bs{u}_0, \bs{u}_{obs})  }{q(\bm{y};\bm\phi)} ~d\by \\ 
 & \ge \int q(\bm{y};\bm\phi)  \log \cfrac{p(\bs{y}, \bs{\lambda}|  \bs{U}_b=\bs{u}_0, \bs{u}_{obs})  }{q(\bm{y};\bm\phi)} ~d\by \quad \textrm{(from Jensen's inequality)} \\
 & = \mathcal{F}_{inv}(q(\bm{y};\bm\phi), ~\bs{\lambda}),
\end{array}
\label{eq:logpost}
\ee
where $q(\bm{y};\bm\phi)$ is a density with respect to the latent state variables. It can be readily shown \cite{bishop_pattern_2006} that the optimal $q^{opt}$, for which the inequality above becomes an equality, corresponds to the posterior $p( \by | \bs{\lambda}, \bs{U}_b=\bs{u}_o, \bs{u}_{obs})$. The latter does not necessarily belong to the family of approximating densities $q(\bm{y};\bm\phi)$ selected.  Nevertheless, it suggests the following iterative scheme, where one alternates (until convergence) between the steps: 
\bi
\item \textbf{E(xpectation)-step:} The model parameters $\bm\lambda$ are fixed, and $\mathcal{F}_{inv}$ is maximized with respect to $\bm \phi$ (so as it approximates as close as possible $ \log p(\bs{\lambda} |  \bs{U}_b=\bs{u}_0, \bs{u}_{obs})$). 
\item \textbf{M(aximization)-step:}  The parameters $\bm\phi$, and therefore $q$, remain fixed and $\mathcal{F}_{inv}$ is maximized with respect to the  parameters $\bm\lambda$.
\ei
As with all Expectation-Maximization schemes \cite{dempster_maximum_1977,neal_view_1998} several relaxations of the aforementioned steps are possible such as improving, rather than maximizing, $\mathcal{F}_{inv}$ during any of the steps, partial updates of the parameters, etc. 

Before we present the algorithmic details, we look more closely at the terms involved in $\mathcal{F}_{inv}$. In particular, based on Equations (\ref{eq:fullpost}) and (\ref{eq:postjointc}), we obtain:
\be
\begin{array}{ll}
 \mathcal{F}_{inv}(q(\bm{y};\bm\phi), ~\bs{\lambda}) & = \int q(\bm{y};\bm\phi)  \log \cfrac{p(\bs{y}, \bs{\lambda}|  \bs{U}_b=\bs{u}_0, \bs{u}_{obs})  }{q(\bm{y};\bm\phi)} ~d\by \\
 & = \int q(\bm{y};\bm\phi)  \log \cfrac{   p(\bs{u}_{obs} | \bs{U}_i)~ p(\bs{\sigma}, \bs{c}, \bs{u}_i | \bs{\lambda}, \bs{U}_b=\bs{u}_0) ~p(\bs{\lambda}) }{  p(\bs{u}_{obs}) ~ q(\bm{y};\bm\phi)}~d\bs{\sigma} ~d \bs{c}~d\bs{u}_i \\
 & =  \int q(\bm{y};\bm\phi)  \log \cfrac{   p(\bs{u}_{obs} | \bs{U}_i) }{q(\bm{y};\bm\phi)}~d\by +
  \eqq \left[ \log p(\bs{\sigma}, \bs{c}, \bs{u}_i | \bs{\lambda}, \bs{U}_b=\bs{u}_0)   \right]  \\ 
  & ~~+\log p(\bs{\lambda}) - \log  p(\bs{u}_{obs}) \\
  & = \eqq \left[\log p(\bs{u}_{obs} | \bs{U}_i) \right]+\mathbb{H} \left[ q(\bm{y};\bm\phi) \right]  \\
  & ~~+\eqq \left[ \log  \cfrac{ \pi_{\bm\lambda} (\bs{\sigma}, \bs{c}, \bs{u}_i, \bs{U}_b=\bs{u}_0 ) }{ \int \pi_{\bm\lambda} (\bs{\sigma}, \bs{c}, \bs{u}_i, \bs{U}_b=\bs{u}_0 ) ~d\bs{\sigma}~d\bs{c}~d\bs{u}_i } \right] \\
  & ~~+\log p(\bs{\lambda})  \\
   & = \eqq \left[\log p(\bs{u}_{obs} | \bs{U}_i) \right]+\mathbb{H} \left[ q(\bm{y};\bm\phi) \right]  \\
   & ~~+ \eqq \left[ \log  \pi_{\bm\lambda} (\bs{\sigma}, \bs{c}, \bs{u}_i, \bs{U}_b=\bs{u}_0 ) \right] -\log \left[ \int \pi_{\bm\lambda} (\bs{\sigma}, \bs{c}, \bs{u}_i, \bs{U}_b=\bs{u}_0 ) ~d\bs{\sigma}~d\bs{c}~d\bs{u}_i \right] \\
    & ~~+\log p(\bs{\lambda}),
\end{array}
\label{eq:elbopost}
\ee
where we have omitted terms (e.g. $\log p(\bs{u}_{obs})$) that do not depend on $q$ nor $\bs{\lambda}$.
The most involved term in the expression above pertains to  $\int \pi_{\bm\lambda} (\bs{\sigma}, \bs{c}, \bs{u}_i, \bs{U}_b=\bs{u}_0 ) ~d\bs{\sigma}~d\bs{c}~d\bs{u}_i$, which in principle needs to be determined in order to compute derivatives with respect to $\bs{\lambda}$. To that end, we advocate the use of an additional variational approximation employing a different class of densities denoted by $r(\by;\bm\xi)$ such that:
\be
\begin{array}{ll}
 \log \left[ \int \pi_{\bm\lambda} (\bs{\sigma}, \bs{c}, \bs{u}_i, \bs{U}_b=\bs{u}_0 ) ~d\bs{\sigma}~d\bs{c}~d\bs{u}_i  \right]& = \log \int \pi_{\bm\lambda} (\by, \bs{U}_b=\bs{u}_0 ) ~d\by \\
 & = \log \int r(\by;\bm\xi) \cfrac{ \pi_{\bm\lambda} (\by, \bs{U}_b=\bs{u}_0 ) }{ r(\bm{y};\bm\xi)} ~d\by  \\
 & \ge \int r(\by;\bm\xi)  \log \cfrac{ \pi_{\bm\lambda} (\by, \bs{U}_b=\bs{u}_0 ) }{ r(\bm{y};\bm\xi)} ~d\by \\
 & = \err \left[ \log \pi_{\bm\lambda} (\by, \bs{U}_b=\bs{u}_0 ) \right]+\mathbb{H} \left[ r(\bm{y};\bm\xi) \right]  \\
 & = \hat{\mathcal{F}}_{inv}(r(\by;\bm\xi), ~\bs{\lambda}).
\end{array}
\label{eq:elboprior}
\ee
We note again that the inequality above becomes an equality iff $r(\by;\bm\xi) = p(\by, \bs{U}_b=\bs{u}_0 )$ given in \refeq{eq:priorjointc}.
 Unfortunately, using the lower bound $ \hat{\mathcal{F}}_{inv}(r(\by;\bm\xi), ~\bs{\lambda})$ in place of $\log \left[ \int \pi_{\bm\lambda} (\bs{\sigma}, \bs{c}, \bs{u}_i, \bs{U}_b=\bs{u}_0 ) ~d\bs{\sigma}~d\bs{c}~d\bs{u}_i  \right]$ in \refeq{eq:elbopost} leads to an upper bound of $\mathcal{F}_{inv}$ (due to the minus sign), i.e.:
 \be
 \begin{array}{ll}
  \mathcal{F}_{inv}(q(\bm{y};\bm\phi), ~\bs{\lambda}) & \le  \eqq \left[\log p(\bs{u}_{obs} | \bs{U}_i) \right]+\mathbb{H} \left[ q(\bm{y};\bm\phi) \right]  \\
   & ~~+ \eqq \left[ \log  \pi_{\bm\lambda} (\bs{\sigma}, \bs{c}, \bs{u}_i, \bs{U}_b=\bs{u}_0 ) \right] - \hat{\mathcal{F}}_{inv}(r(\by;\bm\xi), ~\bs{\lambda}) \\
    & ~~+\log p(\bs{\lambda})  \\
    &= \tilde{\mathcal{F}}_{inv}(q(\bm{y};\bm\phi),~r(\by;\bm\xi), ~\bs{\lambda}) \\
    & = \mathcal{L}_{inv}(\bm\phi, ~\bm\xi, ~\bs{\lambda})
 \end{array}
 \label{eq:elboinv}
 \ee
 Hence, $\mathcal{L}_{inv}$ will not necessarily be a lower bound of the marginal log-posterior $\log p(\bs{\lambda} |  \bs{U}_b=\bs{u}_0, \bs{u}_{obs})$ in \refeq{eq:logpost}. This introduces an additional error, which is smaller the closer $r(\by;\bm\xi)$ is to the optimum.
 We note that the final objective $\tilde{\mathcal{F}}_{inv}$ depends on two sets of approximating densities: a) $q(\bm{y};\bm\phi)$ that attempts to approximate the {\em posterior} (\refeq{eq:logpost}), and b) $r(\by;\bm\xi)$ which attempts to approximate the {\em prior} 
 (\refeq{eq:elboprior}). 
As  a result, the E-step in the Variational EM scheme described previously should involve (with $\bs{\lambda}$ fixed), apart from the updates of  $q(\bm{y};\bm\phi)$, the maximization of $\hat{{F}}_{inv}$ (\refeq{eq:elboprior}) with respect to  $r(\bm{y};\bm\phi)$.

We discuss the form of the approximating densities $q,r$ in   subsection~\ref{sec:formqr} and the associated algorithmic steps (Algorithms \ref{alg:inverse_problem} and \ref{alg:inverse_problem_model_error}) as well as provide additional details for the derivatives  of the objective $\mathcal{L}_{for}$  with respect to $\bm \phi, \bm \xi $ and $\bm \lambda$ in  \ref{sec:appB}.


\subsection{Form of approximating densities}
\label{sec:formqr}

While arbitrary forms of approximating densities $q$ and $r$ are allowed, for reasons of computational efficiency we focus on the Gaussian family of distributions $\mathcal{N}(\bm\mu,\bm S)$, parameterized by the mean $\bm\mu$ and covariance matrix $\bm S$, i.e $\bm \phi$ or $\bm \xi = \{\bm \mu, \bm S\}$ . In terms of the covariance matrix, several parameterizations are possible:
\begin{itemize}
  \item diagonal covariance: this mean-field assumption represents the most efficient choice, where all correlations between latent variables are neglected and $\bm{S}$ becomes a diagonal matrix.
  \item full covariance: taking into account all possible cross-correlations, the covariance matrix can be parameterized using its Cholesky decomposition $\bm S = \bm{L}\bm{L}^T$.
  \item banded covariance: modeling only a few correlations that are assumed to dominate the variance can result in a banded covariance structure, exhibiting significantly less parameters than a full covariance model.
\end{itemize}
Diagonal covariance models scale {\em linearly} with the dimension of the latent variables, however, they tend to underestimate the posterior uncertainty. Full covariance models provide better uncertainty estimates, but scale   {\em quadratically} with the dimension of the latent variables. As a compromise, we chose a banded covariance model and make use of  the spatial character of the latent variables in order to endow a nearest-neighbor correlation structure. More specifically, we assume that off-diagonal entries in $\bm S$ are nonzero if they pertain  to quantities associated with the same or a neighboring (i.e. with a common boundary) element.
We provide numerical evidence on  the relative accuracy of all three covariance models  in section~\ref{subsec:model_comparison}.

\subsection{Stochastic Variational Inference}\label{sec:optim}

The optimization objectives $\mathcal{L}_{for}$ and $\mathcal{L}_{inv}$ presented in subsections \ref{sec:vif} and \ref{sec:vii} for the Gaussian densities $q$ and $r$  parametrized as in subsection \ref{sec:formqr}, is analytically intractable due to the implicit dependence on $\bm \phi$ and $\bm \xi$ of expectations involving $q$ and $r$   (see \refeq{eq:elbof} or \refeq{eq:elboinv}).  To that end we employ  Stochastic Variational Inference (SVI)~\citep{hoffman_stochastic_2013,paisley_variational_2012,titsias_doubly_2014}, which makes use of Monte Carlo estimates of the gradients of the objective functions  in order to iteratively update the parameters $\bm \phi$ and $\bm \xi$ as well as $\bm \lambda$.
Further details  for the gradients  are contained in  \ref{sec:appA} and \ref{sec:appB}.

We note that the use of Gaussian densities for $q$ and $r$ enables the application of the well-known reparametrization trick \citep{kingma_auto-encoding_2013,rezende_stochastic_2014,titsias_doubly_2014}, which can significantly reduce the variance of the Monte Carlo estimators for the gradients. The latter consists of the simple observation that for any function $f(\by)$ its expectation with respect to a Gaussian $q_{\bm \phi}(y)=\mathcal{N}(\by | \bm \mu, \bm S=\bm L \bm L^T )$ can be re-expressed as:
\be
\eqq [f(\by)]=\mathbb{E}_{\mathcal{N}(\bm{0}, \bs{I})} [f(\bm \mu +\bm L \bm \eta)],
\label{eq:reparam1}
\ee
where the latter expectation is with respect to $\bm \eta \sim \mathcal{N}(\bm{0}, \bs{I})$  based on the transformation $\by=\bs{g}_{\bm \phi}(\bm \eta)=\bm \mu +\bm L \bm \eta$. Hence, derivatives with respect to $\bm \phi=\{\bm \mu, \bm L\}$ can be computed by employing the chain rule as follows:
\be
\nabla_{\bm \phi} \mathbb{E}_{q(\bm{y};\bm\phi)}[f(\by)]
= \mathbb{E}_{\mathcal{N}(\bm{0}, \bs{I})}  [\nabla_{\by} f(\bm \mu +\bm L \bm \eta)
 \nabla_{\bm \phi} \bs{g}_{\bm \phi}(\bm \eta) ].
 \label{eq:reparam2}
\ee
We finally  note that the gradient-based updates of the parameters are carried out using the  {\em Adam} scheme \citep{kingma_adam:_2014} (\ref{sec:adam}
). We recap the basic algorithmic steps for the application of SVI for the solution of forward problems in Algorithm \ref{alg:forward_problem} and inverse problems in Algorithm \ref{alg:inverse_problem_model_error} with references to the pertinent equations/sections. We have also included Algorithm \ref{alg:inverse_problem}, which can be employed for the solution of inverse problems in the absence of model errors (i.e. assuming $\bs{\lambda} \to \infty$).
With regards to the overall scalability of the algorithms, we note that due to their first-order nature they scale {\em linearly} with the number of parameters $\bm \phi$ for forward problems and $(\bm \phi, \bm \xi, \bm \lambda)$ for general inverse problems.

\begin{algorithm}[!htpb]
  \caption{Forward problem}\label{alg:forward_problem}
  \begin{algorithmic}
    \State Latent variables: $\bm{Y} = \{ \bm\Sigma, \bm{U}_i \}$
    \State Objective function: $\mathcal{L}_{for}(\bm\phi)$
    \State Initialize $\bm\phi$ randomly
    \While{not converged}
      \State Estimate gradients: $\bm\nabla_{\bm\phi} \mathcal{L}_{for}(\bm\phi^{(k)})$ as in~\ref{sec:appA}
      \State Update parameters: $\bm\phi^{(k+1)} \leftarrow \bm\phi^{(k)} + \rho^{(k)} \bm\nabla_{\bm\phi} \mathcal{L}(\bm\phi^{(k)})$ as in~(\ref{eq:param_update_phi})
      \State Estimate objective function: $\mathcal{L}_{for}(\bm\phi^{(k+1)})$
    \EndWhile
  \end{algorithmic}
\end{algorithm}

\begin{algorithm}[!htpb]
  \caption{Inverse problem with fixed $\bm\lambda$ (no model error)}\label{alg:inverse_problem}
  \begin{algorithmic}
    \State Latent variables: $\bm{Y} = \{ \bm\Sigma, \bm{U}_i,\bm{C} \}$
    \State Objective function: $\mathcal{L}_{inv}(\bm\phi)$
    \State Initialize $\bm\phi$ randomly
    \While{not converged}
      \State Estimate gradients: $\bm\nabla_{\bm\phi} \mathcal{L}_{inv}(\bm\phi^{(k)})$ as in~ \ref{sec:appB}.
      \State Update parameters: $\bm\phi^{(k+1)} \leftarrow \bm\phi^{(k)} + \rho^{(k)} \bm\nabla_{\bm\phi} \mathcal{L}(\bm\phi^{(k)})$ as in~(\ref{eq:param_update_phi})
      \State Estimate objective function: $\mathcal{L}_{inv}(\bm\phi^{(k+1)})$
    \EndWhile
  \end{algorithmic}
\end{algorithm}

\begin{algorithm}[!htpb]
  \caption{Inverse problem with model error}\label{alg:inverse_problem_model_error}
  \begin{algorithmic}
    \State Latent variables: $\bm{Y} = \{ \bm\Sigma, \bm{U}_i, \bm{C}, \bm\Lambda \}$
    \State Objective function:\\{}
    $\mathcal{L}_{inv}(\bm\phi,\bm\xi,\bm\lambda)$
    \State Initialize $\bm\phi,\bm\xi$ randomly, $\bm\lambda$ large
    \While{not converged}
      \State Estimate gradients: $\bm\nabla_{\bm\phi,\bm\xi,\bm\lambda} \mathcal{L}_{inv}(\bm\phi^{(k)},\bm\xi^{(k)},\bm\lambda^{(k)})$ as in~ \ref{sec:appB} and~(\ref{eq:dlambda1})
      \State Update parameters: ${\{ \bm\phi,\bm\xi,\bm\lambda \}^{(k+1)} \leftarrow\{\bm\phi,\bm\xi,\bm\lambda\}^{(k)} + \rho^{(k)} \bm\nabla_{\bm\phi,\bm\xi,\bm\lambda} \mathcal{L}(\bm\phi^{(k)},\bm\xi^{(k)},\bm\lambda^{(k)})}$
      \State as in~(\ref{eq:param_update_phi}), (\ref{eq:param_update_xi}) and~(\ref{eq:param_update_lambda})
      \State Estimate objective function: $\mathcal{L}_{inv}(\bm\phi^{(k+1)},\bm\xi^{(k+1)},\bm\lambda^{(k+1)})$
    \EndWhile
  \end{algorithmic}
\end{algorithm}

 \clearpage

\section{Numerical illustrations}
\label{sec:examples}

\begin{figure}[!ht]
\centering
  \includegraphics[height=5.5cm]{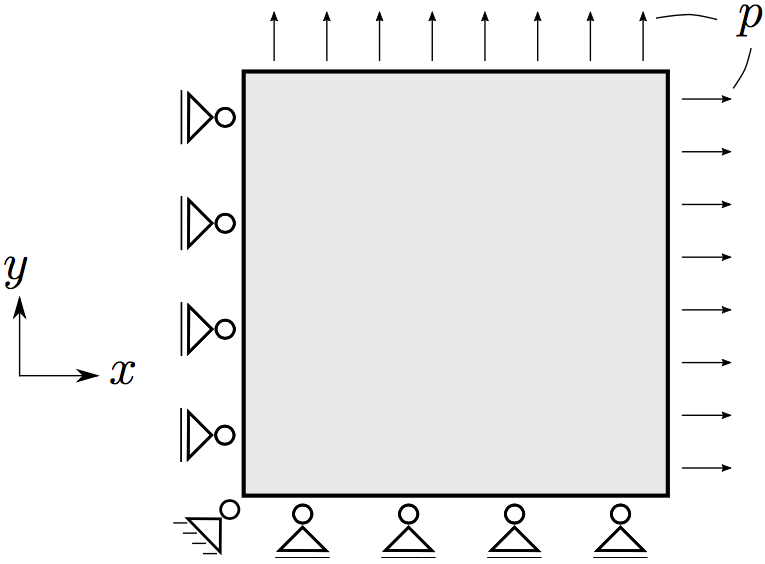}
	\caption{The two dimensional domain considered with dimensions $[0,L]\times[0,L]$, with $L = 10$, thickness $d = 1$ and Young's moduli $E(x,y)$. External normal tractions  (Neumann boundary conditions)  $p=1.0$ are applied along  the boundaries at $x = L$ and $y = L$, while the boundaries at $x = 0$ and $y = 0$ are fixed in the respective normal direction.
}
	\label{fig:2D_problem}
\end{figure}

In the following, we  demonstrate the capabilities and efficacy of the proposed modeling framework for an elliptic problem described by the equations in section \ref{sec:probabilistic_cm} and for the geometry and boundary conditions in  \autoref{fig:2D_problem}.
The linear elastostatic problem that will be considered is motivated by an application to elastography, where the identification of tumors/inclusions in healthy tissue is  sought. 

We consider plane stress conditions and the only unknown parameter is assumed to be the Young's modulus\footnote{A Poisson ratio of $\nu = 0.49$ was used to approximate  incompressible material behavior}, the values of which at various points/elements are represented by  the vector $\bm C$. 
In section \ref{subsec:model_comparison}, we investigate   comparatively the  performance of the three covariance structures discussed in section \ref{sec:formqr} and demonstrate that the banded structure offers a good balance between accuracy and scalability.
Subsequently, we investigate the following cases:
\bi
\item  in section~\ref{subsec:forward_problem},  the solution of a forward problem under the probabilistic inference reformulation proposed.
\item in  section ~\ref{subsec:inverse_problem} the  solution of an inverse problem  under the assumption of no  model errors.
\item in section ~\ref{subsec:model_error_problem}, the solution of an inverse problem with model errors.
\ei
The goal is three-fold: a) to assess the validity of the model-informed augmented prior (section \ref{sec:probabilistic_cm}) in representing the governing equations, b) to assess the accuracy of the proposed inference schemes in identifying material parameters and their spatial variability, and c) to assess the ability of the proposed formulation to identify and quantify model discrepancies.
We note that the results obtained in the latter three subsections were based on a banded covariance structure. Parameters for the Adam stochastic optimization scheme are contained in Appendix \ref{sec:adam}. For the inverse problems (sections \ref{subsec:inverse_problem}, \ref{subsec:model_error_problem}), a total variation prior on the material parameters is employed, penalizing jumps in the material parameters between neighboring elements and corresponding to the factor potential
\be
\psi_3(\bs{C}) = \prod_{e=1}^{n_{el}} e^{-\frac{\theta_{C}}{2} \sum_{i\in\mathcal{N}(c_e)} \left| c_e - c_i \right|},
\ee
with $\theta_{C}$ controlling the penalty. In terms of the hyperparameters for the factor potentials appearing in the augmented prior/posterior densities, we chose for the forward problems $\varepsilon = 10^{-4}, \lambda = 10^{8}, \theta_u = \theta_\sigma = \theta_C = 10^{-6}$. For the inverse problems, these parameters need to be adjusted to the respective likelihood noise, e.g. for $\sigma_n = 10^{-4}$ and for fixed model parameters, we chose $\varepsilon = 10^{-4}, \Lambda = 10^{8}, \theta_u = \theta_\sigma = 10^{-6}, \theta_C = 10^{4} $.


\subsection{Comparison of diagonal, banded and full covariance approximations}\label{subsec:model_comparison}

\begin{figure}[!ht]
  \centering
   \begin{tabular}{ccc} 
        \includegraphics[height=5.5cm]{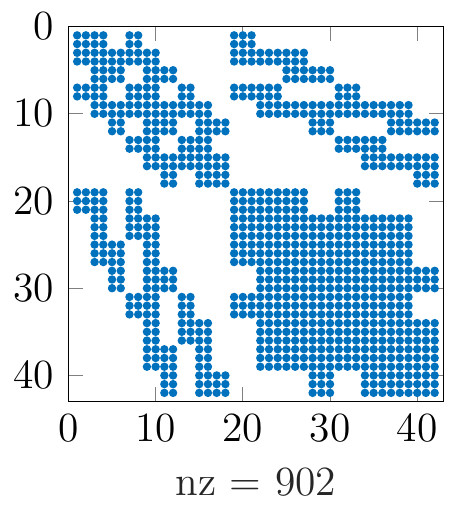} & 
     \includegraphics[height=5.5cm]{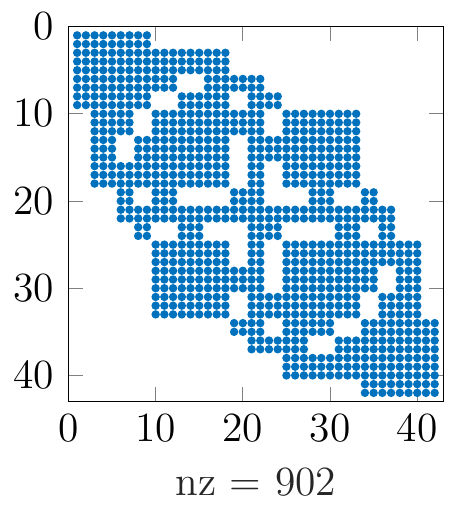} & 
   \includegraphics[height=5.5cm]{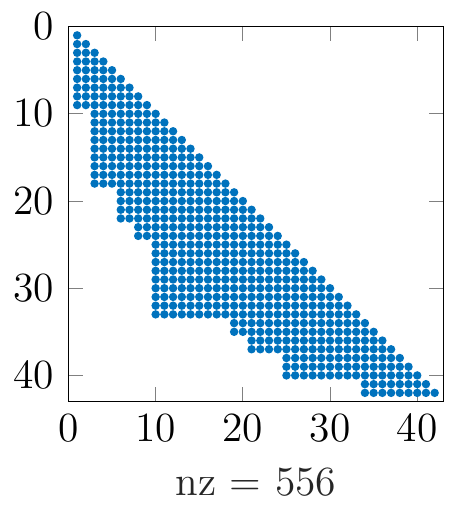} \\
   \end{tabular}
  \caption{The left figure visualizes the covariance structure that accounts for nearest-neighbor correlations (blue dots indicating a non-zero value) and requires $902$ parameters. In the middle, this structure was reordered using a Cuthill-McKee permutation algorithm and on the right hand side, the final Cholesky factor $\bm L$ is shown with the number of non-zero parameters having been reduced to $556$.}\label{fig:banded_covariance_structure}
\end{figure}

Before the larger numerical examples are presented, we compare the accuracy of the proposed banded covariance  model in relation to  the diagonal and full covariance cases. For that purpose, a forward problem on a very small regular  mesh  with only eight triangular elements is considered as well as a uniform distribution of material parameters (i.e. $c_e=1.0$ over all elements). In this case, the number of latent variables is $42$ (i.e. $dim(\bY)=42$) and consist of stresses $\bm \Sigma$ and displacements $\bm U_i$. In order to account for nearest-neighbor correlations,  while minimizing the band-length and therefore the number of unknown parameters, a Cuthill-McKee permutation algorithm\footnote{The following implementation was used: \url{https://de.mathworks.com/help/matlab/ref/symrcm.html}} was used and its effect is illustrated in   \autoref{fig:banded_covariance_structure}. Therein, one can also see the number of non-zero terms of the Cholesky factor $\bs{L}$ that was ultimately used in the parametrization $\bm \phi$ (along with the mean $\bm \mu$ of the Gaussian).

\begin{figure}[!t]
  \centering
       \includegraphics[width=0.6\textwidth]{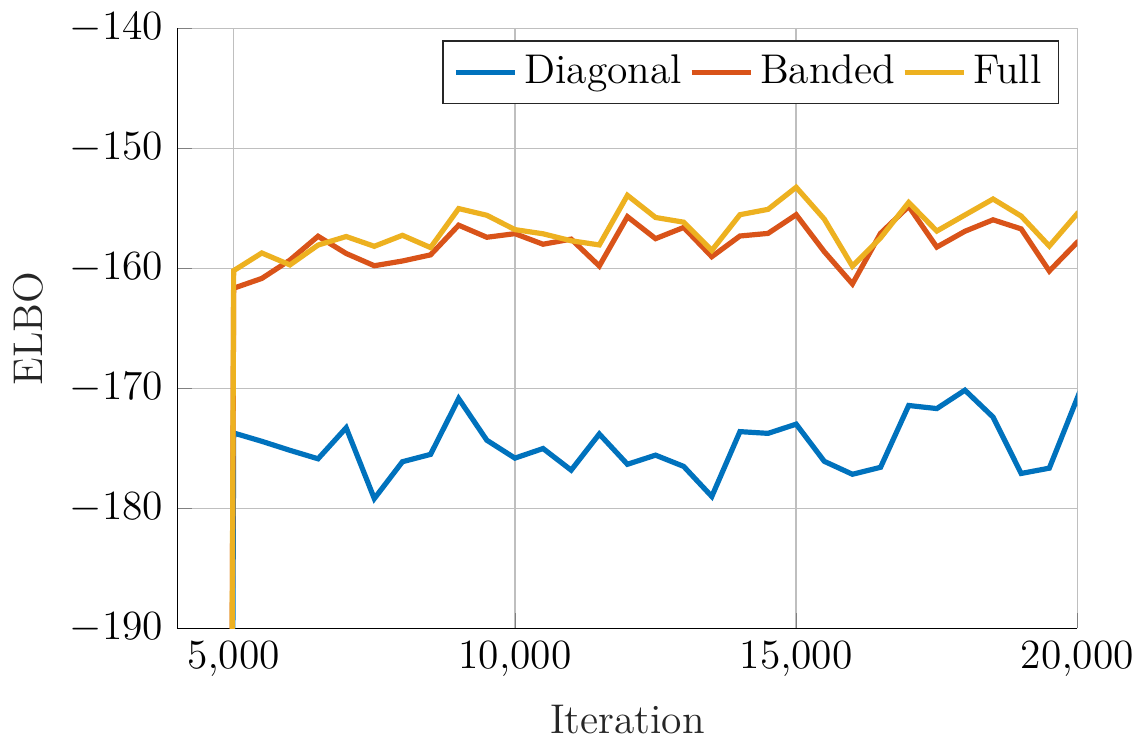} 
  \caption{Estimated ELBO $\mathcal{L}_{for}$  over the number of iterations for the simple forward problem (section \ref{subsec:model_comparison}) and the three considered covariance models: diagonal ($\mathcal{L}_{for} \approx -175.50$), banded ($\mathcal{L}_{for} \approx -156.39$) and full covariance ($\mathcal{L}_{for} \approx -156.2264$).}\label{fig:variational_models_elbo}
\end{figure}

\begin{figure}[!t]
  \centering
  \begin{tabular}{ccc}
  \textbf{Diagonal} & \textbf{Banded} & \textbf{Full} \\
    \includegraphics[height=4cm]{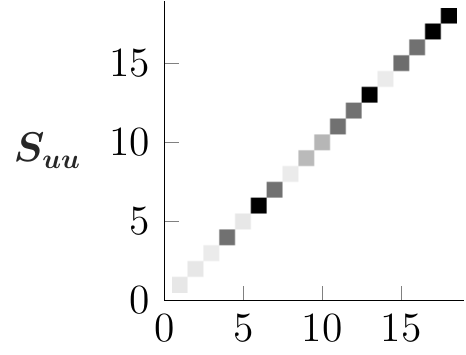} &
      \includegraphics[height=4cm]{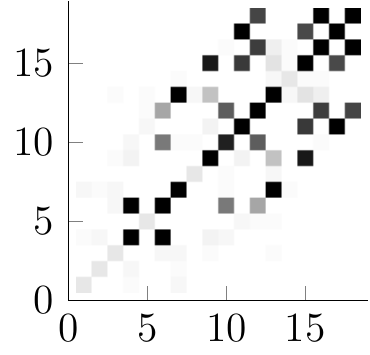} & 
        \includegraphics[height=4cm]{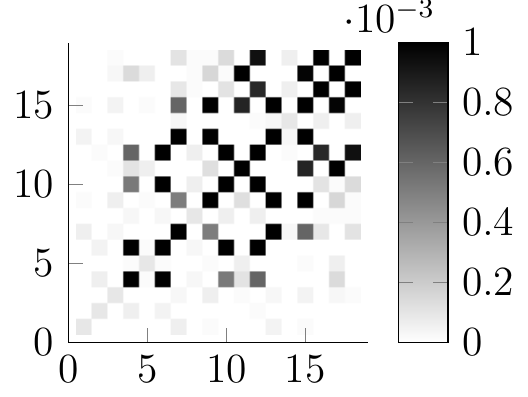} \\
    \includegraphics[height=4cm]{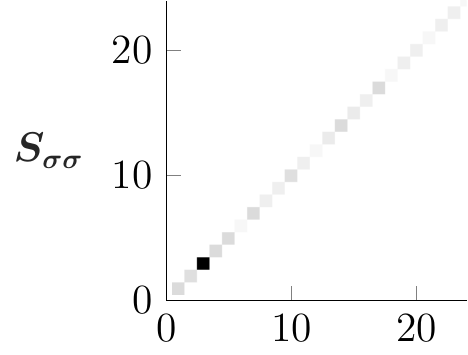} &
      \includegraphics[height=4cm]{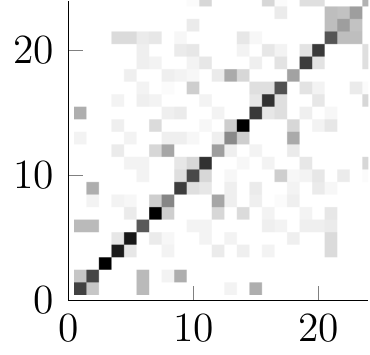} & 
        \includegraphics[height=4cm]{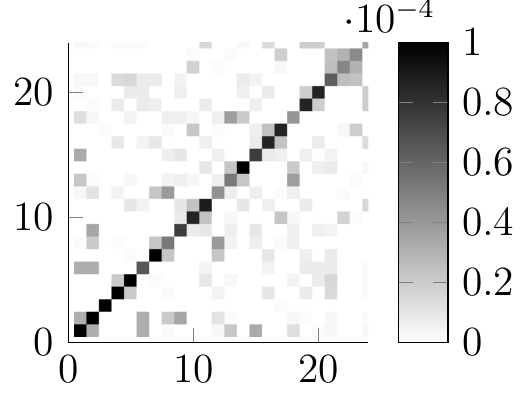} \\
    \includegraphics[height=4cm]{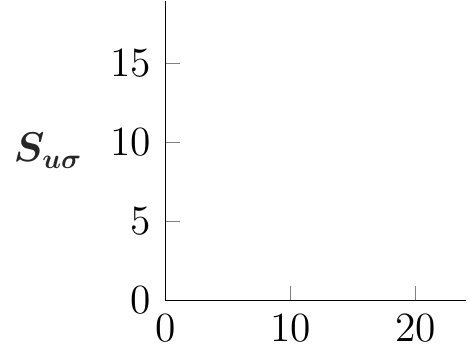} &
      \includegraphics[height=4cm]{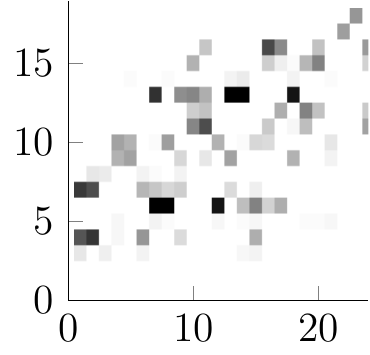} & 
        \includegraphics[height=4cm]{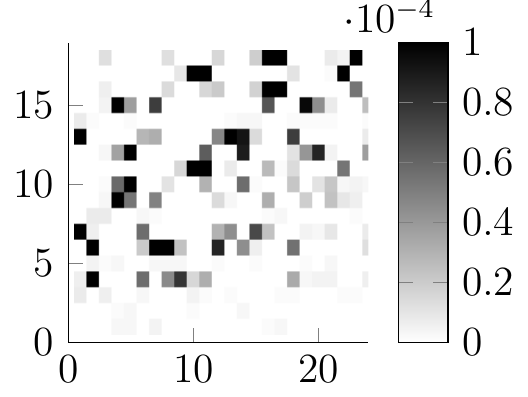} \\
  \end{tabular}
  \caption{Learned covariance matrices $\bm S$ using the  diagonal, banded and full covariance models for the test forward problem. For illustration purposes, the covariance matrix has been partitioned so that in the first row  covariances $\bm S_{\bm{u} \bm{u}}$ associated with the displacements  are depicted, in the second row the  covariances between the stress variables $\bm S_{\bm\sigma \bm\sigma}$ and in the third row the cross-correlations $\bm S_{\bm{u}\bm\sigma}$ between the displacements and stresses.}\label{fig:variational_models_covariances}
\end{figure}

 \autoref{fig:variational_models_elbo} depicts the estimated ELBO $\mathcal{L}_{for}$ (\refeq{eq:elbof}) with three covariance structures for $q$, i.e. diagonal (which requires $42$ parameters), full (which requires $903$ parameters) and banded (which requires $556$ parameters as per  \autoref{fig:banded_covariance_structure}). The diagonal covariance model ($\mathcal{L}_{for} \approx -175.50$) is clearly outperformed by the  banded ($\mathcal{L}_{for} \approx -156.39$) and full covariance ($\mathcal{L}_{for} \approx -156.2264$) approximations. Given that the latter is the most accurate (at least in the class of Gaussian $q$'s considered), we note that the banded covariance can achieve the same level of accuracy with much fewer parameters.  This behavior is also reflected in the learned covariance matrices of the respective models, as illustrated in~\autoref{fig:variational_models_covariances}. The diagonal covariance model clearly misses important correlations captured by the banded and full covariance variational approximations. The latter two mostly differ in the learned cross-correlations $\bm{S}_{\bm{u}\bm\sigma}$, while the displacement covariances $\bm{S}_{\bm{u}\bm{u}}$ and $\bm\Sigma_{\bm\sigma\bm\sigma}$ are very similar.

\begin{figure}[!htpb]
  \centering
      \includegraphics[width=0.75\textwidth]{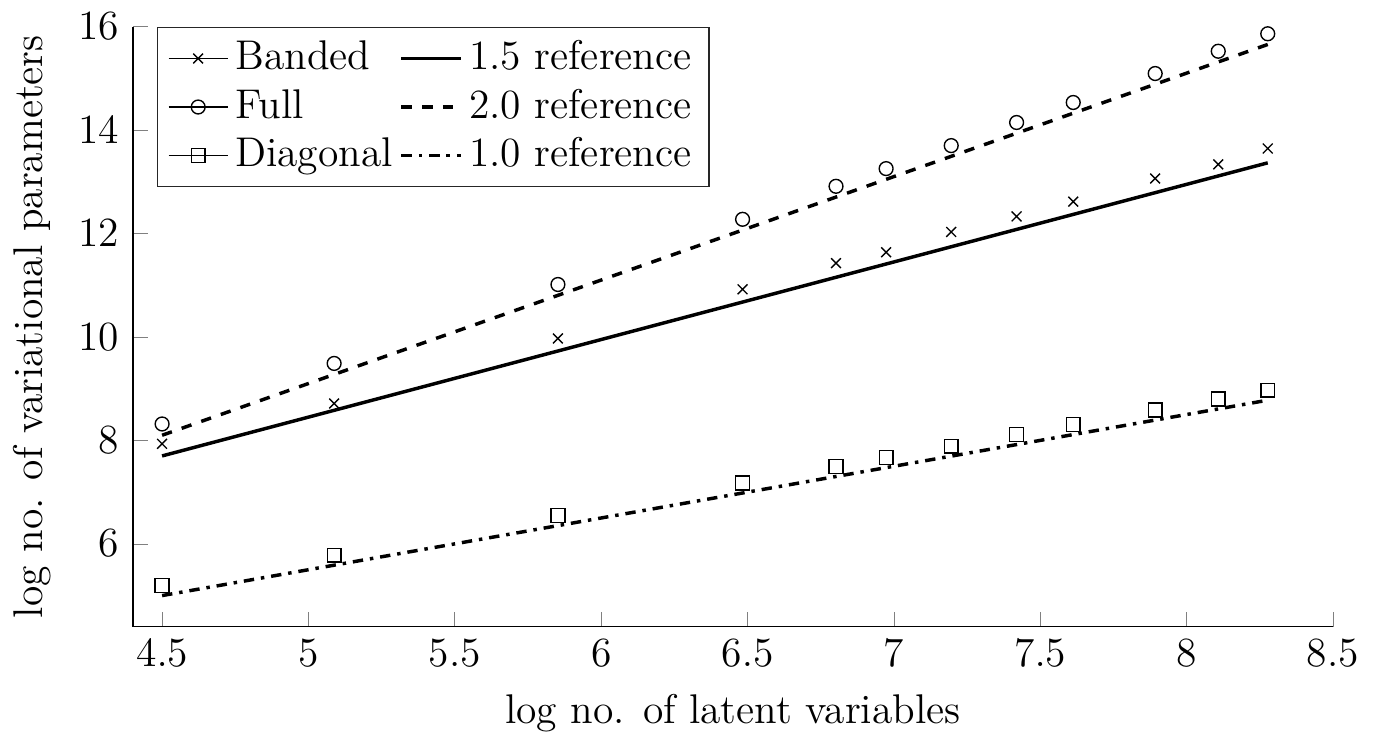}  
  \caption{Number of variational parameters $\bm \phi$ needed for the covariance matrix $\bm S$ as a function of the latent variables $\by$.  The number of optimization parameters of the banded covariance variational model scales approximately to the power of 1.5 with the number of latent variables. For comparison, the full covariance variational model scales to the power of 2, while the diagonal covariance exhibits linear scaling.}\label{fig:variational_model_complexity}
\end{figure}

Lastly,~\autoref{fig:variational_model_complexity} provides a graphical overview over the scalability of the three models. The number of optimization parameters of the banded covariance variational model scales approximately to the power of $1.5$ with the number of latent variables. For comparison, the full covariance variational model scales to the power of $2$ (quadratically), while the diagonal covariance  scales linearly.

\subsection{Forward problem}\label{subsec:forward_problem}

\begin{figure}[!ht]
  \centering
  \begin{tabular}{ll}
         \includegraphics[height=5.8cm]{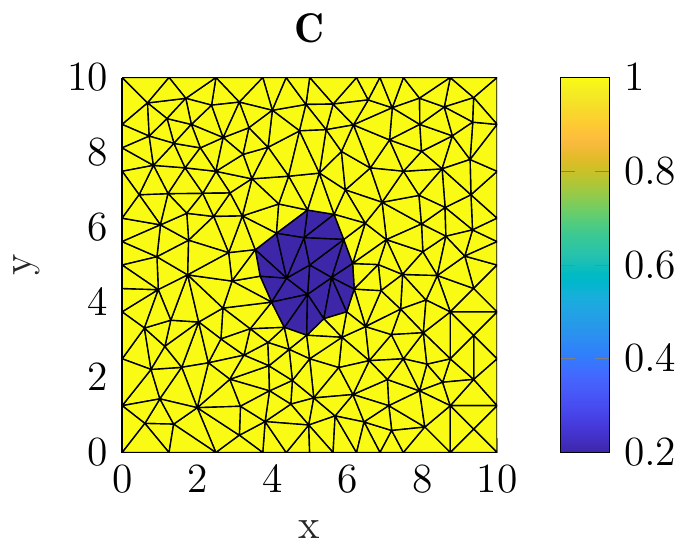}  
 &       \includegraphics[height=5.5cm]{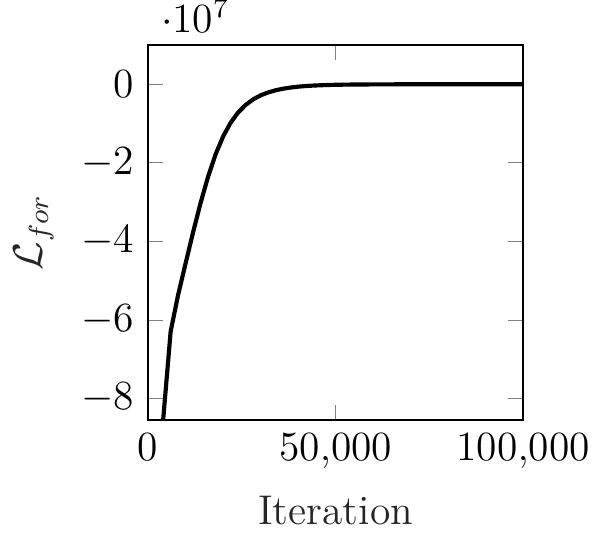}   \\
  \end{tabular}
  \caption{Spatial distribution of Young's modulus for the forward problem (left) and objective function $\mathcal{L}_{for}$ over the number of iterations (right).}
  \label{fig:2d_forward_problem_inclusion_youngs_truth}
\end{figure}

\begin{figure}[!htpb]
  \centering \textbf{Forward problem solution - mean values}
  \begin{tabular}{ll}
    \includegraphics[height=5.5cm]{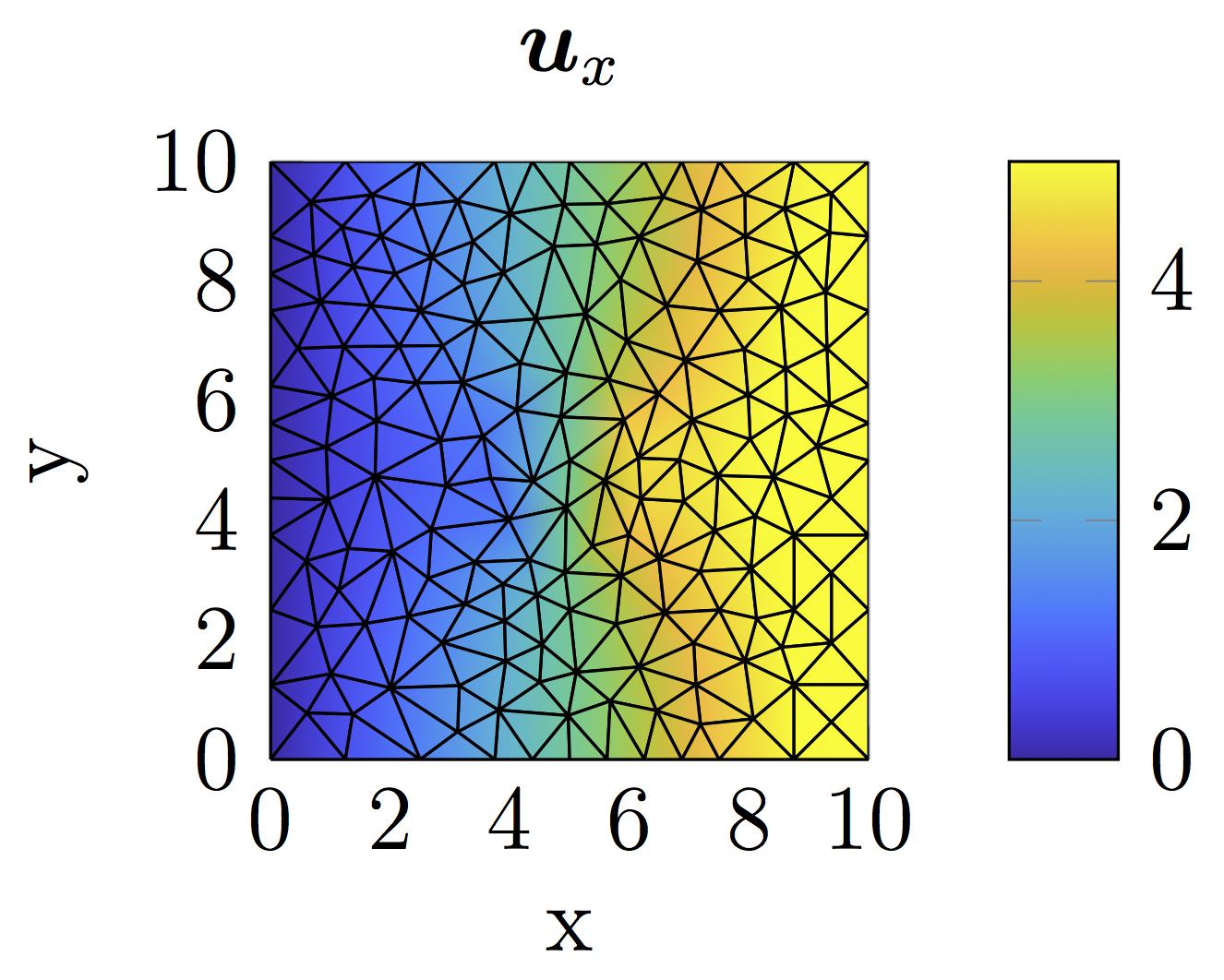} & 
        \includegraphics[height=5.5cm]{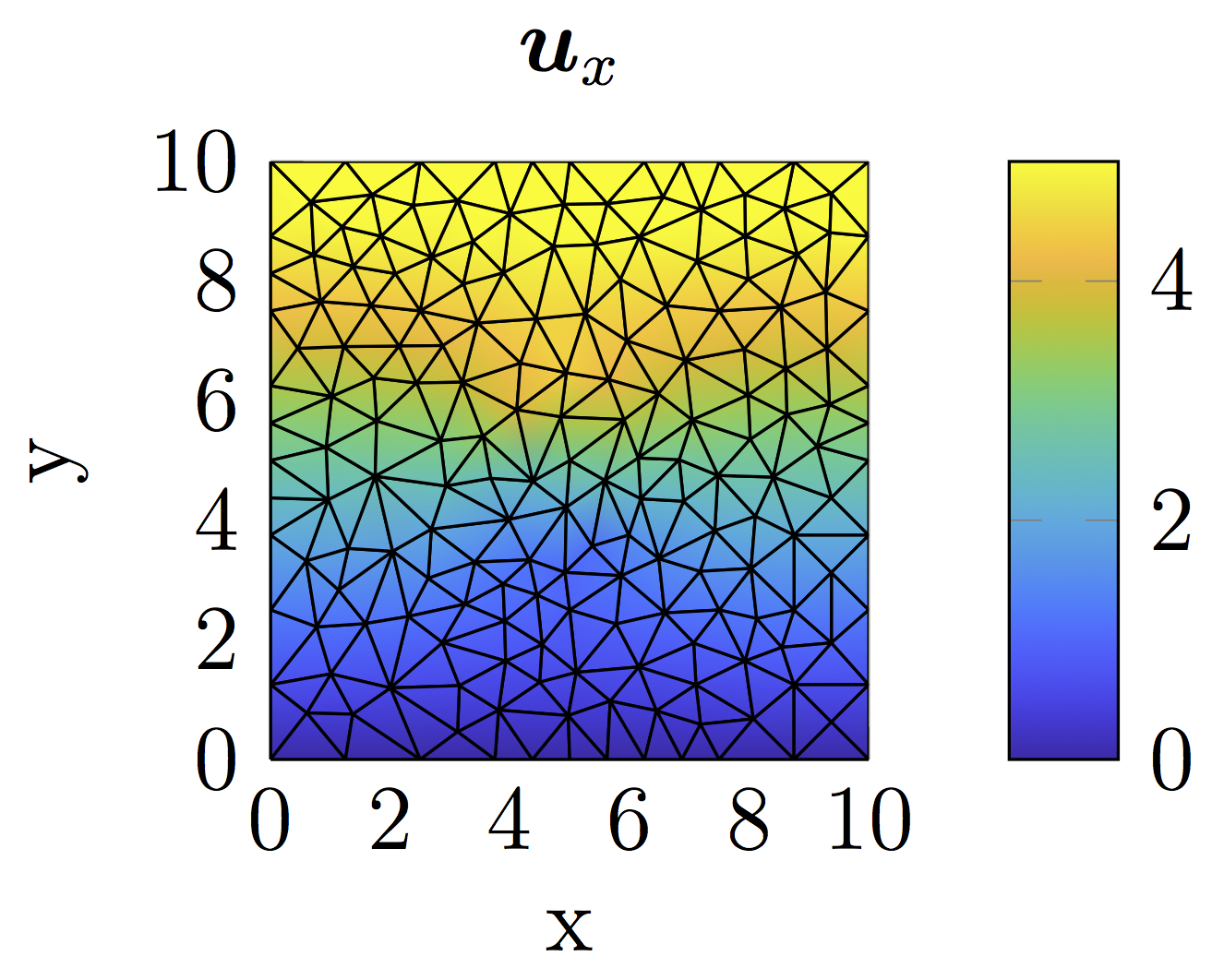}  \\
         \includegraphics[height=5.5cm]{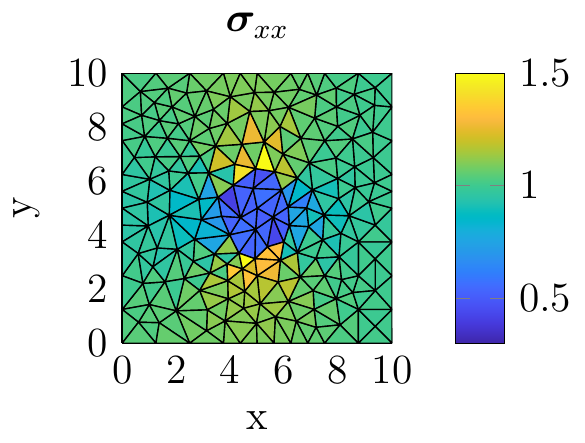} & 
        \includegraphics[height=5.5cm]{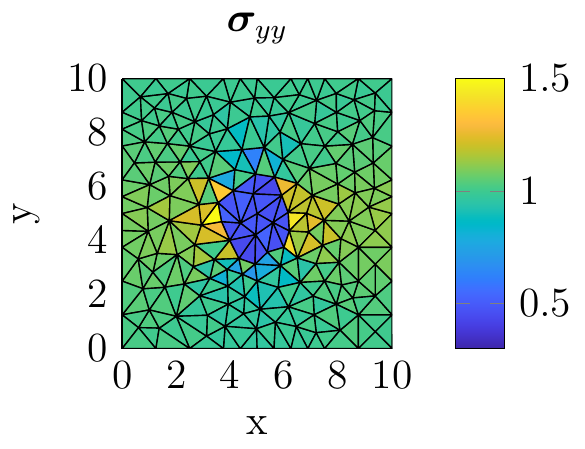}  \\
  \end{tabular}
  \includegraphics[height=5.5cm]{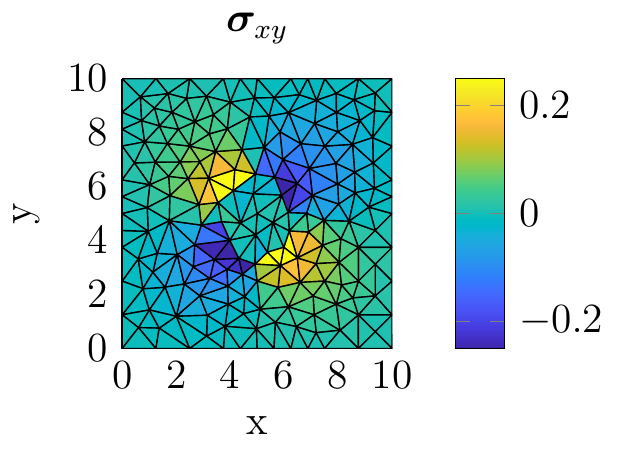} 
  \caption{Inferred  mean values for the displacements and stresses for the  forward problem with a circular inclusion at the domain center.}\label{fig:2d_forward_problem_inclusion}
\end{figure}

\begin{figure}[!htpb]
  \centering \textbf{Forward problem solution - comparison with reference}
  \begin{tabular}{lr}
    \includegraphics[height=5.5cm]{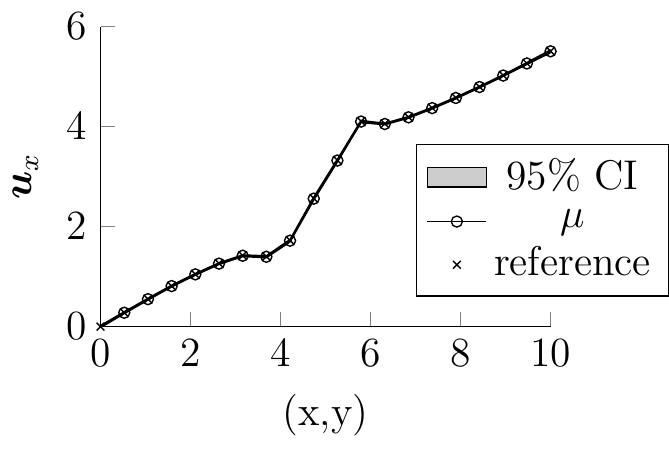} & 
    \includegraphics[height=5.5cm]{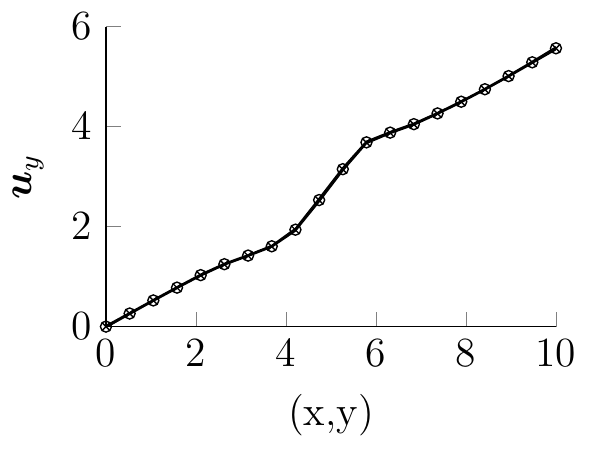} \\ 
    \includegraphics[height=5.5cm]{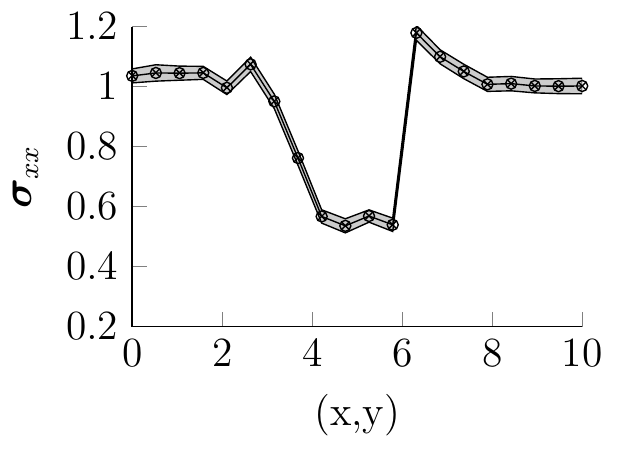} & 
    \includegraphics[height=5.5cm]{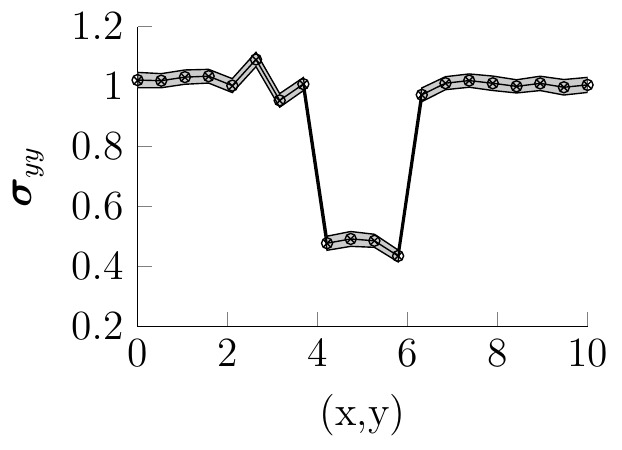}  \\
  \end{tabular}
  \includegraphics[height=5.5cm]{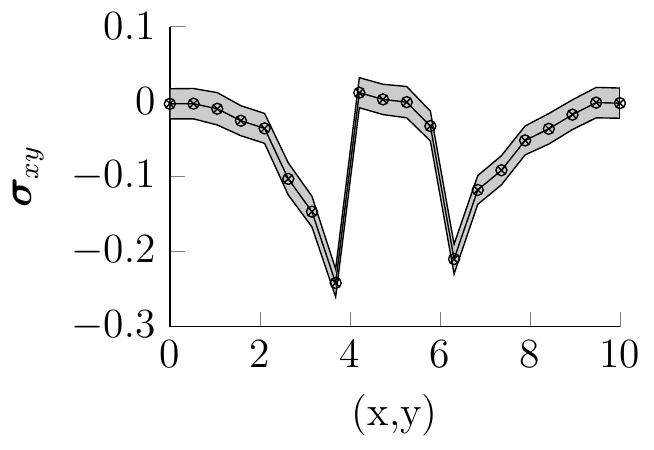}
  \caption{Comparison of reference displacements and stresses obtained from a deterministic FE model along the diagonal from $(0,0)$ to $(10,10)$ with the inferred  means and 95\% confidence intervals.}\label{fig:2d_forward_reference_comparison}
\end{figure}

The problem domain from~\autoref{fig:2D_problem} is discretized using a randomized  Delaunay triangulation\footnote{The Triangle software was used  available from: \url{https://www.cs.cmu.edu/~quake/triangle.html} }   resulting in $n_{el} = 325$ elements, $n_{nodes} = 184$ nodes. The corresponding number of latent variables (i.e. stresses $\bm \Sigma$ and displacements $\bm U_i$) was $1343$ and the number of parameters $\bm \phi$ in the approximating density $146989$ (i.e. $dim(\bm \mu)=1343$ and $dim(\bm L)=145646$). At the  center of the physical domain, an inclusion was considered as shown in  \autoref{fig:2d_forward_problem_inclusion_youngs_truth}. The spatial distribution of the Young's modulus was assumed to be (if $(x_e,y_e)$ are the coordinates of the center of an element $e$):
\be
c_e = \begin{cases} 0.2 & \textrm{if } (x_e-5)^2 + (y_e-5)^2 \leq 1 \\ 1.0 & \mathrm{otherwise.} \end{cases}
\label{eq:refemod}
\ee
 The evolution of the objective function $\mathcal{L}_{for}$ over the iterations for the parameter updates is shown in~\autoref{fig:2d_forward_problem_inclusion_youngs_truth}.  As previously discussed, the output of the proposed formulation for such a forward problem is a density for the latent variables, namely stresses $\bm \Sigma$ and (interior) displacements $\bm U_i$. Two-dimensional plots of  the means of these variables are contained in \autoref{fig:2d_forward_problem_inclusion}. More importantly, in  \autoref{fig:2d_forward_reference_comparison} the means as well as the $95\%$ credible intervals (CI)  along the diagonal line from the lower left corner at $(x,y) = (0,0)$ to the upper right corner at $(x,y) = (10,10)$ are compared with the  values (reference) one would obtain by solving deterministically the governing PDE using the same mesh/discretization.  The excellent agreement between the two indicate that the proposed augmented prior model correctly encapsulates the governing equations and can be employed as suggested in section \ref{sec:probabilistic_cm} instead of the black-box solver in the context of the ensuing inverse problems.

\subsection{Inverse problem without model error}\label{subsec:inverse_problem}

In this section we investigate the performance of the proposed formulation for an inverse problem without model error, i.e. assuming $\bm \lambda \to \infty$. To that end, we assume material properties as in \refeq{eq:refemod} and perform a forward analysis using a deterministic finite element solver which employed a different randomized mesh as compared to the one used for the inverse problem (\autoref{fig:2d_forward_problem_inclusion_youngs_truth}). The displacements computed were contaminated with various levels of additive   Gaussian noise in order to obtain the observables $\bs{u}_{obs}$. The  corresponding  Signal to Noise  Ratios (SNR) considered were   $SNR_{dB} = 60dB, 50 dB$ and $40dB$.

\begin{figure}[!htpb]
  \centering\textbf{Inverse problem solution - mean values}
  \begin{tabular}{ll}
  \includegraphics[height=5.5cm]{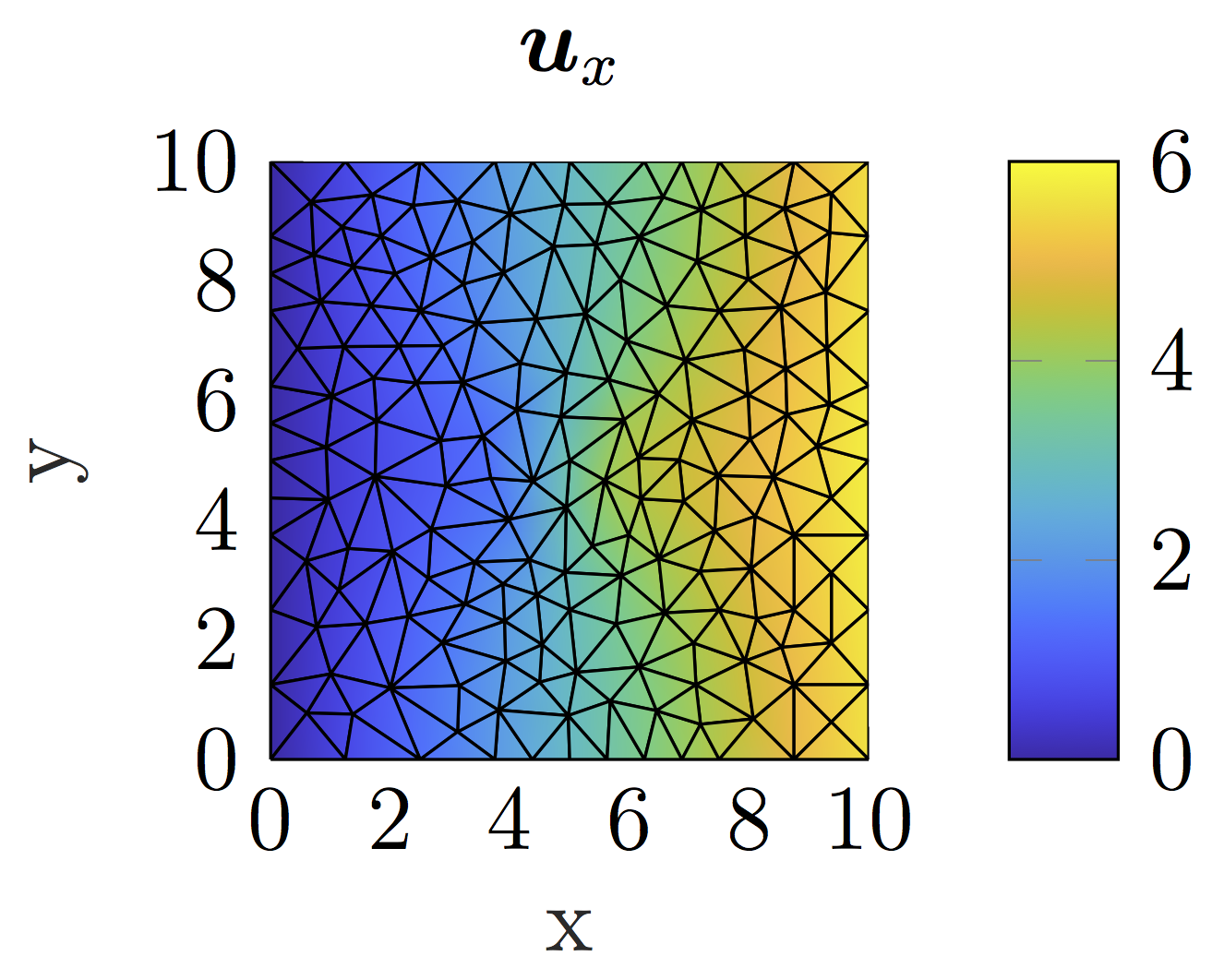} & 
  \includegraphics[height=5.5cm]{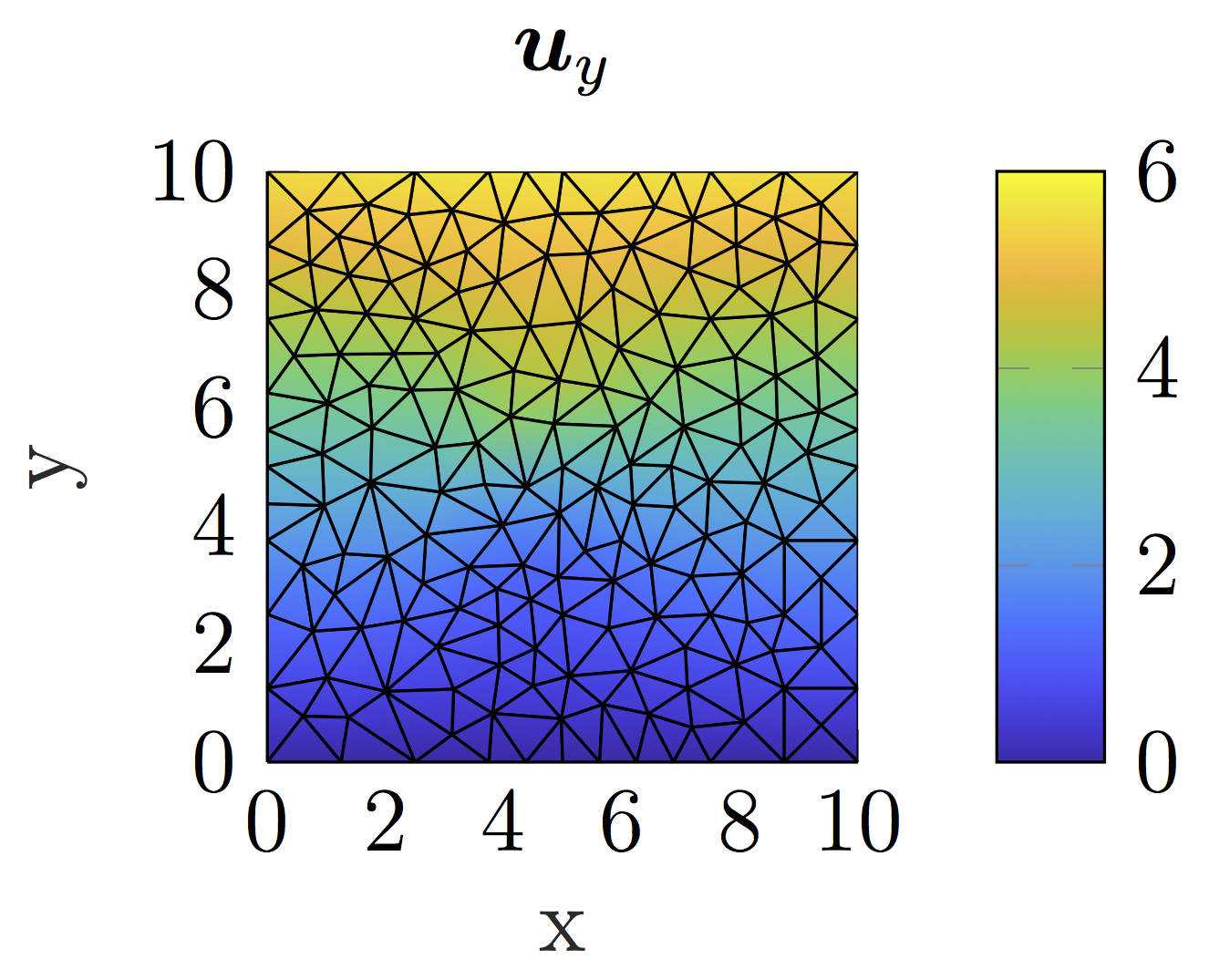} \\
  \includegraphics[height=5.5cm]{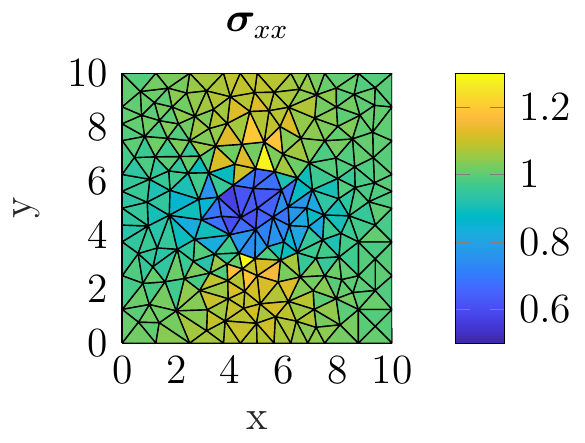} & 
  \includegraphics[height=5.5cm]{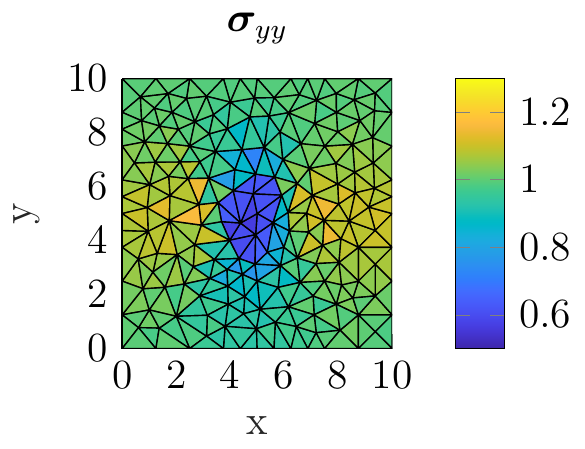} \\
  \includegraphics[height=5.5cm]{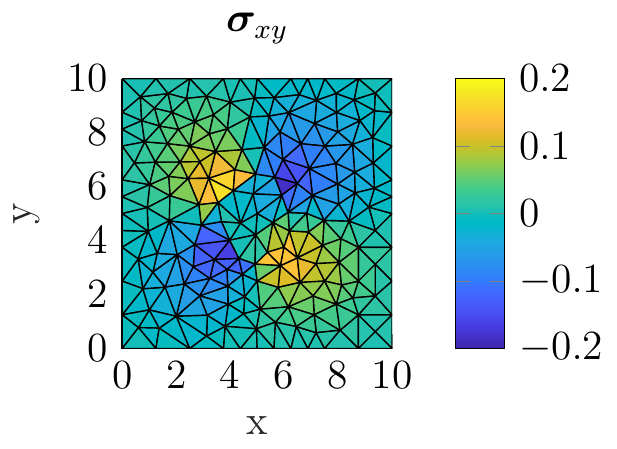} & 
  \includegraphics[height=5.5cm]{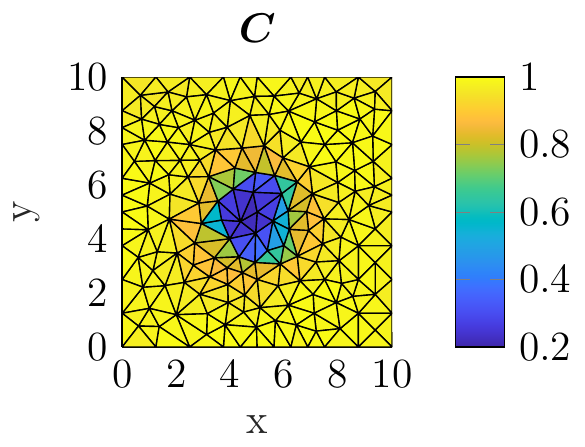} \\
  \end{tabular}
  \caption{Inferred posterior mean values for displacements $ u_x, u_y$, stresses $\sigma_{xx}, \sigma_{yy}, \sigma_{xy}$ and material parameters $\bm C$ for the   inverse problem (without model error - section \ref{subsec:inverse_problem} - $SNR_{dB} = 50 dB$).}\label{fig:2d_forward_problem_inclusion_proj_field_means}
\end{figure}

\begin{figure}[!htpb]
  \centering\textbf{Inverse problem solution - variances}
  \begin{tabular}{ll}
  \includegraphics[height=5.5cm]{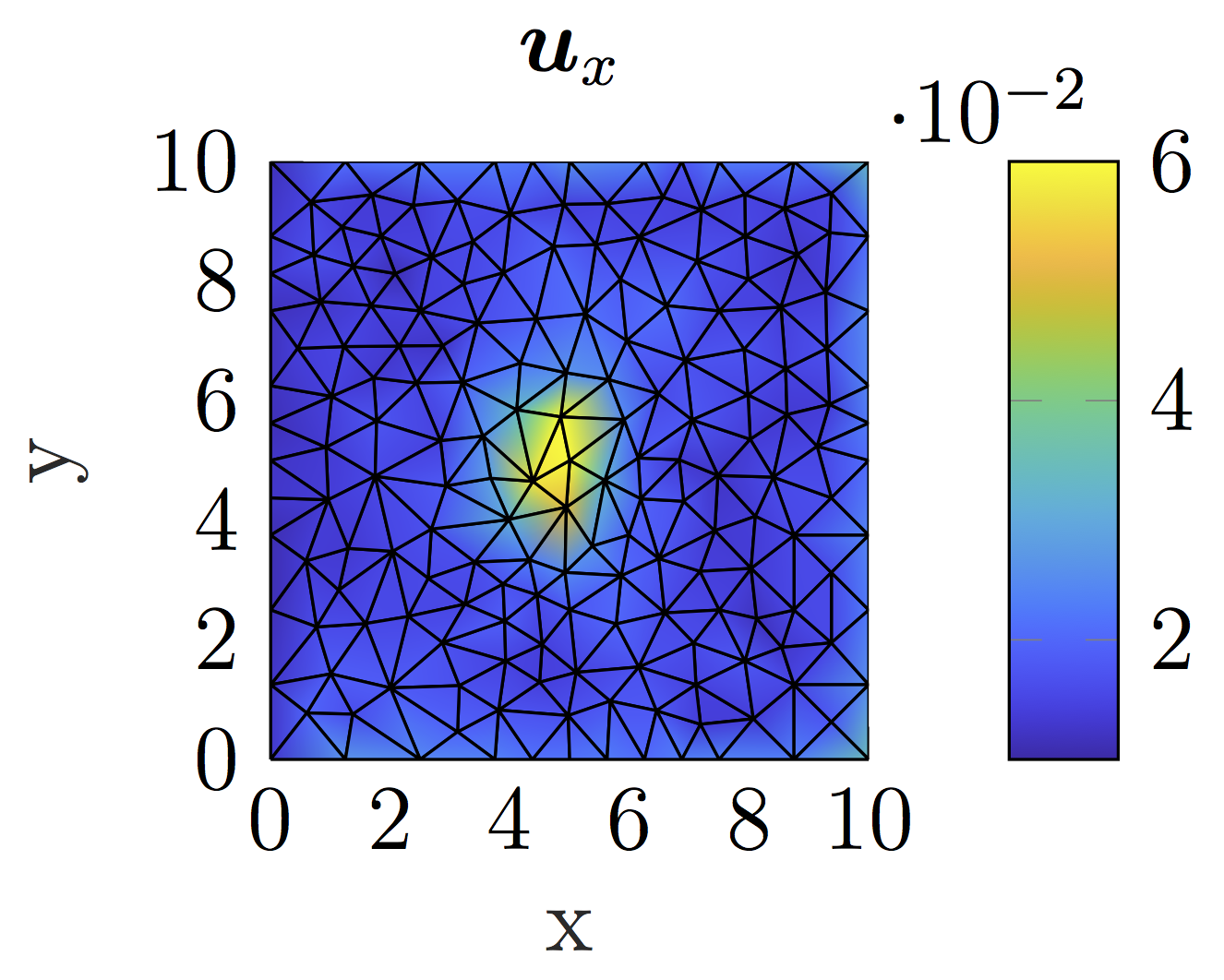} & 
  \includegraphics[height=5.5cm]{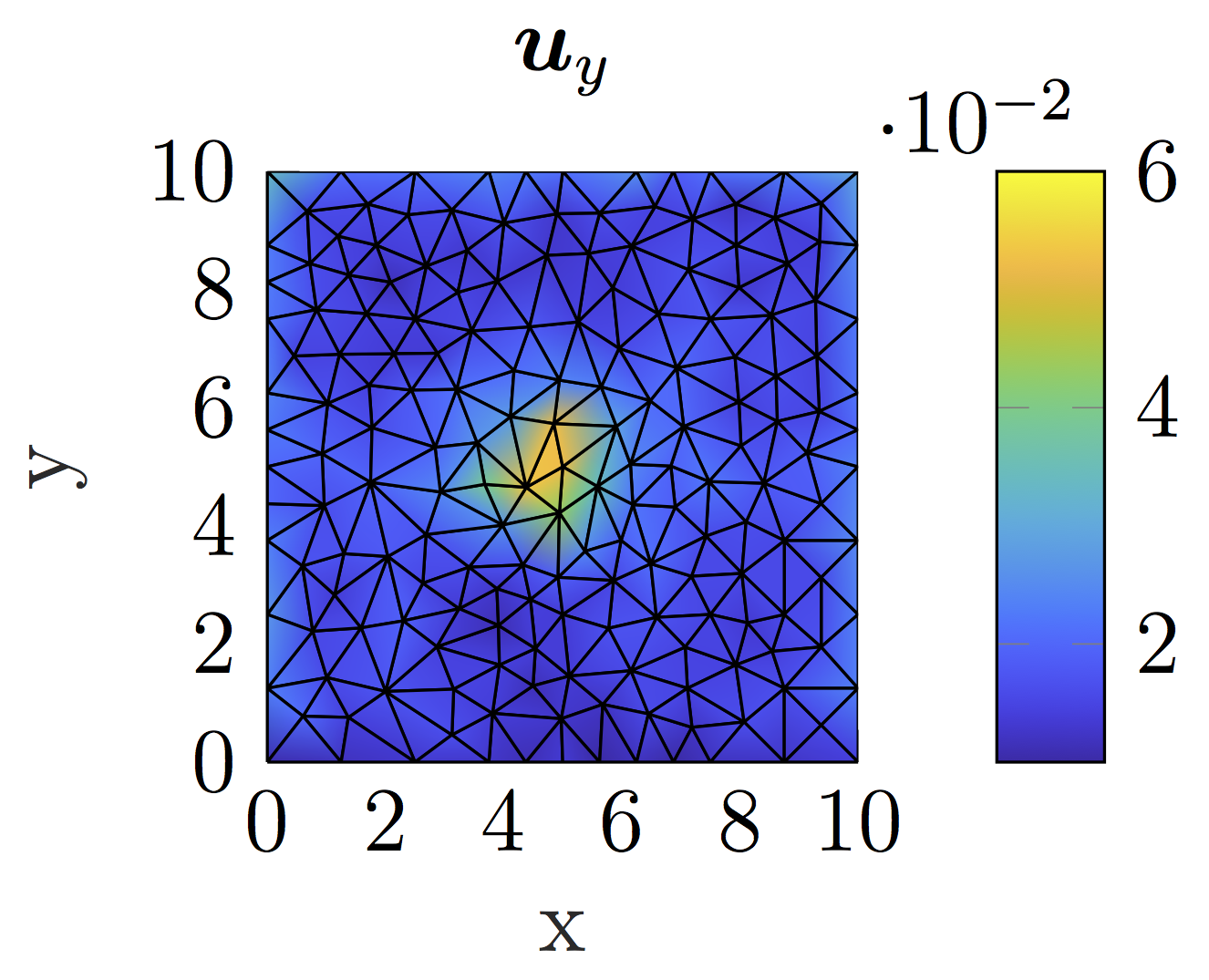} \\
  \includegraphics[height=5.5cm]{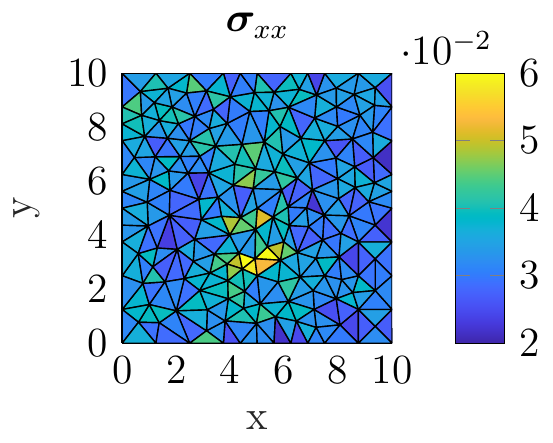} & 
  \includegraphics[height=5.5cm]{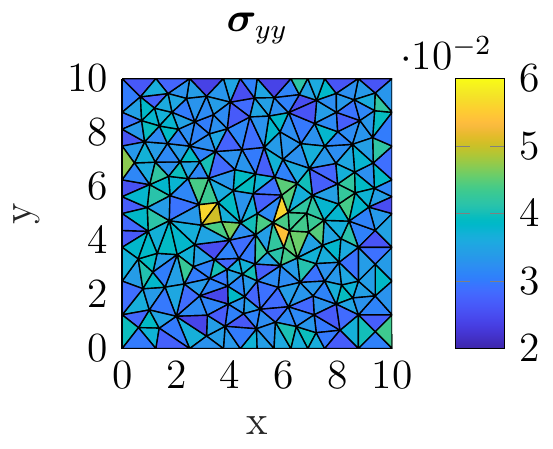} \\
  \includegraphics[height=5.5cm]{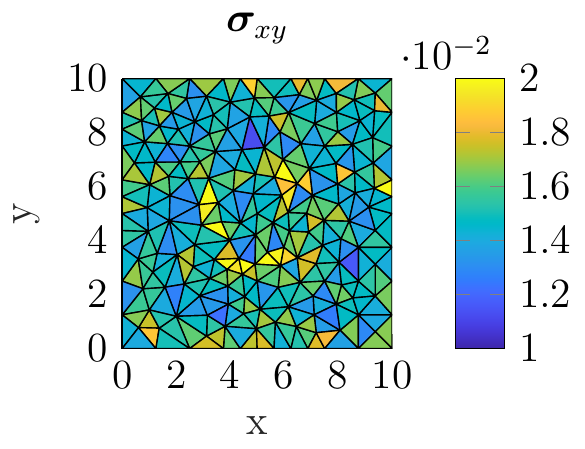} & 
  \includegraphics[height=5.5cm]{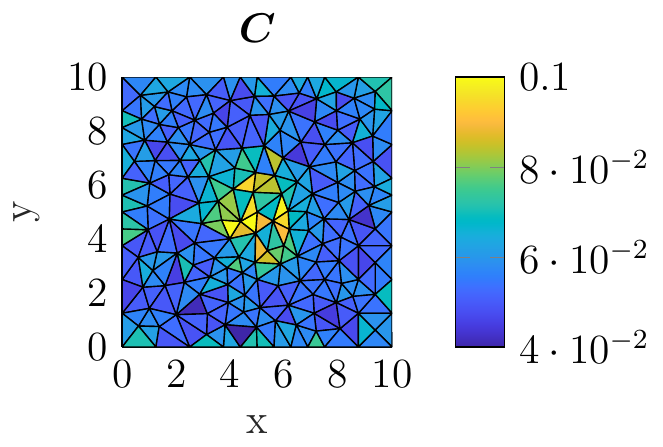} \\
  \end{tabular}
  \caption{Inferred posterior variance values for displacements $ u_x, u_y$, stresses $\sigma_{xx}, \sigma_{yy}, \sigma_{xy}$ and material parameters $\bm C$ for the   inverse problem (without model error - section \ref{subsec:inverse_problem}) - $SNR_{dB} = 50 dB$).}\label{fig:2d_forward_problem_inclusion_proj_field_variances}
\end{figure}

\begin{figure}[!htpb]
  \centering\textbf{Inverse problem solutions - comparison with reference}
  \begin{tabular}{lr}
  \includegraphics[height=5cm]{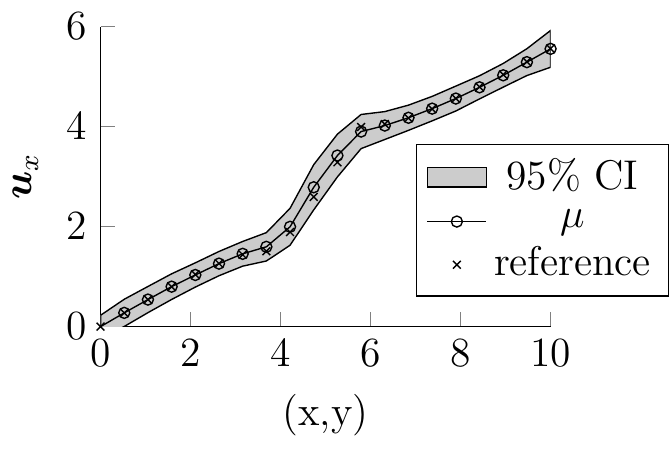} & 
  \includegraphics[height=5cm]{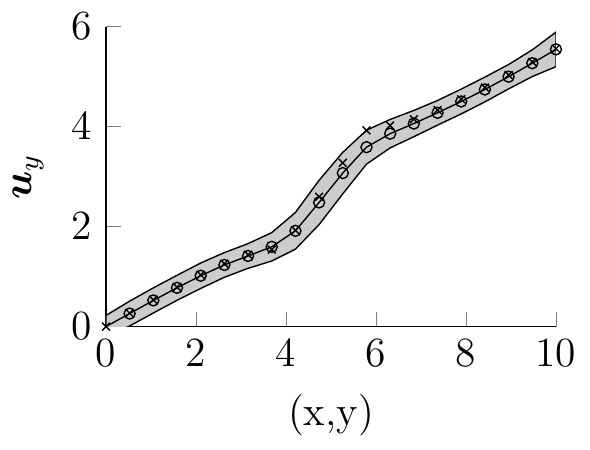} \\
  \includegraphics[height=5cm]{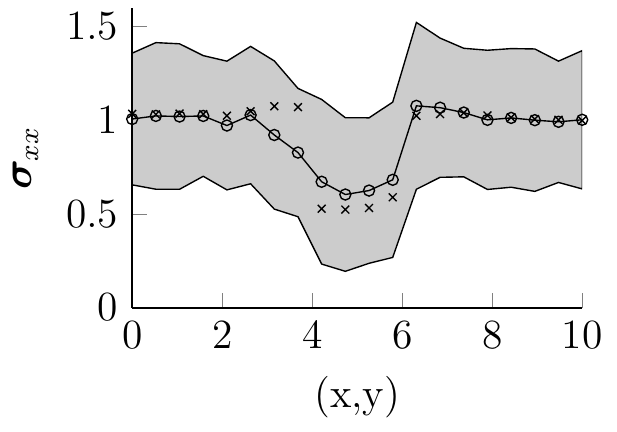} & 
  \includegraphics[height=5cm]{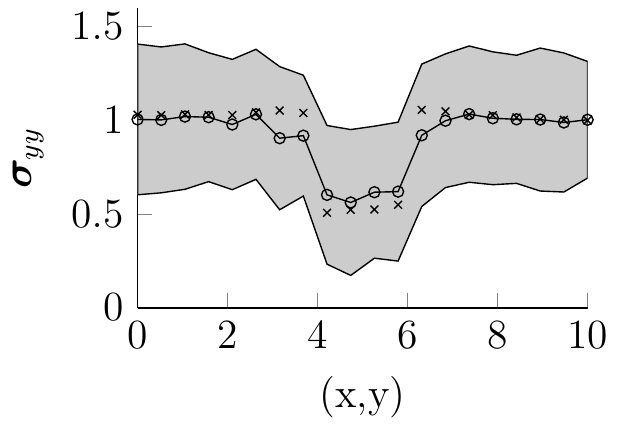} \\
  \includegraphics[height=5cm]{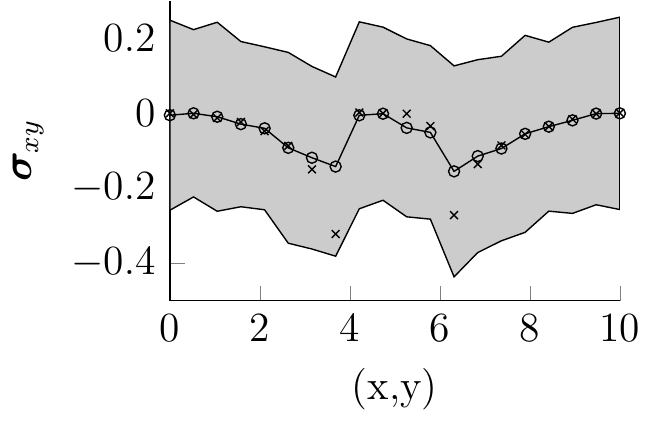} & 
  \includegraphics[height=5cm]{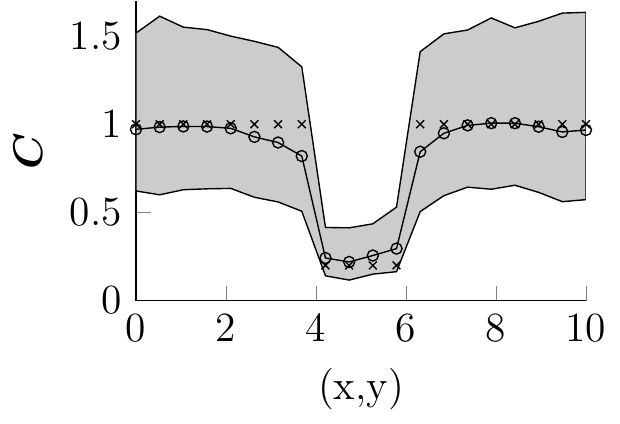} \\
  \end{tabular}
  \caption{Comparison of reference values  for displacements $ u_x, u_y$, stresses $\sigma_{xx}, \sigma_{yy}, \sigma_{xy}$ and material parameters $\bm C$ 
    along the diagonal  (from $(0,0)$ to $(10,10)$) with the inferred posterior means and 95\% confidence intervals ($SNR_{dB} = 50 dB$).}\label{fig:2d_inverse_problem_inclusion_proj_reference}
\end{figure}

\begin{figure}[!htpb]
  \centering\textbf{Inverse problem solutions - convergence behavior}
  \begin{tabular}{rr}
  \includegraphics[width=0.5\textwidth]{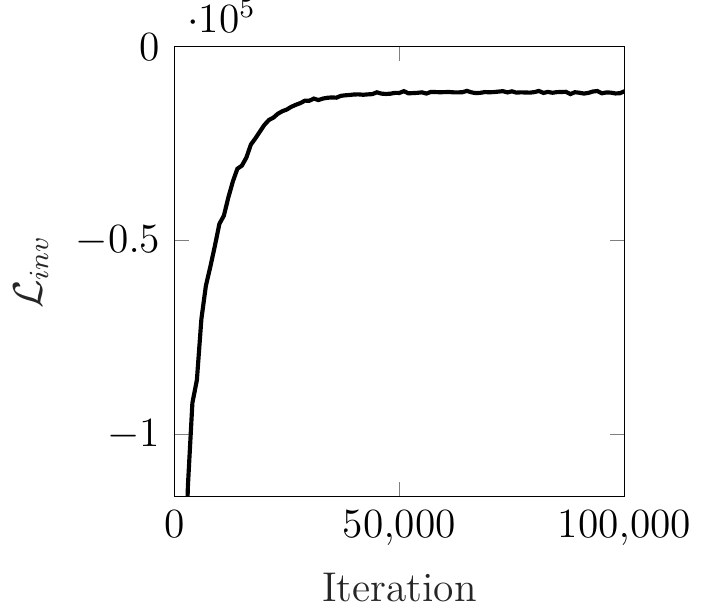} & 
  \includegraphics[width=0.45\textwidth]{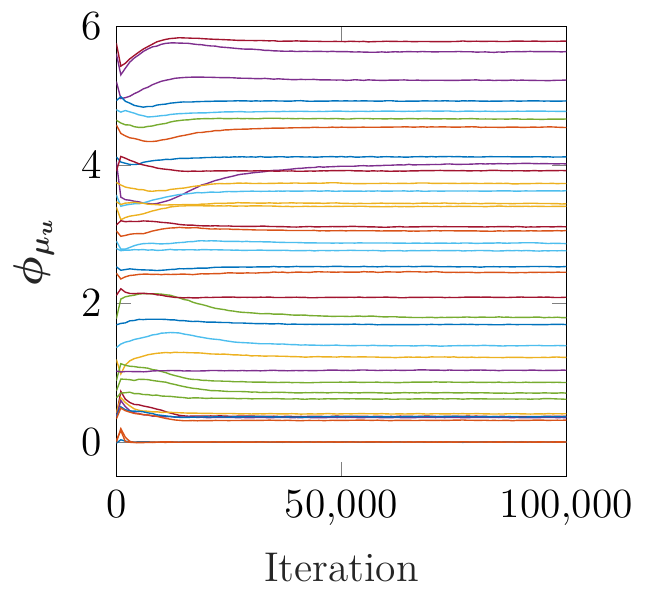} \\
  \includegraphics[width=0.45\textwidth]{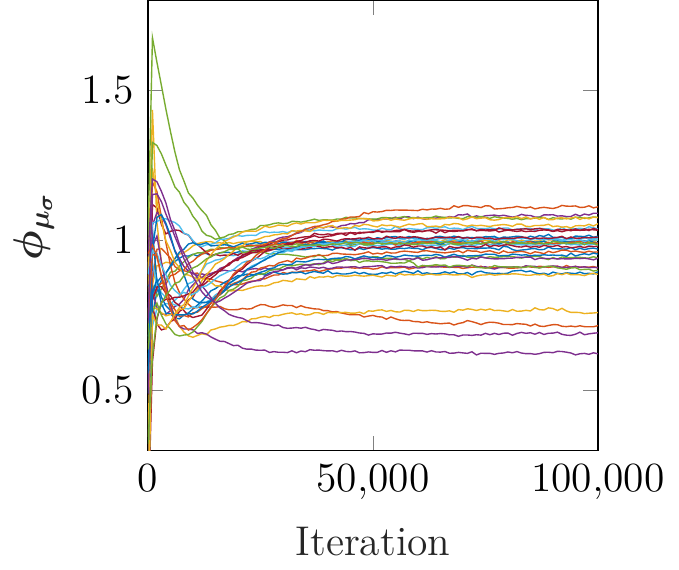} & 
  \includegraphics[width=0.45\textwidth]{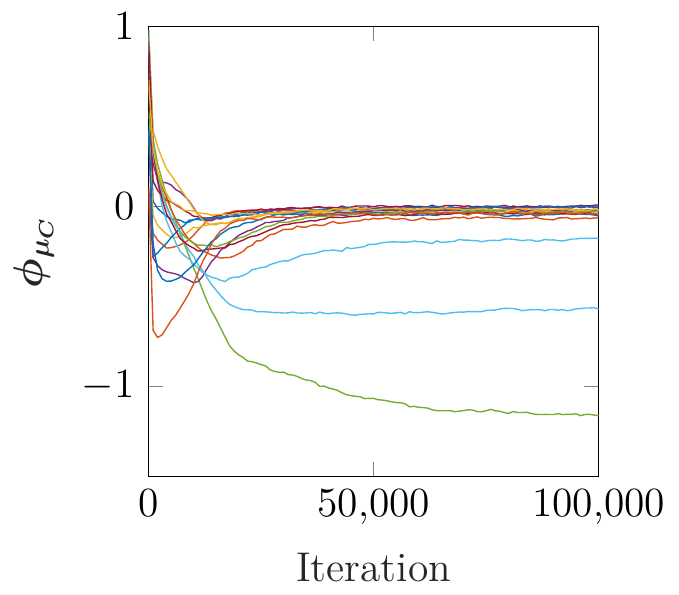} \\
  \end{tabular}
  \includegraphics[width=0.9\textwidth]{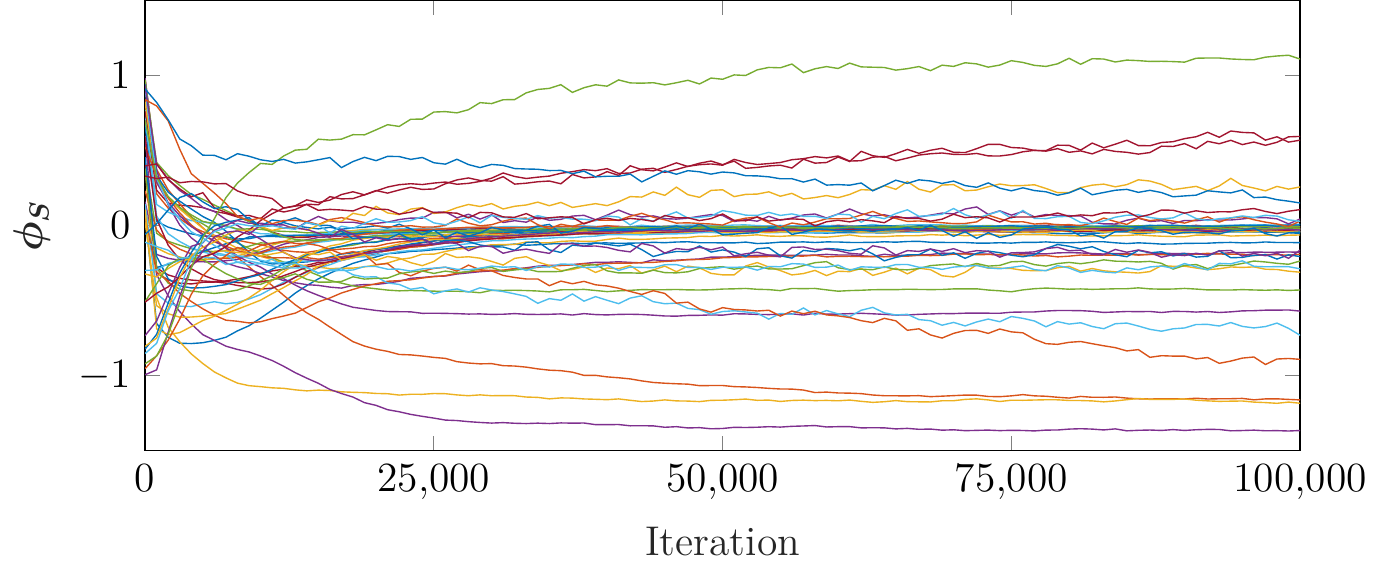} 
  \caption{Evolution of the objective function $\mathcal{L}_{inv}$ (upper left) and convergence behavior of the displacement means $\bm\phi_{\bm\mu_{\bm{u}}}$ (upper right), stress means $\bm\phi_{\bm\mu_{\bm\sigma}}$ (center left) and material parameter means $\bm\phi_{\bm\mu_{\bm{C}}}$ (center right), along with the covariance parameters of the banded covariance model $\bm \phi_{\bm S}$ (bottom). Only a subset of the $\bm \phi$'s are shown. }\label{fig:2d_inverse_problem_inclusion_proj_parameter_convergence}
\end{figure}

In such an inverse  problem, the latent variables are of dimension $1668$ and include stresses $\bs{\Sigma}$, (interior) displacements $\bs{U}_i$ as well as the material parameters $\bs{C}$. Using the banded covariance model  gave rise to  $ 226 578$ parameters $\bm \phi$. Indicative two-dimensional plots for $SNR_{dB} = 50 dB$ of the posterior means and variances  for all latent fields/variables are contained in ~\autoref{fig:2d_forward_problem_inclusion_proj_field_means} and  ~\autoref{fig:2d_forward_problem_inclusion_proj_field_variances} respectively. In \autoref{fig:2d_inverse_problem_inclusion_proj_reference}, posterior means and $95\%$ confidence intervals along the diagonal are depicted and compared with the reference solutions.  In all cases, good agreement is observed and the predicted uncertainty envelops the ground truth.
 The evolution of the ELBO objective $\mathcal{L}_{inv}$ as well as convergence plots of  selected optimization parameters are provided in~\autoref{fig:2d_inverse_problem_inclusion_proj_parameter_convergence}.

\begin{figure}[!htpb]
  \centering \textbf{Inverse problem solution - different noise levels}
  \begin{tabular}{lr}
  \includegraphics[height=4.1cm]{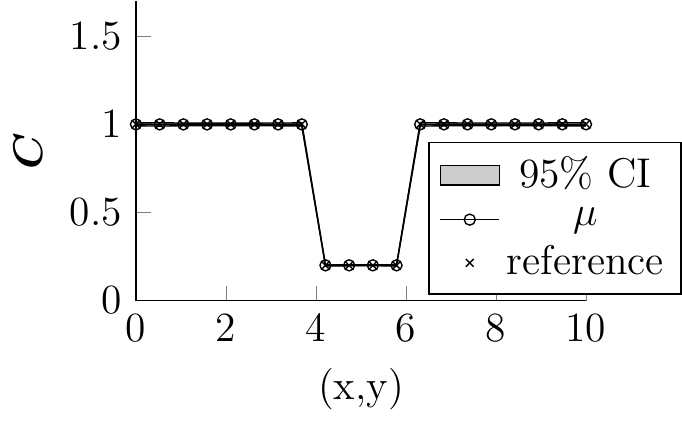} & 
  \includegraphics[height=4.5cm]{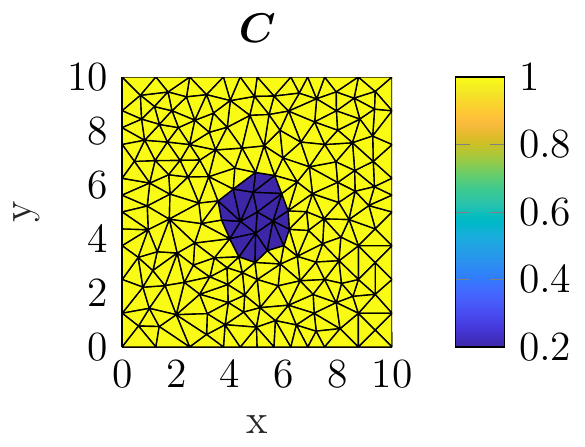} \\
  \includegraphics[height=4.1cm]{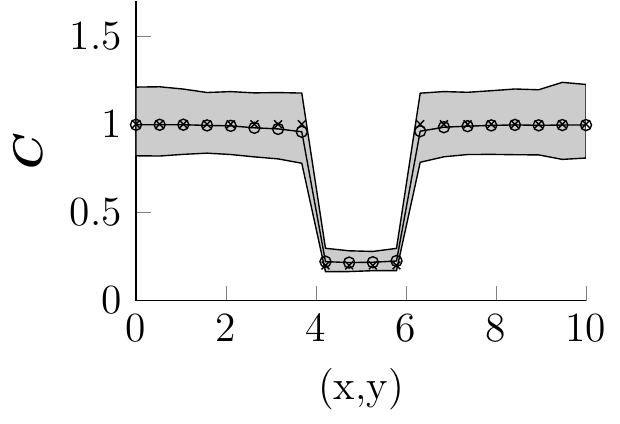} & 
  \includegraphics[height=4.5cm]{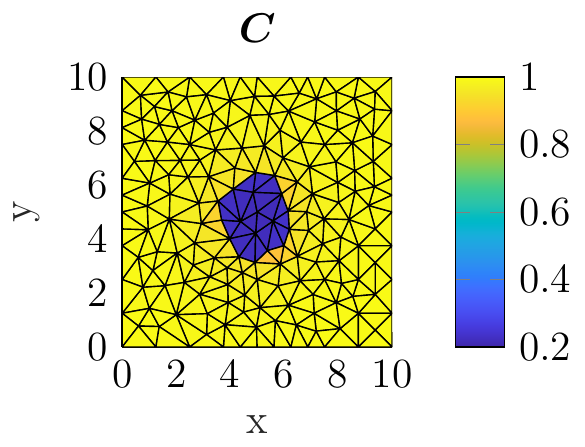} \\
  \includegraphics[height=4.1cm]{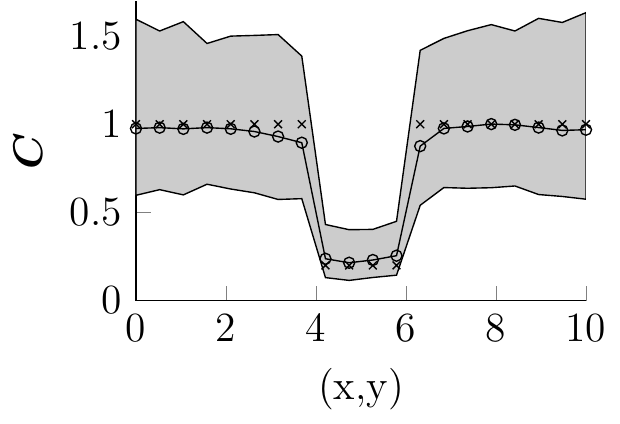} & 
  \includegraphics[height=4.5cm]{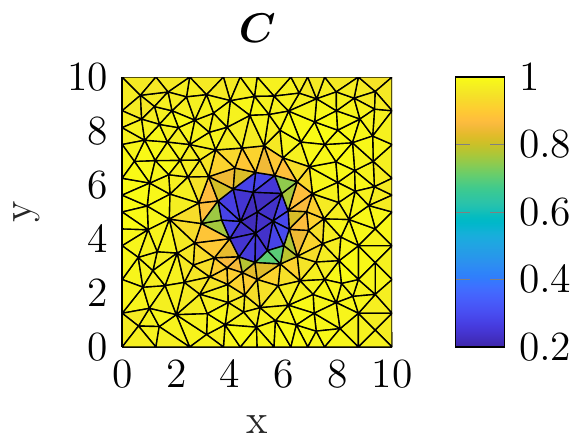} \\
  \includegraphics[height=4.1cm]{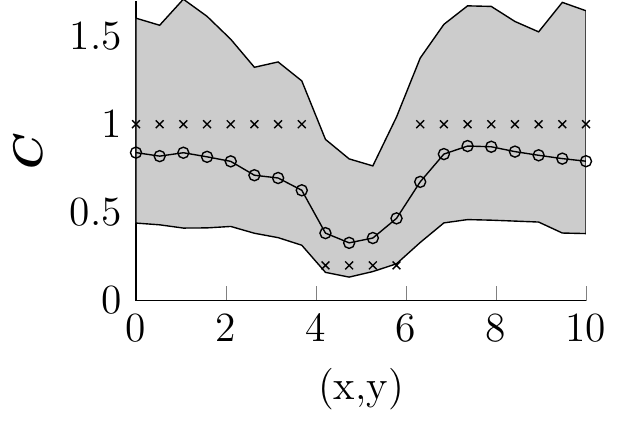} & 
  \includegraphics[height=4.5cm]{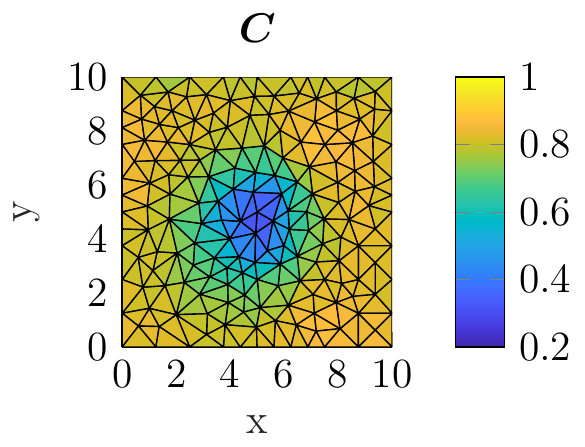} \\
  \end{tabular}
  \caption{Inference results at different SNRs. The plots show the inferred means of the Young's moduli along the diagonal (left column) and over the whole domain (right column)  for no noise (top row) as well as for 60dB (2nd row), 50dB (3rd row) and 40dB (4th row)  noise levels.}\label{fig:2d_inverse_problem_inclusion_comparison}
\end{figure}

For comparison purposes, posterior statistics of the material parameters are depicted  in~\autoref{fig:2d_inverse_problem_inclusion_comparison},  corresponding to various SNRs.  
As expected, and as the noise level increases, the total variation prior for the material parameters plays an essential role as a regularizer. An important feature of the banded covariance variational approximation is that in all cases it is able to capture the truth in the 95\% confidence interval, even when the noise becomes very large.

\subsection{Inverse problem with model error}\label{subsec:model_error_problem}

\begin{figure}[!htpb]
  \centering
  \includegraphics[height=5cm]{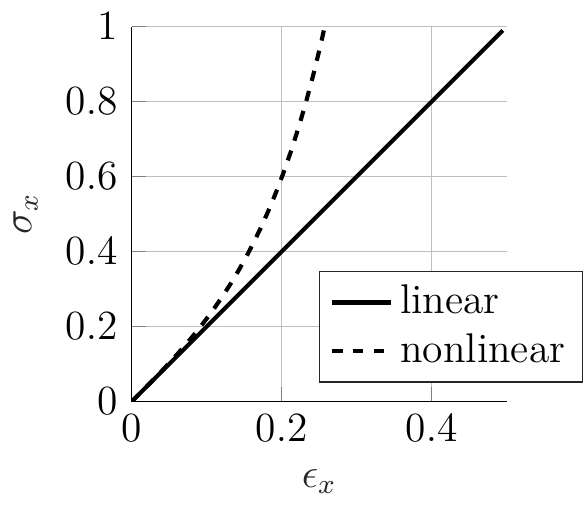}
  \caption{Employed nonlinear stress-strain  relation (in one dimension) for  the inclusion material.}\label{fig:inverse_model_error_stress_strain_relation}
\end{figure}

In this final problem scenario, we consider an inverse problem in the presence of model errors. We employed the same deterministic forward model described in the previous section with one fundamental difference with regards to the constitutive law for the inclusion. In particular, we employed the following stress-strain relations \footnote{The matrix $\bm D=\frac{1}{1 - \nu^2} \begin{bmatrix} 1 & \nu & 0 \\ \nu & 1 & 0 \\ 0 & 0 & \frac{1-\nu}{2} \end{bmatrix}$ represents the standard stress-strain relation in plane stress when the Young's modulus is $1.0$.}: 
\begin{align}
  \bm\sigma_e = \begin{cases} \bm{D} \bm{\epsilon}_e \odot \exp \left[ 10 ~\bm{\epsilon}_e  \odot \bm{\epsilon}_e \right] & \textrm{if the element is in the inclusion}, \\
  \bm{D} \bm{\epsilon}_e,  & \mathrm{otherwise,} \end{cases}
\end{align}
 where $\odot$ denotes component-wise multiplication for vectors and the exponential is taken component-wise. Thus, the constitutive behavior of the material in the inclusion is {\em nonlinear}. This particular  material model is motivated by~\citep{sun_elastography_2009} where it was used it to reconstruct nonlinear breast tissue properties in the context of elastography. A plot of the resulting stress-strain relation compared to the linear one is provided in~\autoref{fig:inverse_model_error_stress_strain_relation}. 
 
 In order to assess the proposed method's ability to identify model errors, for the inverse problem, a linear elastic constitutive law is assumed (as in  the previous subsection) with unknown Young's moduli. The model error parameters $\bm\lambda$ are assumed unknown and one per element is employed, i.e. $325$ in total. In addition, we have $1668$ latent variables which  include, as in section \ref{subsec:inverse_problem}, stresses $\bs{\Sigma}$, (interior) displacements $\bs{U}_i$ as well as the material parameters $\bs{C}$.  We recall from section \ref{sec:variational_approx} that two approximating densities $q$ and $r$ are  needed  with parameters $\bm \phi$ and $\bm \xi$ respectively, each of which is of dimension  $226578$ (i.e. $453156$ in total).
 %

\begin{figure}[!htpb]
  \centering
    \includegraphics[height=5cm]{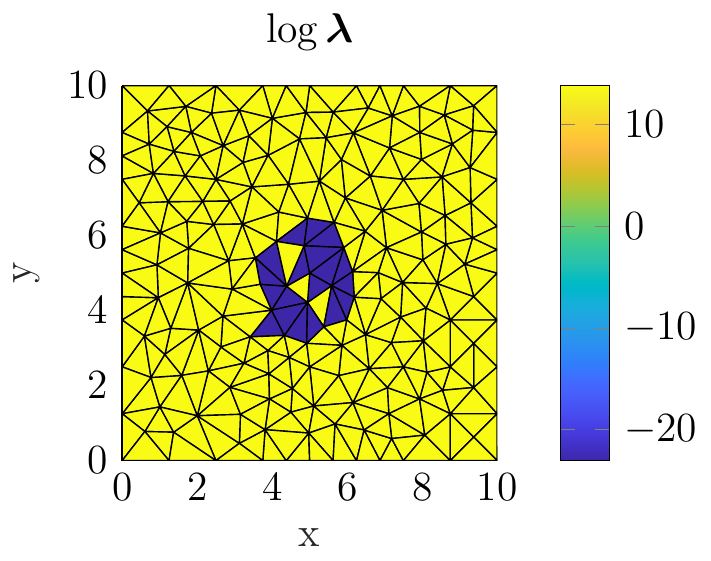} 
  \begin{tabular}{cc}
  \includegraphics[height=5cm]{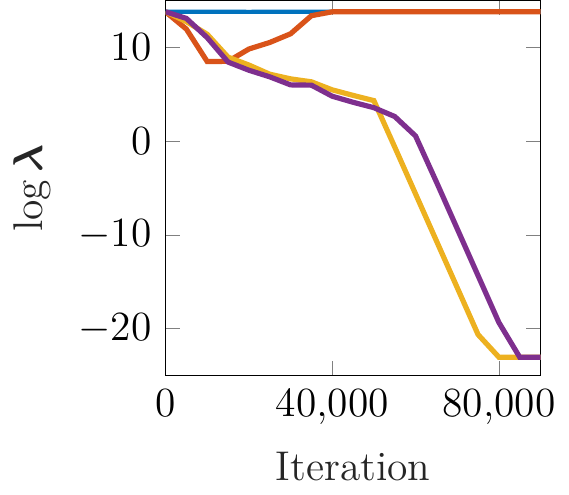} & 
  \includegraphics[height=5.4cm]{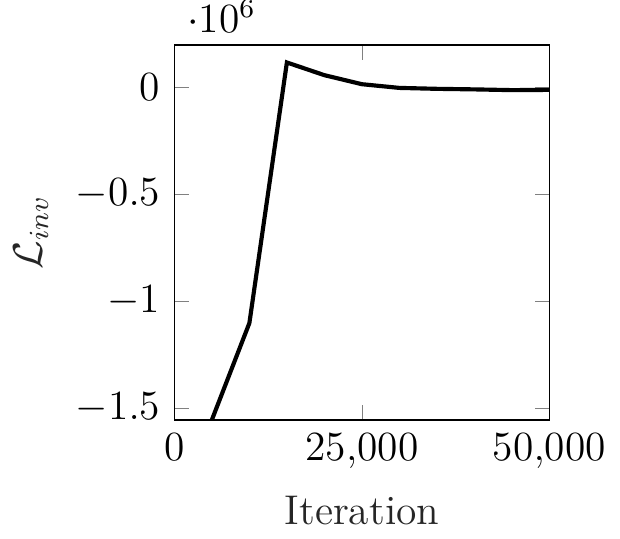} \\
  \end{tabular}
  \caption{Learned values for parameters $\bm\lambda$ in log-scale over the whole domain (top), convergence behavior for some selected $\bm\lambda$'s (lower left) and evolution of the objective function $\mathcal{L}_{inv}$ over the number of iterations (lower right).}\label{fig:model_error_solution_convergence}
\end{figure}

In \autoref{fig:model_error_solution_convergence}, the learned $\bm \lambda$'s are depicted as well as the evolution (of selected ones) over the EM iterations performed.  As it can be clearly seen, for almost all the elements belonging to the inclusion, the algorithm correctly identifies the presence of model error and sets the corresponding precisions  $\bm \lambda$ to very small values indicating large discrepancies between the true stresses and the ones predicted by the linear elastic constitutive law.
In the same Figure we show the evolution of the ELBO objective $\mathcal{L}_{inv}$.
As discussed in section \ref{sec:vii} there is no guarantee that $\mathcal{L}_{inv}$ will be always ascending due to the second variational approximation employed for the normalization constant of the prior.   

\begin{figure}[!htpb]
  \centering\textbf{Model error estimation - inferred mean values, variances and reference}
  \begin{tabular}{lll}
  \includegraphics[height=3cm]{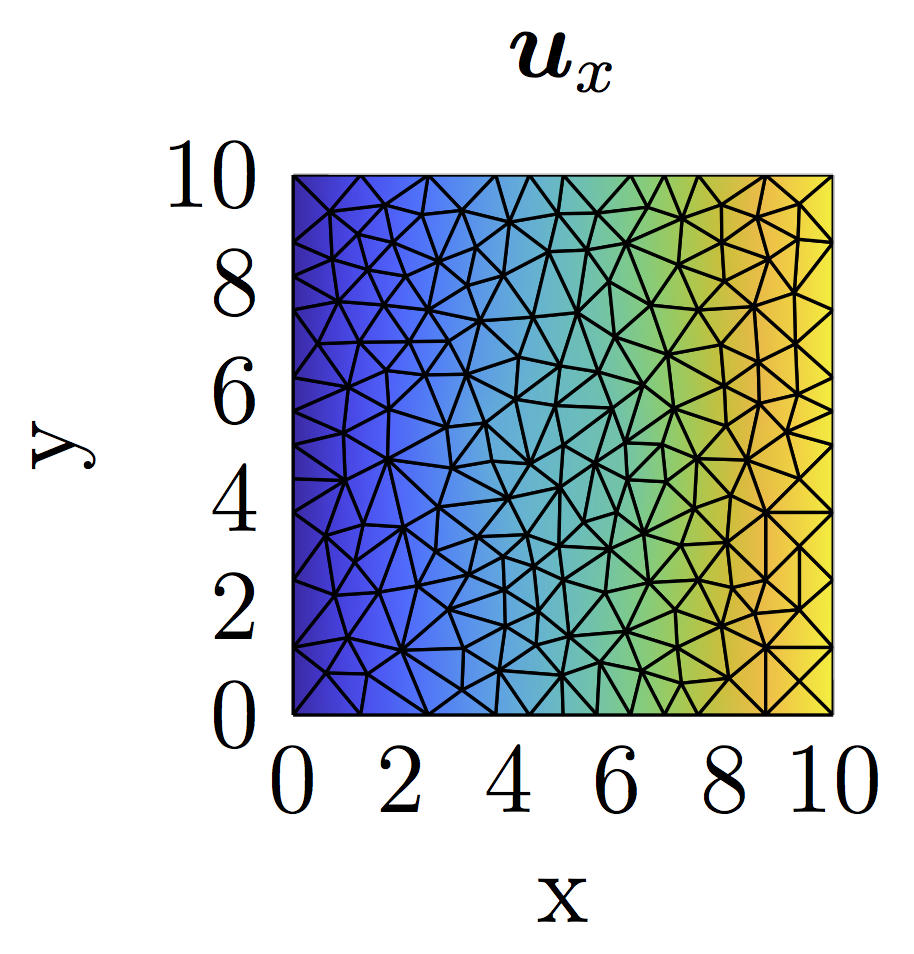} & 
  \includegraphics[height=3cm]{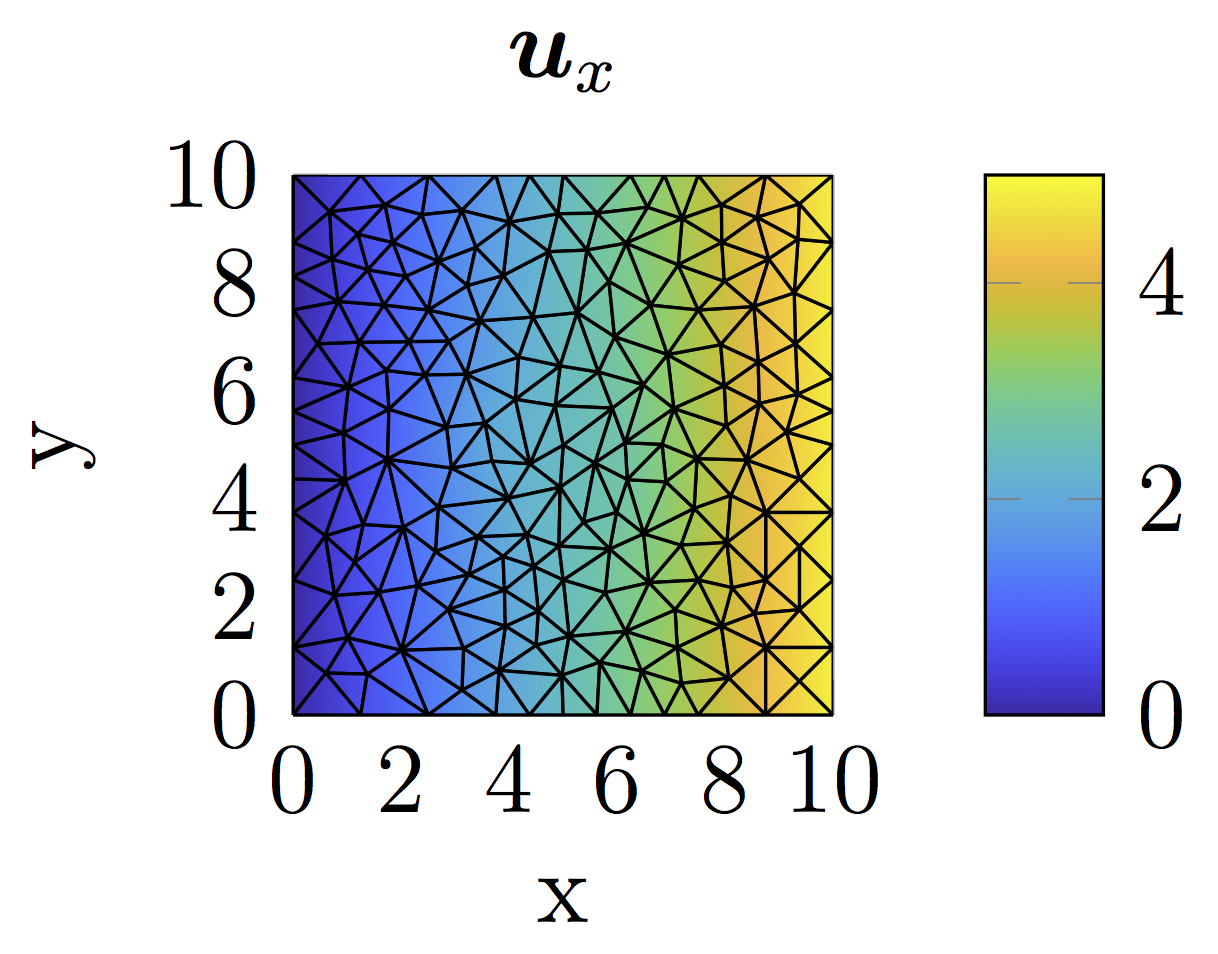} & 
    \includegraphics[height=3cm]{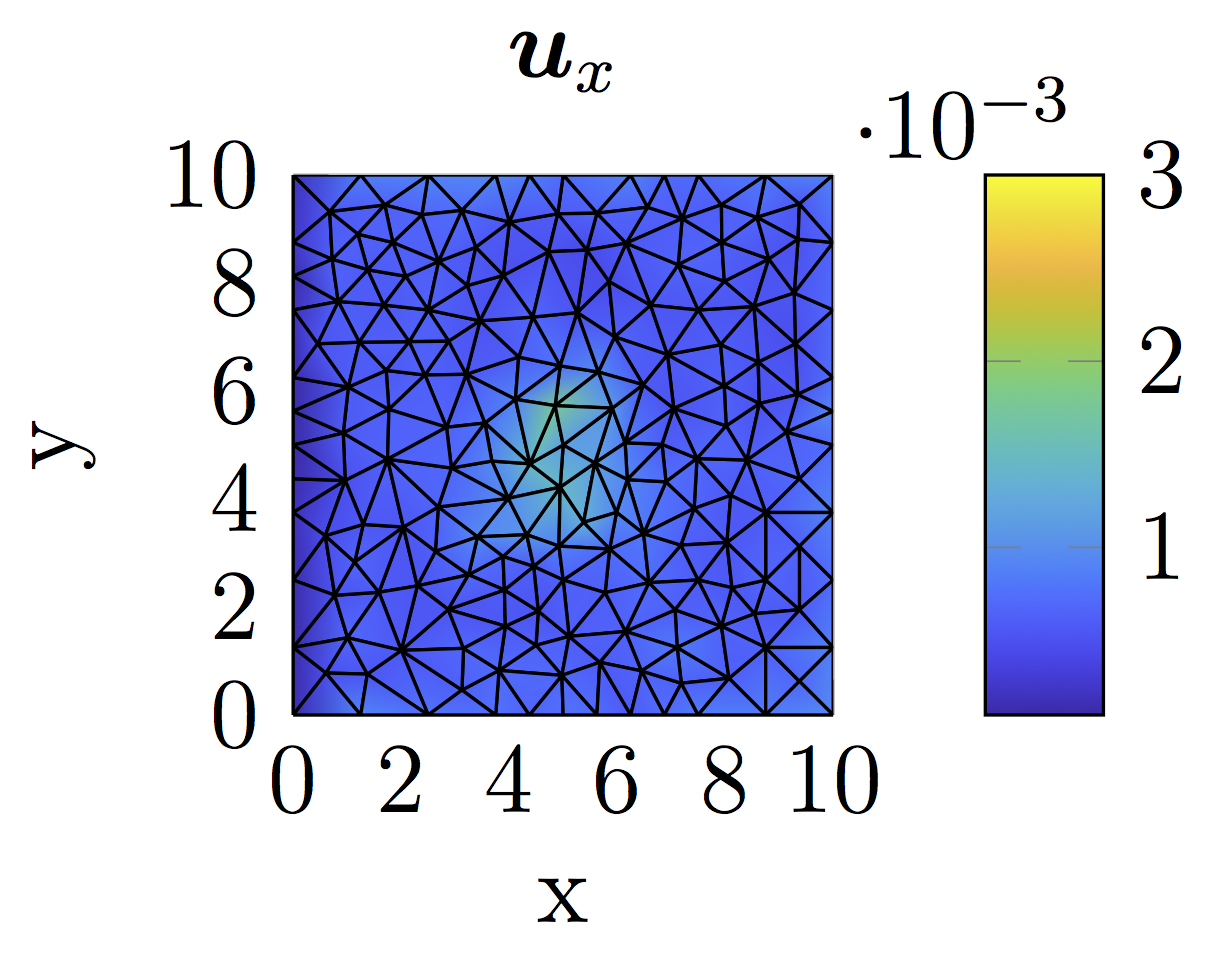} \\
     \includegraphics[height=3cm]{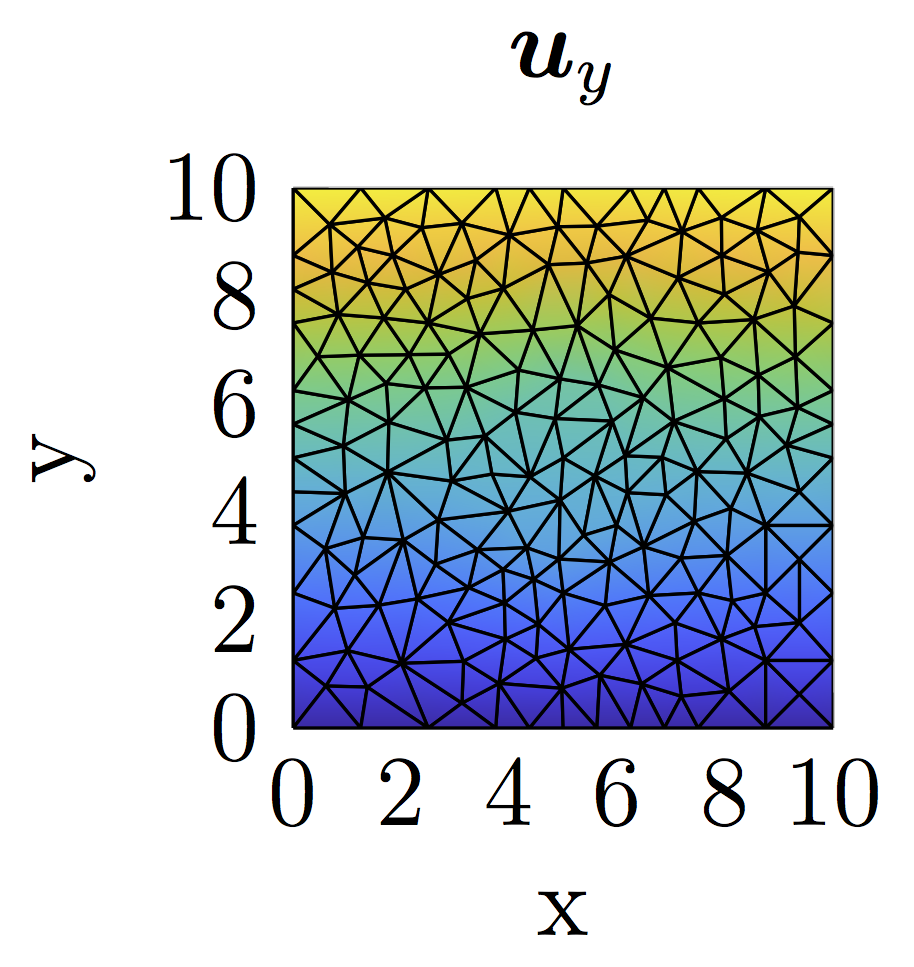} & 
  \includegraphics[height=3cm]{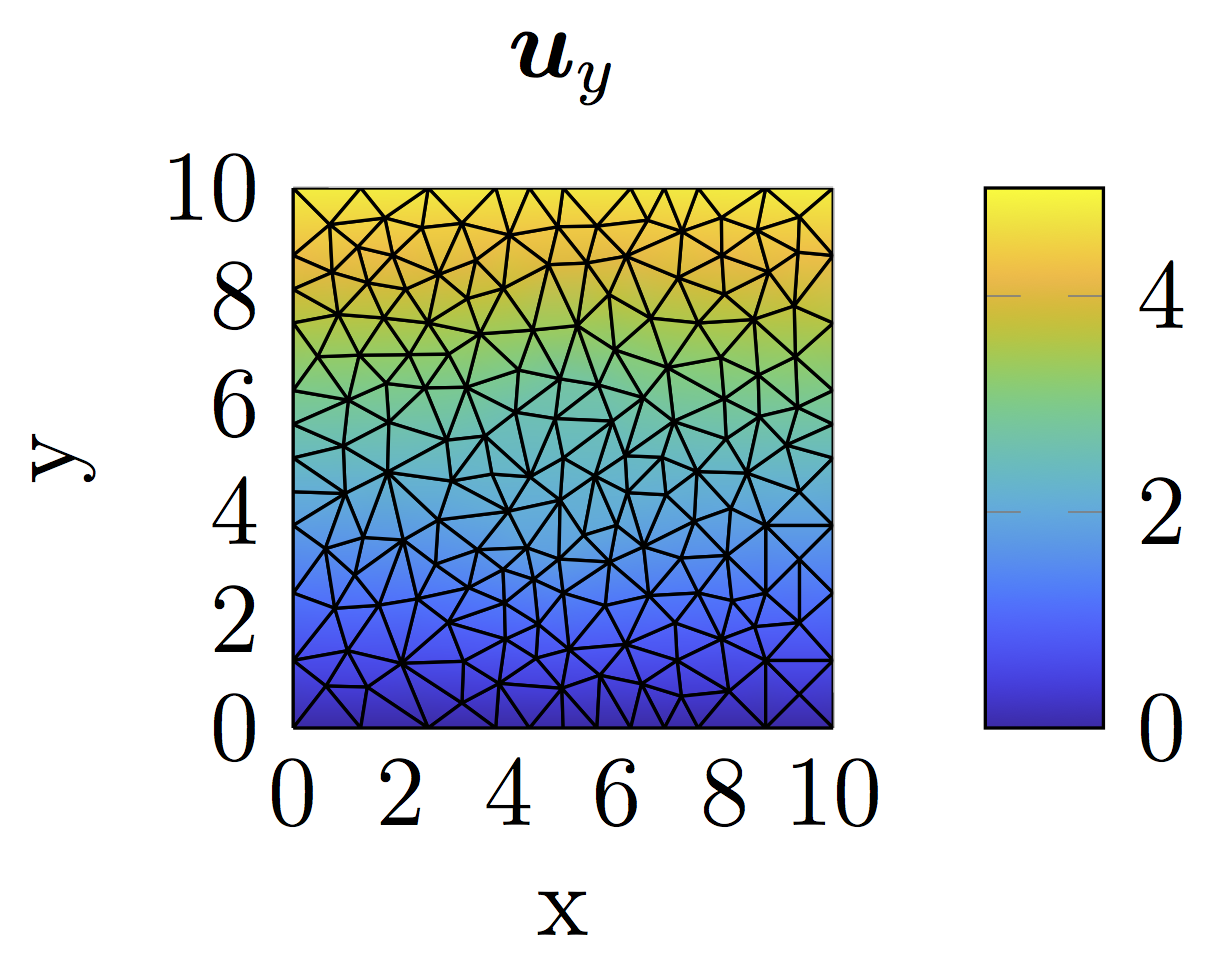} & 
    \includegraphics[height=3cm]{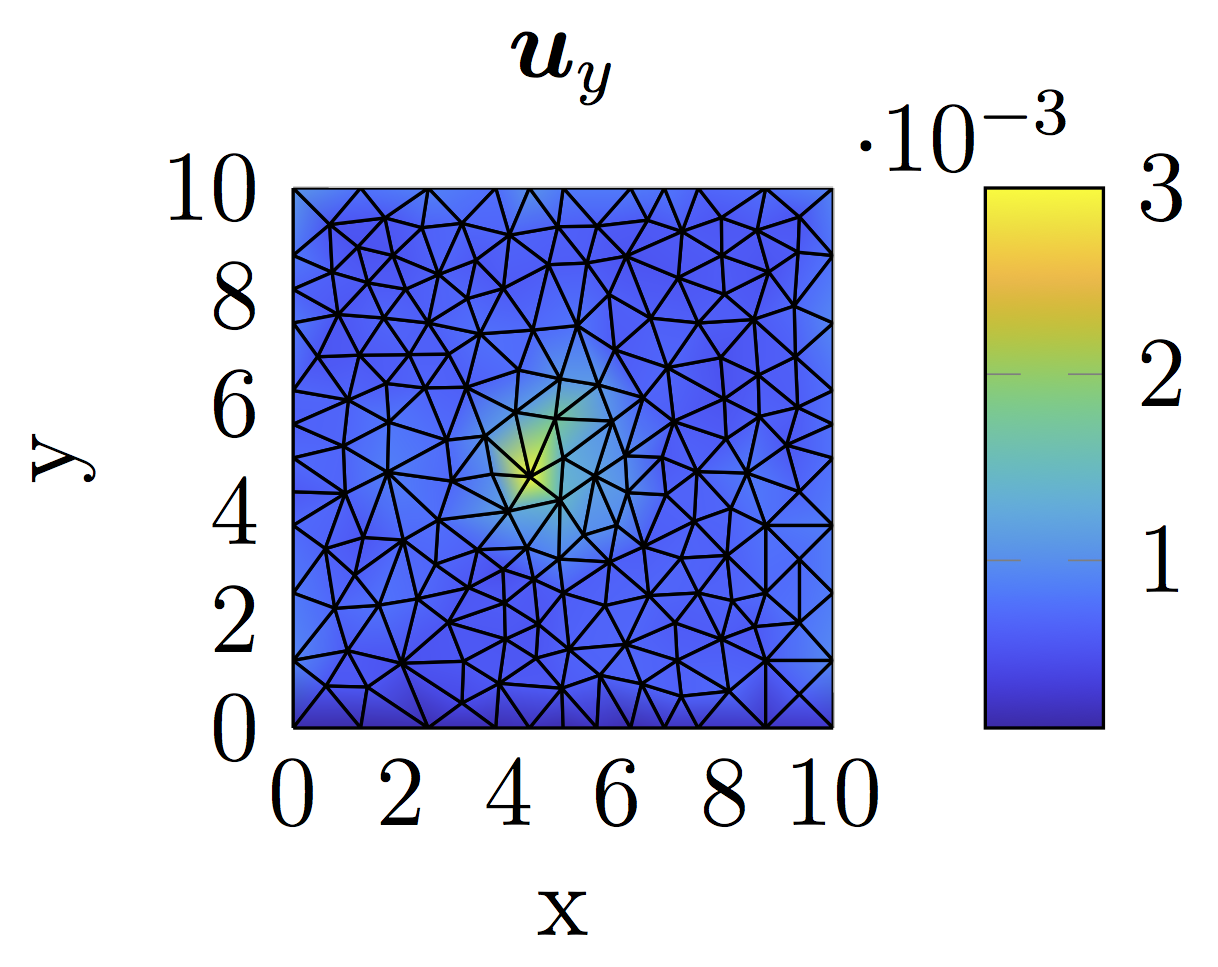} \\
     \includegraphics[height=3cm]{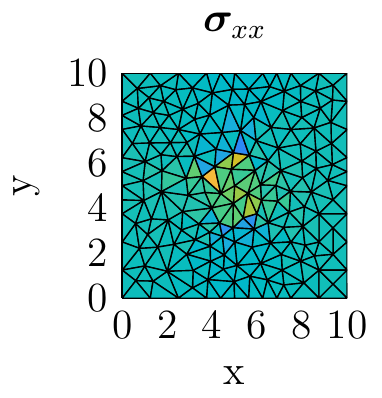} & 
  \includegraphics[height=3cm]{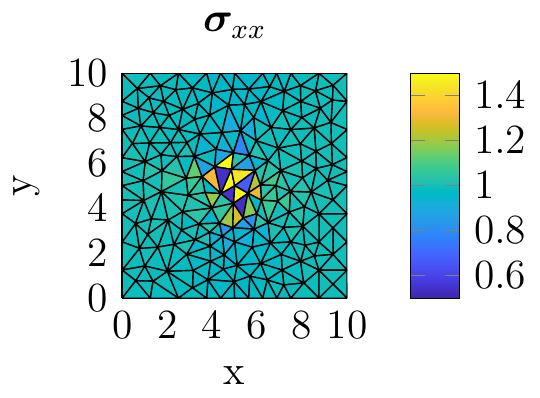} & 
    \includegraphics[height=3cm]{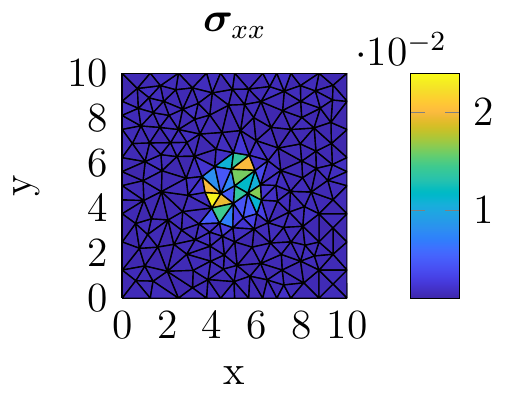} \\    
     \includegraphics[height=3cm]{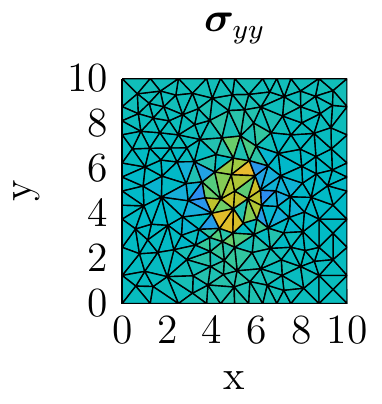} & 
  \includegraphics[height=3cm]{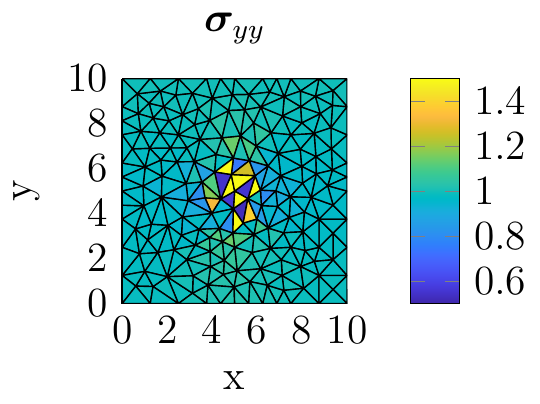} & 
    \includegraphics[height=3cm]{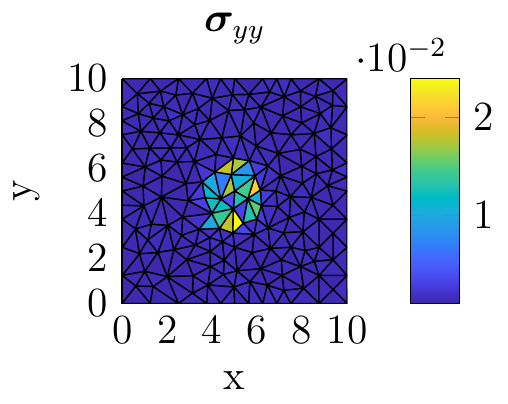} \\    
     \includegraphics[height=3cm]{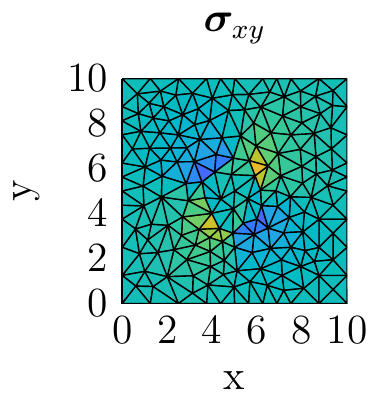} & 
  \includegraphics[height=3cm]{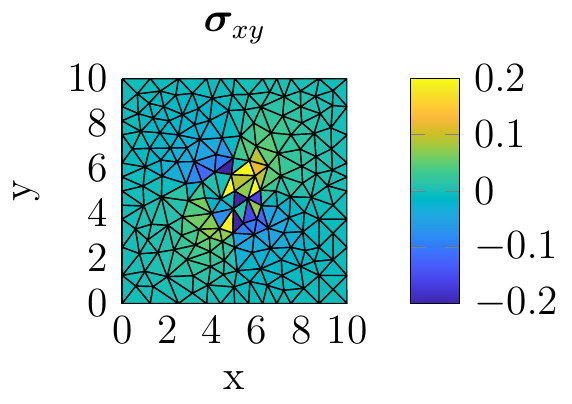} & 
    \includegraphics[height=3cm]{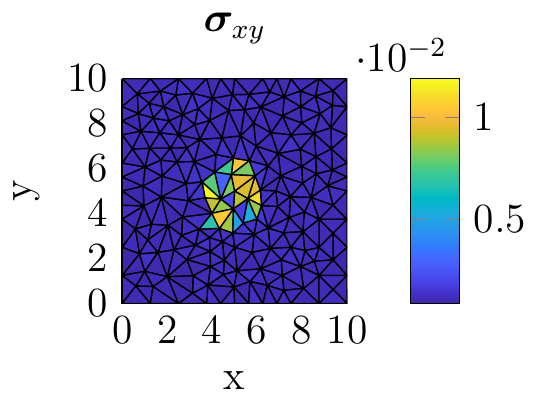} \\
    \includegraphics[height=3cm]{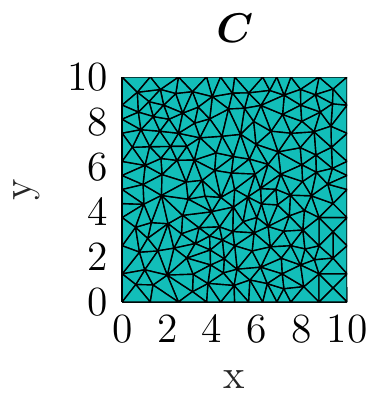} & 
  \includegraphics[height=3cm]{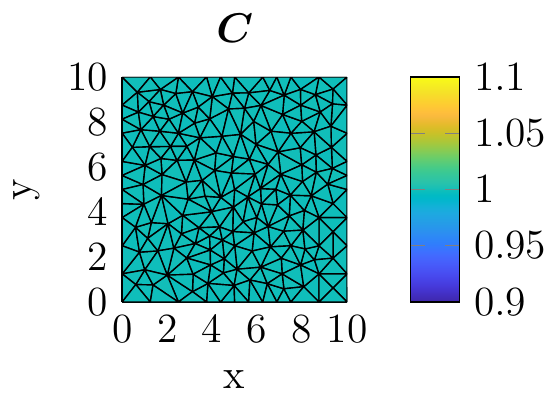} & 
    \includegraphics[height=3cm]{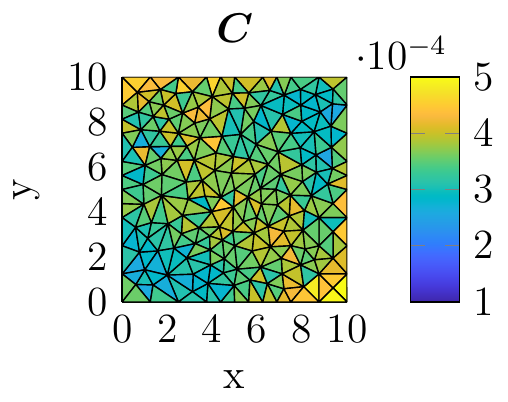} \\
  \end{tabular}
  \caption{ The plots compare the ground truth (left column), the inferred posterior means (center column) and variances (right column) of the displacements and stresses for the inverse problem with model error in section \ref{subsec:model_error_problem}.}\label{fig:2d_inverse_model_error_posterior_and_reference}
\end{figure}

\begin{figure}[!htpb]
  \centering\textbf{Model error estimation - banded covariance vs. diagonal covariance}
  \begin{tabular}{cc}
   \includegraphics[height=5cm]{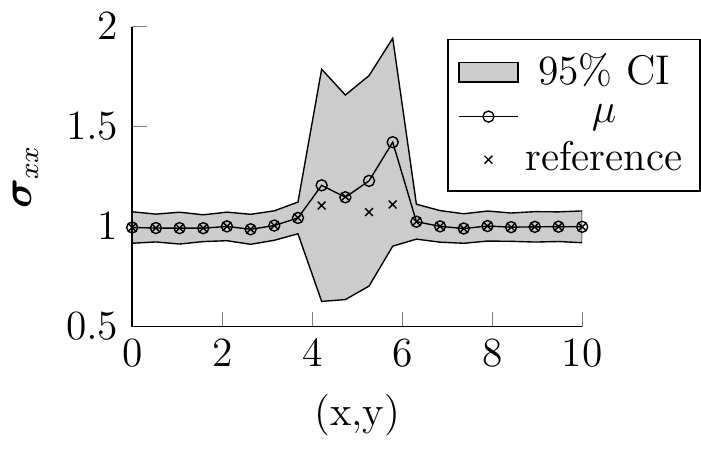} & 
    \includegraphics[height=5cm]{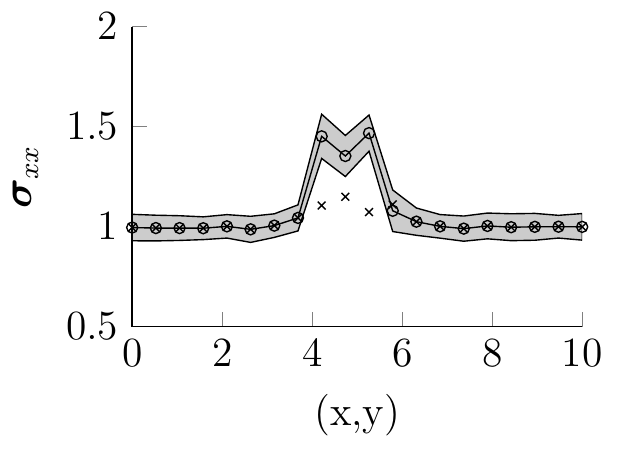} \\ 
     \includegraphics[height=5cm]{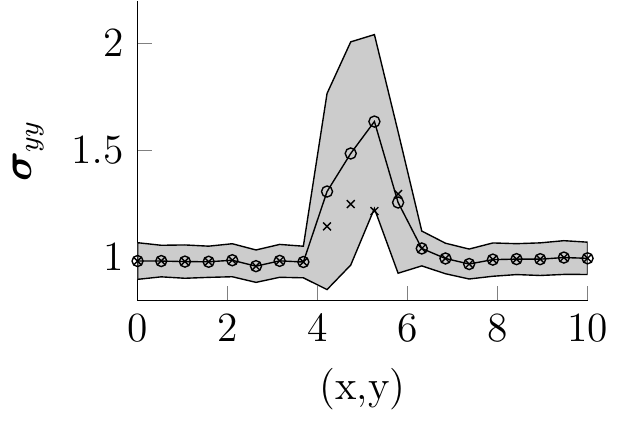} & 
    \includegraphics[height=5cm]{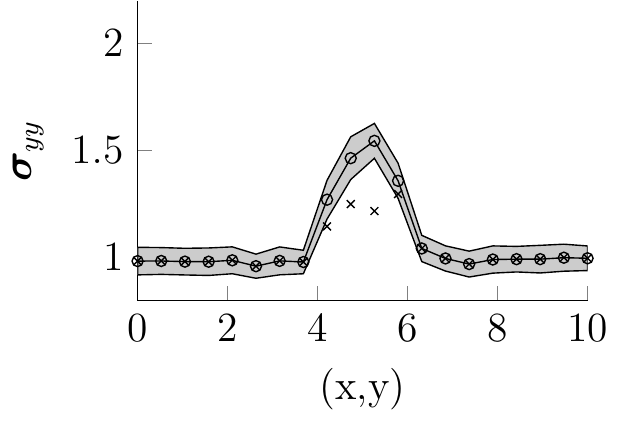} \\ 
     \includegraphics[height=5cm]{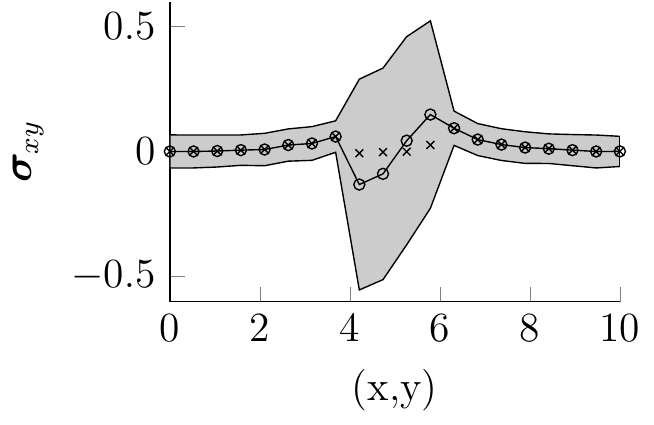} & 
    \includegraphics[height=5cm]{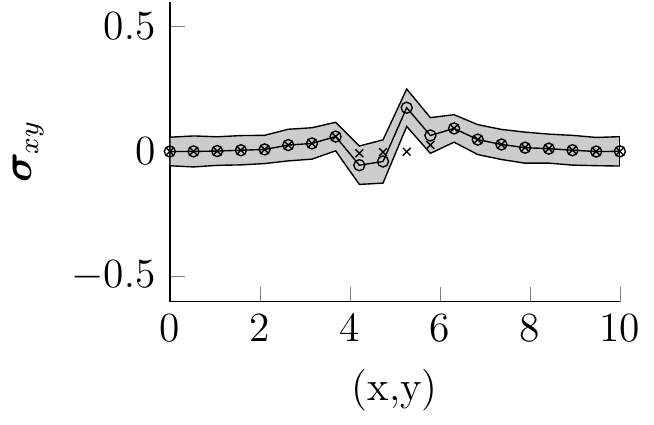} \\ 
  \end{tabular}
  \caption{Comparison of the ground truth values along the diagonal with the inferred posterior means and 95\% credible intervals for displacements and stresses. The results in the left column were obtained using a  banded covariance and the results in the right column using a diagonal covariance model.}\label{fig:2d_inverse_model_error_posterior_solution_diagonals_sgdg_vs_sgbc}
\end{figure}

A comparison between the reference solution  and the posterior means and variances for the stresses and displacements is provided in~\autoref{fig:2d_inverse_model_error_posterior_and_reference}. 
We note that outside of the inclusion, where the constitutive law is found to be adequate, the right material parameters (Young's moduli) are identified. For the elements over which model errors were identified (i.e. small $\bm \lambda$'s) the material parameter values are essentially determined by the smoothing prior. 
Despite the untrustworthiness of the constitutive law in these elements, the model can still correctly identify the true, latent stresses as it can be seen in ~\autoref{fig:2d_inverse_model_error_posterior_solution_diagonals_sgdg_vs_sgbc}, where the posterior means and credible intervals along the diagonal are depicted. For the stresses in the  inclusion elements, it is effectively the (discretized) equations of equilibrium (i.e. the conservation law) that are  the sole source of information, in combination with the stresses from the other elements which, in the presence of a reliable constitutive law, are correctly identified. It is therefore essential to correctly capture these correlations in the approximate posteriors. In the same Figure we also depict the  posterior inference results when a diagonal covariance matrix is used for the approximating densities $q$ and $r$ (section \ref{sec:variational_approx}).
 While the diagonal covariance model offers a much more efficient framework for inference and learning (due to the significantly lower number of optimization parameters), it ignores some of the aforementioned dependencies and can therefore lead to the wrong conclusions.

 \clearpage

\section{Conclusion}
\label{sec_conclusion}

We have presented a  new formulation for model calibration and validation  problems in the context of continuum thermodynamics.
 Rather than using associated forward models as black-boxes, we advocate making use of the governing equations as information sources which  imply constraints for the state variables. We enforce these constraints probabilistically and quantify their validity in a Bayesian fashion. Reliable equations such as conservation laws imply stronger probabilistic constraints, whereas constitutive laws (or phenomenological  closures in general) can be modeled  hierarchically in order to infer their validity from the observational data. 
 The formulation advocated employs  an undirected probabilistic model which   explicitly contains the dependencies between the state variables and can be used in place of the black-box forward model. As a result the solution of, even ill-posed, deterministic forward problems is recast as a well-posed probabilistic inference  task. More importantly, in the context of inverse problems it   leads to an augmented posterior that distinguishes between the different error sources and can quantify model errors in a  interpretable manner that respects known physical invariances.

 One of the challenges associated with the proposed formulation pertains to the dimensionality of the densities involved. While these consists of large products of explicit terms, they involve all the  physical state variables and as such live in much higher dimensions as compared to canonical Bayesian posteriors. We advocate a double-layer Stochastic Variational Inference scheme that scales linearly with the number of optimization parameters and provides closed-form, approximate inference results but much more efficiently than Monte Carlo-based sampling schemes.
We demonstrate the potential of this framework  in the context of biomechanics where the solution of the 
aforementioned issues can  significantly impact  progress in the non-invasive, diagnostic 
capabilities and assist in the development of patient-specific treatment strategies. 

Important extensions in increasing the efficacy of the proposed formulation relate to the representation of the latent variables and adaptivity. In this presentation, we have employed standard finite element basis functions to represent the physical fields. While this enables the use of pre-existing FE machinery, it does not necessarily provide the sparsest representation nor the one that minimizes discretization errors.
We envision that tools from Dictionary Learning/Sparse Coding \cite{olshausen_sparse_1997} could be valuable in this direction. We note also that the infinite dimensional governing equations potentially provide an infinite amount of data (e.g. as $N_e \to \infty$ in \refeq{eq:rese} or $N_c \to \infty$ in \refeq{eq:resc})
 which open the door to very rich and expressive  representations such as those implied by Deep Neural Nets \cite{goodfellow_deep_2016,tripathy_deep_2018,zhu_bayesian_2018}.
 Finally, we envision   that significant scalability improvements can be achieved by  an adaptive scheme which will initiate with a small number of $N_e$ and $N_c$ and
 progressively add weighting functions  (\refeq{eq:rese}) as well as interrogation points for the constitutive
law (\refeq{eq:resc}). This sequential enrichment of the prior/posterior model apart from providing a
natural tempering effect, can be guided by posterior inferences. For example additional weighting
functions of interrogation points $\bx^{(i_c)}$ can be added in areas of the problem domain where the posterior variance of the state variables is largest. Several other metrics of the posterior uncertainty could be used  as drivers of such adaptive schemes which will enable superior allocation of computational
resources in regions that are most informative with regards to the posterior beliefs.

\clearpage

\bibliographystyle{apalike_cust}
\bibliography{paper}

\begin{thebibliography}{}

\bibitem[Arnold et~al., 2010]{ISI:000275756200016}
Arnold, A., Reichling, S., Bruhns, O., and Mosler, J. (2010).
\newblock Efficient computation of the elastography inverse problem by
  combining variational mesh adaption and a clustering technique.
\newblock {\em PHYSICS IN MEDICINE AND BIOLOGY},
\newblock 55(7):2035--2056.

\bibitem[Attias, 2000]{attias_variational_2000}
Attias, H. (2000).
\newblock A {Variational} {Bayesian} {Framework} for {Graphical} {Models}.
\newblock In {\em In {Advances} in {Neural} {Information} {Processing}
  {Systems} 12}, pages 209--215.
\newblock MIT Press.

\bibitem[Barbone et~al., 2010]{ISI:000275699600006}
Barbone, P.~E., Rivas, C.~E., Harari, I., Albocher, U., Oberai, A.~A., and
  Zhang, Y. (2010).
\newblock Adjoint-weighted variational formulation for the direct solution of
  inverse problems of general linear elasticity with full interior data.
\newblock {\em Int. J. Num. Meth. Eng.},
\newblock 81(13):1713--1736.

\bibitem[Bardsley, 2013]{bardsley_gaussian_2013}
Bardsley, J. (2013).
\newblock Gaussian {Markov} random field priors for inverse problems.
\newblock {\em Inverse Problems and Imaging},
\newblock 7(2):397--416.

\bibitem[Bayarri et~al., 2007]{bayarri_framework_2007}
Bayarri, M., Berger, J., Paulo, R., Sacks, J., Cafeo, J., Cavendish, J., Lin,
  C., and Tu, J. (2007).
\newblock A framework for validation of computer models.
\newblock {\em TECHNOMETRICS},
\newblock 49(2):138 -- 154.

\bibitem[Beal, 2003]{beal_variational_2003}
Beal, M.~J. (2003).
\newblock {\em Variational algorithms for approximate {Bayesian} inference}.
\newblock
\newblock University of London United Kingdom.

\bibitem[Berliner et~al., 2008]{berliner_modeling_2008}
Berliner, L., Jezek, K., Cressie, N., Kim, Y., Lam, C., and van~der Veen, C.
  (2008).
\newblock Modeling dynamic controls on ice streams: a {Bayesian} statistical
  approach.
\newblock {\em Journal of Glaciology},
\newblock 54(187):705--714.

\bibitem[Biegler et~al., 2010]{biegler_large-scale_2010}
Biegler, L., Biros, G., Ghattas, O., Heinkenschloss, M., Keyes, D., Mallick,
  B., Tenorio, L., Waanders, B. v.~B., Willcox, K., and Marzouk, Y., editors
  (2010).
\newblock {\em Large-{Scale} {Inverse} {Problems} and {Quantification} of
  {Uncertainty}}.
\newblock Wiley, Chichester, West Sussex,
\newblock 1 edition edition.

\bibitem[Bilionis and Koutsourelakis, 2012]{bilionis_free_2012}
Bilionis, I. and Koutsourelakis, P.~S. (2012).
\newblock Free energy computations by minimization of {Kullback}–{Leibler}
  divergence: {An} efficient adaptive biasing potential method for sparse
  representations.
\newblock {\em Journal of Computational Physics},
\newblock 231(9):3849 -- 3870.

\bibitem[Bilionis and Zabaras, 2014]{bilionis_solution_2014}
Bilionis, I. and Zabaras, N. (2014).
\newblock Solution of inverse problems with limited forward solver evaluations:
  a {Bayesian} perspective.
\newblock {\em Inverse Problems},
\newblock 30(1):015004.

\bibitem[Bishop, 2006]{bishop_pattern_2006}
Bishop, C.~M. (2006).
\newblock {\em Pattern recognition and machine learning}.
\newblock Information science and statistics. Springer,
\newblock New York.

\bibitem[Bishop and Tipping, 2000]{bishop_variational_2000}
Bishop, C.~M. and Tipping, M.~E. (2000).
\newblock Variational relevance vector machines.
\newblock {\em Proceedings of the Sixteenth conference on Uncertainty in
  artificial intelligence},
\newblock (1):46--53.

\bibitem[Blei et~al., 2017]{blei_variational_2017}
Blei, D.~M., Kucukelbir, A., and McAuliffe, J.~D. (2017).
\newblock Variational {Inference}: {A} {Review} for {Statisticians}.
\newblock {\em Journal of the American Statistical Association},
\newblock 112(518):859--877.

\bibitem[Brynjarsdottir and O'Hagan, 2014]{brynjarsdottir_learning_2014}
Brynjarsdottir, J. and O'Hagan, A. (2014).
\newblock Learning about physical parameters: {The} importance of model
  discrepancy.
\newblock {\em Inverse Problems},
\newblock 30(11):114007.

\bibitem[Bui-Thanh and Girolami, 2014]{bui-thanh_solving_2014}
Bui-Thanh, T. and Girolami, M. (2014).
\newblock Solving large-scale {PDE}-constrained {Bayesian} inverse problems
  with {Riemann} manifold {Hamiltonian} {Monte} {Carlo}.
\newblock {\em Inverse Problems},
\newblock 30(11):114014.

\bibitem[Bui-Thanh et~al., 2008]{bui-thanh_model_2008}
Bui-Thanh, T., Willcox, K., and Ghattas, O. (2008).
\newblock Model {Reduction} for {Large}-{Scale} {Systems} with
  {High}-{Dimensional} {Parametric} {Input} {Space}.
\newblock {\em SIAM Journal on Scientific Computing},
\newblock 30(6):3270.

\bibitem[Candes et~al., 2006]{candes_robust_2006}
Candes, E., Romberg, J., and Tao, T. (2006).
\newblock Robust uncertainty principles: {Exact} signal reconstruction from
  highly incomplete frequency information.
\newblock {\em IEEE Trans. Information Theory},
\newblock 52:489--509.

\bibitem[Carvalho et~al., 2009]{carvalho_handling_2009}
Carvalho, C.~M., Polson, N.~G., and Scott, J.~G. (2009).
\newblock Handling sparsity via the horseshoe.
\newblock In {\em International {Conference} on {Artificial} {Intelligence} and
  {Statistics}},
\newblock pages 73--80.

\bibitem[Chkrebtii et~al., 2016]{chkrebtii_bayesian_2016}
Chkrebtii, O.~A., Campbell, D.~A., Calderhead, B., and Girolami, M.~A. (2016).
\newblock Bayesian {Solution} {Uncertainty} {Quantification} for {Differential}
  {Equations}.
\newblock {\em Bayesian Analysis},
\newblock 11(4):1239--1267.

\bibitem[Chopin et~al., 2012]{chopin_free_2012}
Chopin, N., Lelièvre, T., and Stoltz, G. (2012).
\newblock Free {Energy} {Methods} for {Bayesian} {Inference}: {Efficient}
  {Exploration} of {Univariate} {Gaussian} {Mixture} {Posteriors}.
\newblock {\em Statistics and Computing},
\newblock 22(4):897--916.

\bibitem[Cockayne et~al., 2016]{cockayne_probabilistic_2016}
Cockayne, J., Oates, C., Sullivan, T., and Girolami, M. (2016).
\newblock Probabilistic {Numerical} {Methods} for {Partial} {Differential}
  {Equations} and {Bayesian} {Inverse} {Problems}.
\newblock
\newblock {\em arXiv:1605.07811 [cs, math, stat]}.

\bibitem[Cockayne et~al., 2017]{cockayne_bayesian_2017}
Cockayne, J., Oates, C., Sullivan, T., and Girolami, M. (2017).
\newblock Bayesian {Probabilistic} {Numerical} {Methods}.
\newblock
\newblock {\em arXiv:1702.03673 [cs, math, stat]}.

\bibitem[Cover and Thomas, 1991]{cover_elements_1991}
Cover, T. and Thomas, J. (1991).
\newblock {\em Elements of {Information} {Theory}}.
\newblock
\newblock John Wiley \& Sons.

\bibitem[Cui et~al., 2016]{cui_dimension-independent_2016}
Cui, T., Law, K.~J., and Marzouk, Y.~M. (2016).
\newblock Dimension-independent likelihood-informed {MCMC}.
\newblock {\em Journal of Computational Physics},
\newblock 304:109--137.

\bibitem[Cui et~al., 2014]{cui_likelihood-informed_2014}
Cui, T., Martin, J., Marzouk, Y.~M., Solonen, A., and Spantini, A. (2014).
\newblock Likelihood-informed dimension reduction for nonlinear inverse
  problems.
\newblock {\em Inverse Problems},
\newblock 30(11):114015.

\bibitem[Cui et~al., 2015]{cui_scalable_2015}
Cui, T., Marzouk, Y.~M., and Willcox, K.~E. (2015).
\newblock Scalable posterior approximations for large-scale {Bayesian} inverse
  problems via likelihood-informed parameter and state reduction.
\newblock
\newblock {\em arXiv preprint arXiv:1510.06053}.

\bibitem[Cullen et~al., 2013]{cullen_large_2013}
Cullen, M., Freitag, M.~A., Kindermann, S., and Scheichl, R. (2013).
\newblock {\em Large {Scale} {Inverse} {Problems}, {Computational} {Methods}
  and {Applications} in the {Earth} {Sciences}}.
\newblock De Gruyter,
\newblock Berlin, Boston.

\bibitem[Curtis and et~al., 2012]{cur12gen}
Curtis, C. and et~al. (2012).
\newblock The genomic and transcriptomic architecture of 2,000 breast tumours
  reveals novel subgroups.
\newblock
\newblock {\em Nature}.

\bibitem[Del~Moral et~al., 2012]{del_moral_adaptive_2012}
Del~Moral, P., Doucet, A., and Jasra, A. (2012).
\newblock On adaptive resampling strategies for sequential {Monte} {Carlo}
  methods.
\newblock {\em Bernoulli},
\newblock 18(1):252--278.

\bibitem[Della~Pietra et~al., 1997]{della_pietra_inducing_1997}
Della~Pietra, S., Della~Pietra, V., and Lafferty, J. (1997).
\newblock Inducing {Features} of {Random} {Fields}.
\newblock {\em IEEE Transactions on Pattern Analysis and Machine Intelligence},
\newblock 19(4):380--393.

\bibitem[Dempster et~al., 1977]{dempster_maximum_1977}
Dempster, A.~A., Laird, N.~N., and Rubin, D. D.~B. (1977).
\newblock Maximum likelihood from incomplete data via the {EM} algorithm.
\newblock {\em Journal of the Royal Statistical Society Series B
  Methodological},
\newblock 39(1):1--38.

\bibitem[Diaconis, 1988]{diaconis_bayesian_1988}
Diaconis, P. (1988).
\newblock Bayesian numerical analysis.
\newblock {\em Statistical decision theory and related topics IV},
\newblock 1:163--175.

\bibitem[Dodwell et~al., 2015]{dodwell_hierarchical_2015}
Dodwell, T.~J., Ketelsen, C., Scheichl, R., and Teckentrup, A.~L. (2015).
\newblock A {Hierarchical} {Multilevel} {Markov} {Chain} {Monte} {Carlo}
  {Algorithm} with {Applications} to {Uncertainty} {Quantification} in
  {Subsurface} {Flow}.
\newblock {\em SIAM/ASA Journal on Uncertainty Quantification},
\newblock 3(1):1075--1108.

\bibitem[Donoho, 2006]{donoho_compressed_2006}
Donoho, D. (2006).
\newblock Compressed sensing.
\newblock {\em IEEE Transactions Information Theory},
\newblock 52(4):1289--1306.

\bibitem[Doyley et~al., 2006]{doyley_enhancing_2006}
Doyley, M., Srinivasan, S., Dimidenko, E., Soni, N., and Ophir, J. (2006).
\newblock Enhancing the performance of model-based elastography by
  incorporating additional a priori information in the modulus image
  reconstruction process.
\newblock {\em PHYSICS IN MEDICINE AND BIOLOGY},
\newblock 51(1):95--112.

\bibitem[Doyley, 2012]{doyley_model-based_2012}
Doyley, M.~M. (2012).
\newblock Model-based elastography: a survey of approaches to the inverse
  elasticity problem.
\newblock {\em Physics in Medicine and Biology},
\newblock 57(3):R35.

\bibitem[Draper, 1995]{draper_assessment_1995}
Draper, D. (1995).
\newblock Assessment and {Propagation} of {Model} {Uncertainty}.
\newblock {\em Journal of the Royal Statistical Society Series
  B-Methodological},
\newblock 57(1):45--97.

\bibitem[Ellam et~al., 2016]{ellam_bayesian_2016}
Ellam, L., Zabaras, N., and Girolami, M. (2016).
\newblock A {Bayesian} approach to multiscale inverse problems with on-the-fly
  scale determination.
\newblock {\em Journal of Computational Physics},
\newblock 326:115--140.

\bibitem[Finlayson, 1972]{finlayson_method_1972}
Finlayson, B., editor (1972).
\newblock {\em The method of weighted residuals and variational principles,
  with application in fluid mechanics, heat and mass transfer, {Volume} 87}.
\newblock Academic Press,
\newblock New York.

\bibitem[Flath et~al., 2011]{flath_fast_2011}
Flath, H.~P., Wilcox, L.~C., Akçelik, V., Hill, J., van Bloemen~Waanders, B.,
  and Ghattas, O. (2011).
\newblock Fast {Algorithms} for {Bayesian} {Uncertainty} {Quantification} in
  {Large}-{Scale} {Linear} {Inverse} {Problems} {Based} on {Low}-{Rank}
  {Partial} {Hessian} {Approximations}.
\newblock {\em SIAM J. Sci. Comput.},
\newblock 33(1):407--432.

\bibitem[Franck and Koutsourelakis, 2016]{franck_sparse_2016}
Franck, I.~M. and Koutsourelakis, P. (2016).
\newblock Sparse {Variational} {Bayesian} approximations for nonlinear inverse
  problems: {Applications} in nonlinear elastography.
\newblock {\em Computer Methods in Applied Mechanics and Engineering},
\newblock 299:215--244.

\bibitem[Ganne-Carri\'e et~al., 2006]{liver2006}
Ganne-Carri\'e, N., Ziol, M., de~Ledinghen, V., Douvin, C., Marcellin, P.,
  Castera, L., Dhumeaux, D., Trinchet, J., and Beaugrand, M. (2006).
\newblock Accuracy of liver stiffness measurement for the diagnosis of
  cirrhosis in patients with chronic liver diseases.
\newblock {\em Hepatology},
\newblock 44(6):1511--1517.

\bibitem[Girolami and Calderhead, 2011]{girolami_riemann_2011}
Girolami, M. and Calderhead, B. (2011).
\newblock Riemann manifold {Langevin} and {Hamiltonian} {Monte} {Carlo}
  methods.
\newblock {\em Journal of the Royal Statistical Society: Series B (Statistical
  Methodology)},
\newblock 73(2):123--214.

\bibitem[Gockenbach and Khan, 2005]{ISI:000240849100006}
Gockenbach, M.~S. and Khan, A.~A. (2005).
\newblock Identification of lame parameters in linear elasticity: A fixed point
  approach.
\newblock {\em JOURNAL OF INDUSTRIAL AND MANAGEMENT OPTIMIZATION},
\newblock 1(4):487--497.

\bibitem[Goodfellow et~al., 2016]{goodfellow_deep_2016}
Goodfellow, I.~J., Bengio, Y., and Courville, A.~C. (2016).
\newblock {\em Deep {Learning}}.
\newblock Adaptive computation and machine learning.
\newblock MIT Press.

\bibitem[Green et~al., 2015]{green_bayesian_2015}
Green, P.~J., Latuszynski, K., Pereyra, M., and Robert, C.~P. (2015).
\newblock Bayesian computation: a summary of the current state, and samples
  backwards and forwards.
\newblock {\em Statistics and Computing},
\newblock 25(4):835--862.

\bibitem[Hennig and Hauberg, 2014]{hennig_probabilistic_2014}
Hennig, P. and Hauberg, S. (2014).
\newblock Probabilistic {Solutions} to {Differential} {Equations} and their
  {Application} to {Riemannian} {Statistics}.
\newblock In {\em Proc. of the 17th int. {Conf}. on {Artificial} {Intelligence}
  and {Statistics} ({AISTATS})}, volume~33.
\newblock JMLR, W\&CP.

\bibitem[Hennig et~al., 2015]{hennig_probabilistic_2015}
Hennig, P., Osborne, M.~A., and Girolami, M. (2015).
\newblock Probabilistic numerics and uncertainty in computations.
\newblock {\em Proceedings. Mathematical, physical, and engineering sciences /
  the Royal Society},
\newblock 471(2179):20150142.

\bibitem[Higdon et~al., 2008]{higdon_computer_2008}
Higdon, D., Gattiker, J., Williams, B., and Rightley, M. (2008).
\newblock Computer model calibration using high-dimensional output.
\newblock {\em Journal of the American Statistical Association},
\newblock 103(482):570--583.

\bibitem[Higdon et~al., 2004]{higdon_combining_2004}
Higdon, D., Kennedy, M., Cavendish, J., Cafeo, J., and Ryne, R. (2004).
\newblock Combining {Field} {Data} and {Computer} {Simulations} for
  {Calibration} and {Prediction}.
\newblock {\em SIAM Journal on Scientific Computing},
\newblock 26(2):448--466.

\bibitem[Hoffman et~al., 2013]{hoffman_stochastic_2013}
Hoffman, M.~D., Blei, D.~M., Wang, C., and Paisley, J. (2013).
\newblock Stochastic variational inference.
\newblock {\em The Journal of Machine Learning Research},
\newblock 14(1):1303--1347.

\bibitem[Holloman et~al., 2006]{holloman_multi-resolution_2006}
Holloman, C., Lee, H., and Higdon, D. (2006).
\newblock Multi-resolution {Genetic} {Algorithms} and {Markov} {Chain} {Monte}
  {Carlo}.
\newblock {\em Journal of Computational and Graphical Statistics},
\newblock pages 861--879.

\bibitem[Hughes, 2000]{hughes_finite_2000}
Hughes, T. J.~R. (2000).
\newblock {\em The {Finite} {Element} {Method}—{Linear} {Static} and
  {Dynamic} {Finite} {Element} {Analysis}}.
\newblock
\newblock Dover.

\bibitem[Ishwaran and Rao, 2005]{ishwaran_spike_2005}
Ishwaran, H. and Rao, J.~S. (2005).
\newblock Spike and slab variable selection: {Frequentist} and {Bayesian}
  strategies.
\newblock {\em The Annals of Statistics},
\newblock 33(2):730--773.

\bibitem[Jordan et~al., 1999]{jordan_introduction_1999}
Jordan, M.~I., Ghahramani, Z., Jaakkola, T.~S., and Saul, L.~K. (1999).
\newblock An {Introduction} to {Variational} {Methods} for {Graphical}
  {Models}.
\newblock {\em Mach. Learn.},
\newblock 37(2):183--233.

\bibitem[Kass and Raftery, 1995]{kass_bayes_1995}
Kass, R.~E. and Raftery, A.~E. (1995).
\newblock Bayes {Factors}.
\newblock {\em Journal of the American Statistical Association},
\newblock 90(430):773--795.

\bibitem[Kennedy and O'Hagan, 2001]{kennedy_bayesian_2001}
Kennedy, M.~C. and O'Hagan, A. (2001).
\newblock Bayesian calibration of computer models.
\newblock {\em Journal Of The Royal Statistical Society Series B-Statistical
  Methodology},
\newblock 63:425--450.

\bibitem[Khalil et~al., 2005]{khalil_tissue_2005}
Khalil, A.~S., Chan, R.~C., Chau, A.~H., Bouma, B.~E., and Mofrad, M. R.~K.
  (2005).
\newblock Tissue elasticity estimation with optical coherence elastography:
  toward mechanical characterization of in vivo soft tissue.
\newblock {\em Annals of Biomedical Engineering},
\newblock 33(11):1631--1639.

\bibitem[Kingma and Ba, 2014]{kingma_adam:_2014}
Kingma, D. and Ba, J. (2014).
\newblock Adam: {A} {Method} for {Stochastic} {Optimization}.
\newblock {\em arXiv:1412.6980 [cs]},
\newblock pages 1--15.

\bibitem[Kingma and Welling, 2013]{kingma_auto-encoding_2013}
Kingma, D.~P. and Welling, M. (2013).
\newblock Auto-{Encoding} {Variational} {Bayes}.
\newblock
\newblock (Ml):1--14.

\bibitem[Koller and Friedman, 2009]{koller_probabilistic_2009}
Koller, D. and Friedman, N. (2009).
\newblock {\em Probabilistic graphical models: principles and techniques}.
\newblock Adaptive computation and machine learning. MIT Press,
\newblock Cambridge, MA.

\bibitem[Koutsourelakis, 2009]{koutsourelakis_multi-resolution_2009}
Koutsourelakis, P. (2009).
\newblock A multi-resolution, non-parametric, {Bayesian} framework for
  identification of spatially-varying model parameters.
\newblock {\em Journal of Computational Physics},
\newblock 228(17):6184--6211.

\bibitem[Koutsourelakis, 2012]{koutsourelakis_novel_2012}
Koutsourelakis, P.-S. (2012).
\newblock A novel {Bayesian} strategy for the identification of spatially
  varying material properties and model validation: an application to static
  elastography.
\newblock {\em International Journal for Numerical Methods in Engineering},
\newblock 91(3):249--268.

\bibitem[Lan et~al., 2016]{lan_emulation_2016}
Lan, S., Bui-Thanh, T., Christie, M., and Girolami, M. (2016).
\newblock Emulation of higher-order tensors in manifold {Monte} {Carlo} methods
  for {Bayesian} {Inverse} {Problems}.
\newblock {\em Journal of Computational Physics},
\newblock 308:81--101.

\bibitem[Lee et~al., 2002]{lee_markov_2002}
Lee, H., Higdon, D., Bi, Z., Ferreira, M., and West, M. (2002).
\newblock Markov random field models for high-dimensional parameters in
  simulations of fluid flow in porous media.
\newblock {\em TECHNOMETRICS},
\newblock 44(3):230 -- 241.

\bibitem[Lewicki and Sejnowski, 2000]{lewicki_learning_2000}
Lewicki, M. and Sejnowski, T.~J. (2000).
\newblock Learning overcomplete representations.
\newblock {\em Neural Computation},
\newblock 12(2):337--365.

\bibitem[MacKay, 2003]{mackay_information_2003}
MacKay, D.~J. (2003).
\newblock {\em Information theory, inference and learning algorithms}.
\newblock
\newblock Cambridge university press.

\bibitem[Martin et~al., 2012]{martin_stochastic_2012}
Martin, J., Wilcox, L.~C., Burstedde, C., and Ghattas, O. (2012).
\newblock A stochastic {Newton} {MCMC} method for large-scale statistical
  inverse problems with application to seismic inversion.
\newblock {\em SIAM Journal on Scientific Computing},
\newblock 34(3):A1460--A1487.

\bibitem[Marzouk et~al., 2007]{marzouk_stochastic_2007}
Marzouk, Y.~M., Najm, H.~N., and Rahn, L.~A. (2007).
\newblock Stochastic spectral methods for efficient {Bayesian} solution of
  inverse problems.
\newblock {\em J. Comput. Phys},
\newblock 224(2):560--586.

\bibitem[Mattingly et~al., 2012]{mattingly_diffusion_2012}
Mattingly, J.~C., Pillai, N.~S., and Stuart, A.~M. (2012).
\newblock Diffusion limits of the random walk {Metropolis} algorithm in high
  dimensions.
\newblock {\em The Annals of Applied Probability},
\newblock 22(3):881--930.

\bibitem[Moral et~al., 2006]{moral_sequential_2006}
Moral, P.~D., Doucet, A., and Jasra, A. (2006).
\newblock Sequential {Monte} {Carlo} for {Bayesian} computation (with
  discussion).
\newblock
\newblock {\em In Bayesian Statistics 8. Oxford University Press}.

\bibitem[Moser and Oliver, 2015]{moser_validation_2015}
Moser, R.~D. and Oliver, T.~A. (2015).
\newblock Validation of {Physical} {Models} in the {Presence} of {Uncertainty}.
\newblock In Ghanem, R., Higdon, D., and Owhadi, H., editors, {\em Handbook of
  {Uncertainty} {Quantification}}, pages 1--28.
\newblock Springer International Publishing.

\bibitem[Murray and Ghahramani, 2004]{murray_bayesian_2004}
Murray, I. and Ghahramani, Z. (2004).
\newblock
\newblock Bayesian {Learning} in {Undirected} {Graphical} {Models}:
  {Approximate} {MCMC} {Methods}.

\bibitem[Muthupillai et~al., 1995]{mut95mag}
Muthupillai, R., Lomas, D., Rossman, P., Greenleaf, J., Manduca, A., and Ehman,
  R. (1995).
\newblock Magnetic resonance elastography by direct visualization of
  propagating acoustic strain waves.
\newblock {\em Science},
\newblock 269(5232):1854--1857.

\bibitem[Neal and Hinton, 1998]{neal_view_1998}
Neal, R. and Hinton, G.~E. (1998).
\newblock A {View} {Of} {The} {Em} {Algorithm} {That} {Justifies}
  {Incremental}, {Sparse}, {And} {Other} {Variants}.
\newblock In {\em Learning in {Graphical} {Models}}, pages 355--368.
\newblock Kluwer Academic Publishers.

\bibitem[Oberai et~al., 2004a]{ISI:000223500200013}
Oberai, A., Gokhale, N., Doyley, M., and Bamber, J. (2004a).
\newblock Evaluation of the adjoint equation based algorithm for elasticity
  imaging.
\newblock {\em PHYSICS IN MEDICINE AND BIOLOGY},
\newblock 49(13):2955--2974.

\bibitem[Oberai et~al., 2004b]{oberai_evaluation_2004}
Oberai, A., Gokhale, N., Doyley, M., and Bamber, J. (2004b).
\newblock Evaluation of the adjoint equation based algorithm for elasticity
  imaging.
\newblock {\em PHYSICS IN MEDICINE AND BIOLOGY},
\newblock 49(13):2955--2974.

\bibitem[Oberai et~al., 2009]{ISI:000263259100006}
Oberai, A.~A., Gokhale, N.~H., Goenezen, S., Barbone, P.~E., Hall, T.~J.,
  Sommer, A.~M., and Jiang, J. (2009).
\newblock Linear and nonlinear elasticity imaging of soft tissue in vivo:
  demonstration of feasibility.
\newblock {\em PHYSICS IN MEDICINE AND BIOLOGY},
\newblock 54(5):1191--1207.

\bibitem[O'Hagan et~al., 1999]{ohagan_uncertainty_1999}
O'Hagan, A., Kennedy, M., and Oakley, J.~E. (1999).
\newblock Uncertainty analysis and other inference tools for complex computer
  codes (with discussion).
\newblock In Bernardo, J. and al, e., editors, {\em In {Bayesian} {Statistics}
  6}.
\newblock Oxford University Press.

\bibitem[Olshausen and Field, 1997]{olshausen_sparse_1997}
Olshausen, B.~A. and Field, D.~J. (1997).
\newblock Sparse coding with an overcomplete basis set: {A} strategy employed
  by {V}1?
\newblock {\em Vision Research},
\newblock 37(23):3311--3325.

\bibitem[Olson and Throne, 2010]{ISI:000280774700004}
Olson, L.~G. and Throne, R.~D. (2010).
\newblock Numerical simulation of an inverse method for tumour size and
  location estimation.
\newblock {\em Inv. Prob. Sc. Eng.},
\newblock 18(6):813--834.

\bibitem[Ophir et~al., 1991]{Ophir:1991}
Ophir, J., Cespedes, I., Ponnekanti, H., Yazdi, Y., and Li, X. (1991).
\newblock Elastography - a quantitative method for imaging the elasticity of
  biological tissues.
\newblock {\em ULTRASONIC IMAGING},
\newblock 13(2):111 -- 134.

\bibitem[Paisley et~al., 2012]{paisley_variational_2012}
Paisley, J., Blei, D., and Jordan, M. (2012).
\newblock Variational {Bayesian} {Inference} with {Stochastic} {Search}.
\newblock {\em Icml},
\newblock (2000):1367--1374.

\bibitem[Petra et~al., 2014]{petra_computational_2014}
Petra, N., Martin, J., Stadler, G., and Ghattas, O. (2014).
\newblock A {Computational} {Framework} for {Infinite}-{Dimensional} {Bayesian}
  {Inverse} {Problems}, {Part} {II}: {Stochastic} {Newton} {MCMC} with
  {Application} to {Ice} {Sheet} {Flow} {Inverse} {Problems}.
\newblock {\em SIAM Journal on Scientific Computing},
\newblock 36(4):A1525--A1555.

\bibitem[Pillai et~al., 2012]{pillai_optimal_2012}
Pillai, N.~S., Stuart, A.~M., and Thiéry, A.~H. (2012).
\newblock Optimal scaling and diffusion limits for the {Langevin} algorithm in
  high dimensions.
\newblock {\em The Annals of Applied Probability},
\newblock 22(6):2320--2356.

\bibitem[Rezende et~al., 2014]{rezende_stochastic_2014}
Rezende, D.~J., Mohamed, S., and Wierstra, D. (2014).
\newblock Stochastic backpropagation and approximate inference in deep
  generative models.
\newblock {\em Proceedings of The 31st ldots},
\newblock 32:1278--1286.

\bibitem[Robbins and Monro, 1951]{robbins_stochastic_1951}
Robbins, H. and Monro, S. (1951).
\newblock A {Stochastic} {Approximation} {Method}.
\newblock {\em The Annals of Mathematical Statistics},
\newblock 22(3):400--407.

\bibitem[Roberts and Rosenthal, 1998]{roberts_optimal_1998}
Roberts, G.~O. and Rosenthal, J.~S. (1998).
\newblock Optimal scaling of discrete approximations to {Langevin} diffusions.
\newblock {\em Journal of the Royal Statistical Society: Series B (Statistical
  Methodology)},
\newblock 60(1):255--268.

\bibitem[Roberts and Tweedie, 1996]{roberts_exponential_1996}
Roberts, G.~O. and Tweedie, R.~L. (1996).
\newblock Exponential convergence of {Langevin} diffusions and their discrete
  approximations.
\newblock {\em Bernoulli},
\newblock 2:341--364.

\bibitem[Rosic et~al., 2012]{rosic_sampling-free_2012}
Rosic, B.~V., Litvinenko, A., Pajonk, O., and Matthies, H.~G. (2012).
\newblock Sampling-free linear {Bayesian} update of polynomial chaos
  representations.
\newblock {\em J. Comput. Physics},
\newblock 231(17):5761--5787.

\bibitem[Sargsyan et~al., 2015]{sargsyan_statistical_2015}
Sargsyan, K., Najm, H.~N., and Ghanem, R. (2015).
\newblock On the {Statistical} {Calibration} of {Physical} {Models}.
\newblock {\em International Journal of Chemical Kinetics},
\newblock 47(4):246--276.

\bibitem[Sarvazyan and Hall, 2011]{sarvazyan_elasticity_2011}
Sarvazyan, A. and Hall, T., editors (2011).
\newblock {\em Elasticity {Imaging} {Part} {I} \& {II}}, volume 7,8.
\newblock
\newblock Cur. Med. Imag. Rev.

\bibitem[Silva and Ghahramani, 2006]{silva_bayesian_2006}
Silva, R. and Ghahramani, Z. (2006).
\newblock Bayesian {Inference} for \{{G}\}aussian {Mixed} {Graph} {Models}.
\newblock
\newblock {\em Uai}.

\bibitem[Silva and Ghahramani, 2009]{silva_hidden_2009}
Silva, R. and Ghahramani, Z. (2009).
\newblock The {Hidden} {Life} of {Latent} {Variables} : {Bayesian} {Learning}
  with {Mixed} {Graph} {Models}.
\newblock {\em Journal of Machine Learning Research},
\newblock 10:1187--1238.

\bibitem[Spantini et~al., 2015]{spantini_optimal_2015}
Spantini, A., Solonen, A., Cui, T., Martin, J., Tenorio, L., and Marzouk, Y.
  (2015).
\newblock Optimal {Low}-rank {Approximations} of {Bayesian} {Linear} {Inverse}
  {Problems}.
\newblock {\em SIAM Journal on Scientific Computing},
\newblock 37(6):A2451--A2487.

\bibitem[Strong and Oakley, 2014]{strong_when_2014}
Strong, M. and Oakley, J. (2014).
\newblock When {Is} a {Model} {Good} {Enough}? {Deriving} the {Expected}
  {Value} of {Model} {Improvement} via {Specifying} {Internal} {Model}
  {Discrepancies}.
\newblock {\em SIAM/ASA Journal on Uncertainty Quantification},
\newblock 2(1):106--125.

\bibitem[Sun et~al., 2009]{sun_elastography_2009}
Sun, L.~Z., Wang, Z.~G., Liu, Y., and Wang, G. (2009).
\newblock Elastography method for reconstruction of nonlinear breast tissue
  properties.
\newblock {\em International Journal of Biomedical Imaging},
\newblock 2009.

\bibitem[Titsias and Lázaro-Gredilla, 2014]{titsias_doubly_2014}
Titsias, M.~K. and Lázaro-Gredilla, M. (2014).
\newblock Doubly stochastic variational {Bayes} for non-conjugate inference.
\newblock
\newblock ICML.

\bibitem[Tripathy and Bilionis, 2018]{tripathy_deep_2018}
Tripathy, R. and Bilionis, I. (2018).
\newblock Deep {UQ}: {Learning} deep neural network surrogate models for high
  dimensional uncertainty quantification.
\newblock
\newblock {\em arXiv:1802.00850 [physics, stat]}.

\bibitem[Wainwright and Jordan, 2008]{wainwright_graphical_2008}
Wainwright, M. and Jordan, M. (2008).
\newblock Graphical models, exponential families, and variational inference.
\newblock In {\em Foundations and {Trends} in {Machine} {Learning}}, volume~1,
\newblock pages 1--305.

\bibitem[Wainwright et~al., 2002]{wainwright_new_2002}
Wainwright, M.~J., Jaakkola, T.~S., and Willsky, A.~S. (2002).
\newblock A new class of upper bounds on the log partition function.
\newblock In {\em Proceedings of the {Eighteenth} conference on {Uncertainty}
  in artificial intelligence}, pages 536--543.
\newblock Morgan Kaufmann Publishers Inc.

\bibitem[Wipf and Nagarajan, 2008]{wipf_new_2008}
Wipf, D.~P. and Nagarajan, S.~S. (2008).
\newblock A {New} {View} of {Automatic} {Relevance} {Determination}.
\newblock In Platt, J.~C., Koller, D., Singer, Y., and Roweis, S.~T., editors,
  {\em Advances in {Neural} {Information} {Processing} {Systems} 20}, pages
  1625--1632.
\newblock Curran Associates, Inc.

\bibitem[Zhu and Zabaras, 2018]{zhu_bayesian_2018}
Zhu, Y. and Zabaras, N. (2018).
\newblock Bayesian {Deep} {Convolutional} {Encoder}-{Decoder} {Networks} for
  {Surrogate} {Modeling} and {Uncertainty} {Quantification}.
\newblock
\newblock {\em arXiv:1801.06879 [physics, stat]}.

\end{thebibliography}
\clearpage
\appendix

\section{Appendix} 
This Appendix contains details for the derivatives of the ELBO objectives formulated in section \ref{sec:variational_approx} for the solution of forward and inverse problems under the framework proposed.

\subsection{Derivatives of $\mathcal{L}_{{for}}$}
\label{sec:appA}
We recall from \refeq{eq:elbof} that:
\be
\begin{array}{ll}
 \mathcal{L}_{for}(\bm \phi)  & =  \mathbb{E}_{q(\by;\bm\phi)} \left[ \log \psi_1(\bs{\sigma})  \right]+ \eqq \left[ \log \psi_2(\bs{\sigma}, \bs{u}_i, \bs{C}=\bs{c} ; \bs{\lambda})\right]  \\
  &~~ + \eqq \left[ \log \psi_4(\bs{\sigma}) \right] + \eqq \left[ \log \psi_5(\bs{u}_i)  \right] +\mathbb{H} \left[ q(\bm{y};\bm\phi) \right] 
\end{array}
\ee
where $ q(\bm{y};\bm\phi) =\mathcal{N}(\by | \bm 0, \bm S= \bm L \bm L^T)$ and $\bm \phi=\{ \bm \mu, \bm L\}$. 
As discussed in section \ref{sec:optim} derivatives with respect to $\bm \phi$ of all the terms involving expectations with respect to $q$ are computed with Monte Carlo using {\em reparameterization trick} (see \refeq{eq:reparam1} and \refeq{eq:reparam2}).
As for the entropy term (up to a constant):
\be
\mathbb{H} \left[ q(\bm{y};\bm\phi) \right] =-\frac{1}{2} \log |\bm S| -\frac{1}{2}dim(\by)=-\log |\bm L| 
\label{eq:ent1}
\ee
Hence:
\be
\nabla_{\bm u} \mathbb{H} \left[ q(\bm{y};\bm\phi) \right]=0, \qquad \nabla_{\bm L} \mathbb{H} \left[ q(\bm{y};\bm\phi) \right]=-\bm L^{-1}
\label{eq:ent2}
\ee

\subsection{Derivatives of $\mathcal{L}_{inv}$}
\label{sec:appB}

We recall from \refeq{eq:elboinv} that:
\be
\begin{array}{ll}
 \mathcal{L}_{inv} & =\eqq \left[\log p(\bs{u}_{obs} | \bs{U}_i) \right]+\mathbb{H} \left[ q(\bm{y};\bm\phi) \right]  \\
   & ~~+ \eqq \left[ \log  \pi_{\bm\lambda} (\bs{\sigma}, \bs{c}, \bs{u}_i, \bs{U}_b=\bs{u}_0 ) \right] - \hat{\mathcal{F}}_{inv}(r(\by;\bm\xi), ~\bs{\lambda}) \\
    & ~~+\log p(\bs{\lambda}) 
\end{array}
\ee
where $\hat{\mathcal{F}}_{inv}(r(\by;\bm\xi), ~\bs{\lambda})$ is given in \refeq{eq:elboprior} as:
\be
\begin{array}{ll}
 \hat{\mathcal{F}}_{inv}(r(\by;\bm\xi), ~\bs{\lambda}) & = \err \left[ \log \pi_{\bm\lambda} (\by, \bs{U}_b=\bs{u}_0 ) \right]+\mathbb{H} \left[ r(\bm{y};\bm\xi) \right]  \\
\end{array}
\ee
Both $q(\bm{y};\bm\phi)$ and  $r(\by;\bm\xi), ~\bs{\lambda})$ are multivariate Gaussian with banded covariance represented by $\bm \phi$ and $\bm \xi$ respectively.
Derivatives of terms involving expectations with respect to $q$ or $r$ are needed in the E-step and are computed with Monte Carlo and the reparametrization trick. The entropy terms and their derivatives can be computed as in \refeq{eq:ent1} and \refeq{eq:ent2}.

For the M-step,  derivatives of the objective $\mathcal{L}_{inv}$  with respect to $\bm \lambda$ are needed. We note that these consist of three terms:
\be
\begin{array}{ll}
 \cfrac{\pa \mathcal{L}_{inv}}{\pa \bs{\lambda}} & = \eqq \left[ \cfrac{\pa \log \pi_{\bm\lambda} (\bs{\sigma}, \bs{c}, \bs{u}_i, \bs{U}_b=\bs{u}_0 ) }{\pa \bs{\lambda} }  \right] - \cfrac{\pa \hat{\mathcal{F}}_{inv}(r(\by;\bm\xi), ~\bs{\lambda}) }{\pa \bs{\lambda} } -\cfrac{\pa   \log p(\bs{\lambda})}{\pa \bs{\lambda}} \\
 & = \eqq \left[ \cfrac{\pa \log \pi_{\bm\lambda} (\bs{\sigma}, \bs{c}, \bs{u}_i, \bs{U}_b=\bs{u}_0 ) }{\pa \bs{\lambda} }  \right]  \\
 & ~~-\err \left[ \cfrac{\pa \log \pi_{\bm\lambda} (\bs{\sigma}, \bs{c}, \bs{u}_i, \bs{U}_b=\bs{u}_0 ) }{\pa \bs{\lambda} }  \right] \quad \textrm{(from \refeq{eq:elboprior})} \\
 & ~~+\cfrac{ \pa  \log p(\bs{\lambda})}{\pa \bs{\lambda}}
\end{array}
\label{eq:dlambda1}
\ee
From \refeq{eq:pi} and \refeq{eq:psi2}, we have:
\be
\begin{array}{ll}
 \cfrac{\pa \log \pi_{\bm\lambda} (\bs{\sigma}, \bs{c}, \bs{u}_i, \bs{U}_b=\bs{u}_0 ) }{\pa \lambda_e }  & = -\frac{1}{2} || \bs{\sigma}_e - \bs{c}_e \bs{B}_e \bs{u}_e||^2
\end{array}
\ee
which leads to (in combination with \refeq{eq:ard}) to:
\be
\begin{array}{ll}
 \cfrac{\pa \mathcal{L}_{inv}}{\pa \lambda_e} & = -\frac{1}{2}  \eqq \left[  || \bs{\sigma}_e - \bs{c}_e \bs{B}_e \bs{u}_e||^2 \right] +\frac{1}{2}  \err \left[  || \bs{\sigma}_e - \bs{c}_e \bs{B}_e \bs{u}_e||^2 \right] \\
 & ~~+\underbrace{\frac{\alpha_0-1}{\lambda_e}-\beta_0}_{\frac{ \pa  \log p(\bs{\lambda})}{\pa \lambda_e} }
 \end{array}
\label{eq:dlambda1}
\ee
The first two terms are computed with Monte Carlo and the reparametrization trick.

\subsection{Parameter update using Adam}
\label{sec:adam}

Parameter updates for stochastic optimization are performed as
\begin{align}\label{eq:param_update_phi}
  \bm\phi^{(k+1)} &= \bm\phi^{(k)} + \rho_{\bm\phi}^{(k)} \bm\nabla_{\bm\phi} \mathcal{L}(\bm\phi^{(k)},\bm\xi^{(k)},\bm\lambda^{(k)}), \\ \label{eq:param_update_xi}
  \bm\xi^{(k+1)} &= \bm\xi^{(k)} - \rho_{\bm\xi}^{(k)} \bm\nabla_{\bm\xi} \mathcal{L}(\bm\phi^{(k)},\bm\xi^{(k)},\bm\lambda^{(k)}), \\ \label{eq:param_update_lambda}
  \bm\lambda^{(k+1)} &= \bm\lambda^{(k)} + \rho_{\bm\lambda}^{(k)} \bm\nabla_{\bm\lambda} \mathcal{L}(\bm\phi^{(k)},\bm\xi^{(k)},\bm\lambda^{(k)}),
\end{align}
where $k \in \mathbb{N}_0$ and $\rho_{\bm\xi}^{(k)}$, $\rho_{\bm\phi}^{(k)}$, $\rho_{\bm\lambda}^{(k)}$ are the respective step sizes at iteration $k$. According to the Robbins-Monro criteria~\citep{robbins_stochastic_1951}, these update rules guarantee convergence to a local optimum, if the step size sequences satisfy the conditions
\begin{align}
  \sum_{k=0}^\infty \rho^{(k)} = \infty \ \ \ \mathrm{and} \ \ \ \sum_{k=0}^\infty \rho^{(k)^2} < \infty.
\end{align}

For our purposes, we employ the update scheme ADAM proposed by~\citep{kingma_adam:_2014}. It features an adaptive moment estimation for faster convergence by calculating exponentially decaying averages of the first and second moments of the gradient components $g_j$
\begin{align}
  m_j^{(k)} &= \beta_1 m_j^{(k-1)} + \left( 1-\beta_1 \right) g_j^{(k)},\\
  v_j^{(k)} &= \beta_2 v_j^{(k-1)} + \left( 1-\beta_2 \right) g_j^{(k)^2},
\end{align}
with hyperparameters $\beta_1$, $\beta_2$ and the moment estimates initialized to $m_j^{(0)} = 0$ and $v_j^{(0)} = 0$, respectively. Bias correction is applied as
\begin{align}
  \hat{m}_j^{(k)} = \frac{m_j^{(k)}}{1-\beta_1^k} \ \ \ \mathrm{and} \ \ \ \hat{v}_j^{(k)} = \frac{v_j^{(k)}}{1-\beta_2^k},
\end{align}
resulting in the component-wise parameter update rule
\begin{align}
  \phi_j^{(k+1)} = \phi_j^{(k)} + \eta \frac{\hat{m}_j^{(k)}}{\sqrt{\hat{v}_j^{(k)}}+\varepsilon}.
\end{align}
Hyperparameters need to be chosen problem dependent. We set them to $\eta = 0.001$, $\beta_1 = 0.6$, $\beta_2 = 0.6$ and $\varepsilon = 1e^{-8}$ in all our experiments.

\end{document}